\documentclass[fleqn,usenatbib]{mnras}

\usepackage{newtxtext,newtxmath}
\usepackage{url}
\usepackage{soul}
\usepackage[normalem]{ulem}

\usepackage[T1]{fontenc}

\DeclareRobustCommand{\VAN}[3]{#2}
\let\VANthebibliography\thebibliography
\def\thebibliography{\DeclareRobustCommand{\VAN}[3]{##3}\VANthebibliography}

\usepackage{graphicx}	
\usepackage{amsmath}	
\usepackage{booktabs}

\title[GVILC: interferometric foreground mitigation]{Generalised Visibility ILC: a new foreground mitigation strategy for interferometric HI intensity mapping in the low-redshift Universe}

\author[a,b]{M. Ruiz-Granda,}
\author[c,d]{A. Mazumder,}
\author[c]{L. Wolz,}
\author[a]{M. Remazeilles,}
\author[a,b]{\\ D. Herranz,}
\author[a]{P. Vielva}

\author[M. Ruiz-Granda et al.]{
M. Ruiz-Granda,$^{1,2,3}$\thanks{E-mail: ruizm@ifca.es}
A. Mazumder,$^{3,4}$
L. Wolz,$^{3}$
M. Remazeilles,$^{1}$
D. Herranz$^{1,2}$
and P. Vielva$^{1}$
\\
$^{1}$Instituto de F\'isica de Cantabria (CSIC-Universidad de Cantabria), Avda. de los Castros s/n, E-39005 Santander, Spain\\
$^{2}$Dpto. de F\'isica Moderna, Universidad de Cantabria, Avda. los Castros s/n, E-39005 Santander, Spain\\
$^{3}$Jodrell Bank Centre for Astrophysics, Department of Physics and Astronomy, The University of Manchester, Manchester M13 9PL, UK\\
$^{4}$State Key Laboratory of Radio Astronomy and Technology, Shanghai Astronomical Observatory, CAS, 80 Nandan Road, Shanghai 200030, China\\
}

\date{Accepted XXX. Received YYY; in original form ZZZ}

\pubyear{\the\year{}}

\begin{document}
\label{firstpage}
\pagerange{\pageref{firstpage}--\pageref{lastpage}}
\maketitle

\begin{abstract}
HI intensity mapping is a powerful probe of the large-scale matter distribution and of the underlying astrophysics and cosmology. However, its application is limited by the challenge of separating the faint neutral hydrogen (HI) signal from much brighter foreground emission. We present a new non-parametric foreground-cleaning method in visibility space for low-redshift HI intensity mapping and apply it to interferometric MeerKAT and SKA-Mid-like simulations. Our newly-developed Generalised Visibility Internal Linear Combination (GVILC) method combines Principal Component Analysis (PCA) for foreground subspace identification, informed by a prior HI signal-plus-noise covariance matrix, with a multi-dimensional ILC in visibility space to suppress foregrounds and recover the HI signal with reduced contamination. The number of modes to remove in the PCA is calculated using a data-driven statistical approach, leading to significant foreground residual suppression with limited HI signal loss. Moreover, GVILC in combination with foreground avoidance can prevent most of the signal loss, while simultaneously recovering modes that were initially contaminated by foregrounds. The mean squared error statistic shows that for both MeerKAT and SKA-Mid-like simulations, GVILC outperforms the foreground avoidance only strategy. We find that including a more realistic HI model with line broadening and spatial correlations between HI and continuum sources leads to increased signal loss after foreground removal, without affecting the number of modes removed by GVILC. This paper demonstrates that GVILC is an effective method to clean foreground contamination in HI intensity mapping with radio interferometric data, enabling the recovery of cosmological information from previously inaccessible regions of Fourier space.
\end{abstract}




\section{Introduction}
Mapping the matter distribution in the Universe is one of the main objectives of observational cosmology. It has traditionally been done with a wide variety of dark matter tracers, from galaxy surveys via galaxy clustering~\citep[see e.g.][]{eBOSS_2021MNRAS, DESI_clustering_2025JCAP} and weak lensing~\citep[see e.g.][]{KIDS100_2021AA, DESY6_2026arXiv} to Cosmic Microwave Background (CMB) probes such as CMB lensing~\citep[see e.g.][]{Planck_lensing_2022JCAP, ACT_lensing_2024ApJ}. An alternative method is to use the most abundant element of the Universe, hydrogen, as a tracer. The hyperfine transition of neutral hydrogen (HI) occurs at wavelengths around 21 cm (or equivalently at 1420 MHz), which is redshifted to lower frequencies due to the expansion of the Universe. This emission can be observed using radio telescopes, providing a tomographic view of the Universe. However, detecting HI emission from individual galaxies is challenging due to the faintness of the signal, as well as limitations in angular resolution and instrumental sensitivity, which have historically restricted HI galaxy surveys to the local Universe \citep{Jones_2018MNRAS}. HI Intensity Mapping (IM) overcomes this problem by measuring the integrated emission of unresolved HI sources, without the need to resolve the HI emission from single galaxies, allowing for an efficient measurement of the distribution of matter in the Universe \citep[e.g.][]{Battye_2004MNRAS, Chang_2008PhRvL, Kovetz_2017}.

Neutral hydrogen is present from the last scattering surface at redshift $z\approx 1100$ to the present day. It is one of the few luminous sources during the Cosmic Dark Ages ($25\lesssim z\lesssim 1100$), when no galaxies and stars were formed yet, opening a window to the physics of this epoch. At the Cosmic Dawn, around redshift $z\approx 25$, the first stars and galaxies started forming, and their ultraviolet emission started ionising the neutral hydrogen in the Intergalactic Medium (IGM). This marked the Epoch of Reionisation (EoR), when the HI is ionised in a patchy manner, giving rise to a completely ionised IGM around $z \approx 6$. During the post-EoR era ($z\lesssim 6$), the neutral hydrogen left is shielded in clumps inside galaxies, tracing the cosmological matter distribution up to today \citep[e.g.][]{Furlanetto_2006PhR, Pritchard_2008PhRvD}. 

Information extracted from 21 cm IM can be used to constrain both cosmology and astrophysics \citep{Bull_2015ApJ, Wolz_2017MNRAS, Wolz_2019MNRAS, Zhaoting_astrophysics_2021MNRAS}. However, detecting the weak HI IM signal is no easy feat because it is several orders of magnitude fainter than foregrounds at low frequencies, mainly consisting of diffuse Galactic synchrotron and free-free emissions~\citep{Oliveira_Costa_2008MNRAS, Zheng_2017MNRAS}, as well as radio-continuum sources~\citep{TRECS_2019MNRAS, TRECS_2023MNRAS}. On top of that, instrumental thermal noise, radio frequency interference (RFI)~\citep[e.g.][]{Engelbrecht_2025MNRAS}, and instrumental systematics \citep[e.g.][]{Barry_2016MNRAS} complicate the reconstruction of the HI signal irrespective of the redshift and scales targeted.

For dish arrays, like MeerKAT and SKA-Mid, the entire set of observable scales in HI intensity mapping in the low-redshift Universe can be split into large (linear) and small scales (quasi-linear and non-linear scales). The former are generally observed in single-dish mode~\citep{MeerKLASS_2016}, while the latter are probed using visibilities~\citep{Paul_a_first_detection}. There are statistically significant detections of the HI signal through cross-correlation of the single-dish HI IM maps with galaxy surveys. These detections have been achieved, for instance, using the Green Bank Telescope (GBT) HI maps cross-correlated with the distribution of galaxies from the WiggleZ~\citep{Masui_2013ApJ, Switzer_2013MNRAS, Wolz2017MNRAS} and eBOSS~\citep{Wolz_2022MNRAS} spectroscopic surveys at redshifts $0.6<z<1$, as well as using the Parkes HI maps cross-correlated with the 2dF Galaxy Survey at redshifts $0.057 < z < 0.098$ \citep{Anderson_2018MNRAS}. The MeerKLASS survey \citep{MeerKLASS_2016, Wang_2021MNRAS} aims to access the cosmological scales with the Square Kilometre Array Observatory (SKAO) precursor MeerKAT, in single-dish mode. Significant statistical detections have been measured through cross-correlations of the MeerKLASS HI IM maps at redshifts $0.39<z<0.46$ with WiggleZ and GAMA surveys \citep{Cunnington_2023MNRAS, MeerKLASS_2025MNRAS, Carucci_2025AA}, as well as stacking the 21 cm intensity maps around galaxy positions from the GAMA survey \citep{Chen_2025ApJS}. There is no detection of the autocorrelation of HI using single-dish mode to date.

For the interferometric observing mode, there are two statistically significant detections in autocorrelation to date. First, detections in autocorrelation have been reported using MeerKAT observations of the 1 deg$^2$ DEEP2 field at redshifts $z\sim 0.32$ and $z\sim 0.44$~\citep{Paul_a_first_detection}. More recently, the CHIME collaboration has reported a statistically significant detection of the cosmological HI signal at redshift $z\sim 1$ from observations of a region of $2.2\times 10^3$ deg$^2$ near the North Galactic Cap \citep[NGC; ][]{CHIME_2025arXiv}. The feasibility of IM across different redshift ranges was also demonstrated in~\citet{Townsend_2026arXiv}, who detected HI power spectrum from intensity maps, consistent with that obtained using individually observed HI galaxies. Their analysis used the 1.6 deg$^2$ central pointing of the COSMOS field from the MIGHTEE survey, covering the redshift range $0.02\lesssim z \lesssim 0.07$. Additionally, upper limits on the HI power spectrum have been obtained at redshifts $1.96<z<3.58$ with uGMRT~\citep{UGMRT_2021ApJ} and at redshifts $5.6< z < 24.5$ with EoR experiments including uGMRT \citep{Paciga_2011MNRAS}, MWA \citep{Trott_2020MNRAS}, HERA \citep{Abdurashidova2026ApJ}, and LOFAR \citep{Mertens_2020MNRAS}.

In this paper, we focus on the challenge of removing the bright emission from diffuse Galactic foregrounds and radio-continuum sources in the context of low-redshift HI intensity mapping with a small field-of-view interferometer, such as MeerKAT. This radiation is smooth in frequency, and this property is largely used to disentangle them from the HI signal. By Fourier transforming the observed visibilities along the frequency axis, we can separate the smooth foregrounds from the rapidly oscillating HI signal \citep{Morales_2004, Parsons_2012ApJ}. The foregrounds are concentrated in the lowest-frequency Fourier modes, corresponding to the largest frequency scales, whereas the HI signal is distributed across the full range of Fourier modes. When computing the 2D power spectrum from the visibilities measured by radio interferometers, we can observe a part of the spectrum where the foregrounds lie, named foreground wedge, and a clean observation window outside these modes \citep{EoR_window_2014PhRvD}. We can reconstruct the HI signal by avoiding the foreground wedge, named foreground avoidance.

However, foreground avoidance is not optimal as the discarded modes also remove the HI signal in them. An alternative approach is foreground removal, which exploits the spectral correlation of the foregrounds to remove them from the data, thus reducing the reconstruction errors as well as enabling analyses beyond the power spectrum. While foreground avoidance restricts analyses to the power spectrum level, foreground removal extends them to the visibility and map levels and to higher-order statistics.

We can divide the foreground cleaning methods into parametric and non-parametric methods. For the parametric approach, some methods use polynomial fits to the extracted foregrounds \citep{Santos_2005ApJ, Bonaldi_2015MNRAS} and forward-modelling methods \citep{Amiri_2023ApJ, Kennedy_2023ApJS, Burba_2024MNRAS, Murphy_2026MNRAS}. The latter are based on using Gaussian constrained realisations to handle missing data, and Gibbs sampling to sample the parameters of the model. For the non-parametric approach, some methods exploit the frequency covariance of the data such as Principle Component Analysis \citep[PCA; see e.g.][]{Alonso_2015MNRAS,Zhaoting_2023MNRAS},  and Generalised Needlet Internal Linear Combination \citep[GNILC; see e.g.][]{Olivari_GNILC_2016MNRAS}. In contrast, Gaussian Process Regression \citep[GPR; see e.g.][]{Mertens_2018MNRAS, Mertens_2024MNRAS} models the frequency correlations of the foregrounds using a covariance kernel, while Fast Independent Component Analysis \citep[FastICA; see e.g.][]{Chapman_2012MNRAS} and Generalised Morphological Component Analysis \citep[GMCA; see e.g.][]{Chapman_2016_mnras, Carucci_2020MNRAS_GMCA} rely on the statistical properties of the foregrounds, such as independence and sparsity, respectively.

In this paper, we present a new non-parametric method adapted from GNILC to work on visibility space: Generalised Visibility Internal Linear Combination (GVILC). It uses the signal-plus-noise covariance matrix as a prior for running a PCA-based method on $uv$ annuli. The method automatically determines the number of modes to remove using a data-driven statistical approach, and minimises the total variance of the cleaned visibilities using an ILC. The method is tested using simulations of the DEEP2 field observations by MeerKAT and SKA-Mid telescopes, including the HI signal, diffuse Galactic foregrounds, radio-continuum sources and noise components. This paper also includes line broadening effects on the HI signal and spatial correlation between HI and continuum sources for foreground removal analyses for the first time. The main aim is to show the potential of applying this technique to MeerKAT and future SKA-Mid data. To date, cosmological HI power spectrum analyses have been performed using MIGHTEE data~\citep{Taylor_2017, Mazumder_2025MNRAS, Townsend_2026arXiv} and DEEP2 data~\citep{Mauch_PB, Paul_a_first_detection}, although other deep MeerKAT observations, such as those from the LADUMA survey~\citep{Baker18}, could also be leveraged for HI intensity mapping science.

The paper is structured as follows. In Section~\ref{sec:simulation} we present the DEEP2 simulation for MeerKAT and SKA-Mid-like observations. In Section~\ref{sec:power spectrum estimation} we explain how to compute the power spectrum from the visibilities using the gridded power spectrum estimator. The methodology to perform foreground removal in visibility space using GVILC is explained in Section~\ref{sec:foreground_cleaning_methods}. In Section~\ref{sec:results} we present the results on the performance of the GVILC method in terms of foreground residuals, signal loss and mode loss. Finally, in Section~\ref{sec:conclusions} we detail the conclusions of the paper. Throughout this paper, we assume Lambda cold dark matter cosmology from Table 2 (TT, TE, EE + lowE + lensing + BAO) of \citet{Planck_VI_2020}.

\section{Simulation of the radio sky}
\label{sec:simulation}
In this section, we describe the simulation of the different components of the radio sky and the generation of interferometric observations for MeerKAT and SKA-Mid. The MeerKAT array consists of 64 dishes of diameter $13.5\,\mathrm{m}$, while the SKA-Mid telescope extend MeerKAT with an additional 133 dishes of diameter $15\,\mathrm{m}$ dishes, making a total of 197 dishes. MeerKAT observes the radio sky from 544 to 3500 MHz in three frequency bands ($L$-band, $UHF$ band and $S$-band)\footnote{\url{https://skaafrica.atlassian.net/wiki/spaces/ESDKB/pages/277315585/MeerKAT+specifications}}, whereas SKA-Mid will observe the radio sky from 350 to 15400 MHz in six frequency bands (Bands 1--5b)\footnote{\url{https://www.skao.int/en/science-users/118/ska-telescope-specifications}}.

We consider a $\rm 12~h$ observation in the DEEP2 region~\citep{Mauch_PB} centred at $\text{RA}=63.36$ deg and $\text{Dec}=-80.00$ deg. We simulate 200 frequency channels centred at 972.85 MHz ($z=0.46$), consistent with the frequency coverage of the $L$- (900-1670\,MHz) and $UHF$ (580-1015\,MHz) MeerKAT bands and the upcoming SKA-Mid Bands 1 (350-1050\,MHz) and 2 (950-1760\,MHz). DEEP2 is a deep observation of a southern field with excellent $uv$ coverage and low foreground levels. The detection of the HI power spectrum reported in \citet{Paul_a_first_detection} used MeerKAT $L$-band observations of this particular patch of the sky. We chose to run our simulations on this field due to these considerations. The input sky image spans $4 \times 4$ deg$^2$, considerably larger than the MeerKAT dish field of view (FoV), which has a full width at half maximum (FWHM) of 1.52 degrees at 972.85 MHz. Table~\ref{tab:parameters} summarises the observational and instrumental parameters used in the simulations.

\begin{table} 
\begin{center}
\begin{tabular}[\columnwidth]{ll}
\hline
\textbf{Parameter} & \textbf{Value}\\
\hline
Central redshift &  0.46 \\
Redshift range &  0.444--0.476 \\
Central frequency &  972.85 MHz \\
Frequency range &  962.51--983.30 MHz \\
Channel width & 104.5 kHz\\
Number of frequency channels & 200 \\
Dish size & 13.5 m\\
Field of view at central frequency & 1.52$^\circ$ \\
\hline
Pointing centre (RA, Dec) & (63.36$^\circ$, 80.00$^\circ$) \\
Observation length & 12 h\\
Starting time [UTC] & 07-07-2018 21:40:20.7 \\
Time resolution & 8 s\\
Total number of time-steps & 720\\
\hline
Number of MeerKAT dishes & 64 \\
Number of SKA-Mid dishes & 197 \\
MeerKAT noise equivalent & $100~\mathrm{h}$ \\
SKA-Mid noise equivalent & $5000~\mathrm{h}$ \\
\hline
\end{tabular}
\end{center}
\caption{Observational and telescope parameters used for the MeerKAT and SKA-Mid-like simulations.}
\label{tab:parameters}
\end{table}

\subsection{HI signal}
\label{sec:HI model}
In the low-redshift Universe, the majority of atomic hydrogen (HI) is confined within galaxies and can thus be used as a tracer of the underlying dark matter distribution. We use the \texttt{T-RECS}\footnote{\url{https://github.com/abonaldi/TRECS}} \citep{TRECS_2019MNRAS, TRECS_2023MNRAS} code to simulate the galaxies containing HI, which we label hereafter as HI sources. This code consistently simulates HI and extragalactic radio sources, allowing us to investigate the effects of correlations between the HI and radio-continuum sources described in Section~\ref{sec: extragalactic point sources}.

Initially, \texttt{T-RECS} generates a catalogue of HI sources by sampling a HI Mass Function (HIMF) and placing the sources in random positions (latitude, longitude, and redshift) within the light cone.  The HIMF is determined empirically using best fit values from observations presented in \citet{Jones_2018MNRAS} at $z=0$, and in \citet{Paul_a_first_detection} at $z=0.32$. Thus, currently \texttt{T-RECS} is limited to the low redshift Universe.

We simulate the HI sources with a minimum flux limit of 1 Jy\,Hz (see Section~\ref{sec: extragalactic point sources} for a discussion of the impact of the flux limit). Although HI sources are intrinsically extended and their sizes can be extracted from the \texttt{T-RECS} catalogue, we model them as point sources as we do not resolve those scales with our observations. To simulate the clustering component of the HI sources, \texttt{T-RECS} associates each HI source with a dark matter sub-halo of the P-millennium N-body cosmological simulation \citep{Baugh_2019MNRAS}. Clustering is simulated in 3 dimensions, changing the initial latitude, longitude, and redshift coordinates. 

\begin{figure}
    \centering
    \includegraphics[width=\columnwidth]{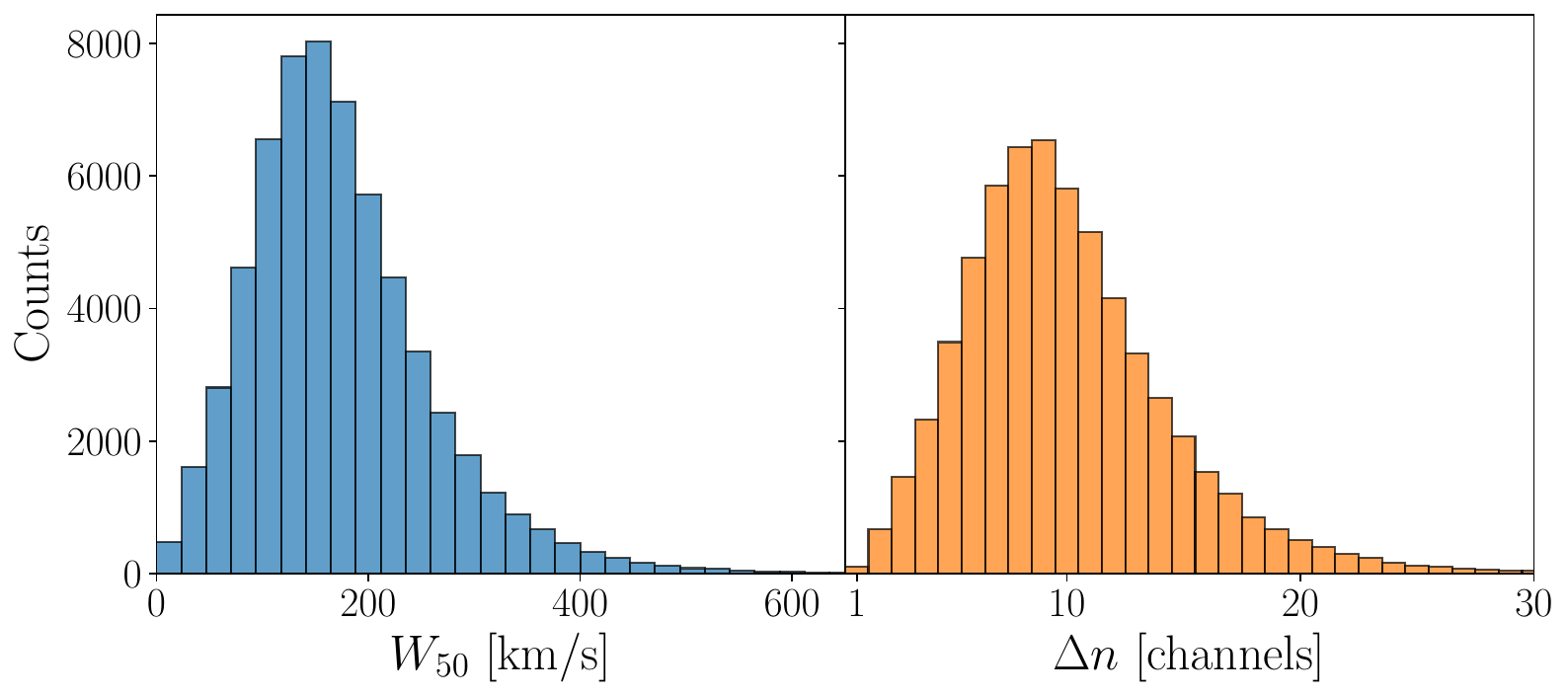}
    \caption{Left panel: histogram showing the counts of $W_{50}$, the FWHM of the HI sources emission profile. Right panel: histogram of the counts of $\Delta n$, the number of frequency channels spanned by the HI sources.
    }
    \label{fig:line_broadening}
\end{figure}

We follow a methodology similar to that of \citet{Zhixing_2024MNRAS} to simulate line broadening at the catalogue level, by adding a line profile to each of the simulated HI sources. Line broadening arises from the rotation velocity of the galaxy projected along the line of sight (LOS), and therefore depends on both its circular velocity and inclination. Given the high frequency resolution of MeerKAT, some of the broad emission profiles of the HI sources can be resolved, making the commonly adopted Dirac delta approximation no longer valid because HI emission from individual galaxies spans several frequency channels. The line profile is characterised by its shape and width. \citet{Zhixing_2024MNRAS} showed that the impact of flat, double Horn and Gaussian profiles is similar; therefore, we decided to use the Gaussian profile. Its FWHM at the rest-frame of the source $i$ is given by the parameter $W_{50,i}$. To simulate line broadening effects, we first compute the standard deviation $\sigma_i$ of the Gaussian profile in redshift space for each source:
\begin{equation} \label{eq: sigma_i}
    \sigma_i = \frac{W_{50,i}(1+z_i)}{2\sqrt{2\ln2}\, c},
\end{equation}
where $c$ is the speed of light and $z_i$ is the redshift of the source $i$. Using equation~\eqref{eq: sigma_i}, we can calculate how the total flux of source $i$ is distributed across frequency channels:
\begin{equation} \label{eq: line-broadening}
    F_i(n)=\frac{F_i}{N_i}\int_{z(n)}^{z(n+1)} \exp\left(-\frac{(z'-z_i)^2}{2\sigma_i^2}\right)\, \text{d}z',
\end{equation}
where $F_i(n)$ is the flux of source $i$ assigned to the $n$th frequency channel, $N_i$ is the normalization factor of source $i$, $F_i$ is the total flux corresponding to source $i$, and $z(n)$ ($z(n+1)$) is the redshift corresponding to the left (right) edge of the frequency channel $n$. In principle, equation~\eqref{eq: line-broadening} can be integrated over all the channels and HI sources. However, this increases the number of HI sources drastically and becomes computationally expensive. Hence, we consider the $\pm 2\sigma_i$ range of the Gaussian profile because it provides a sufficiently large span to capture the line broadening effect. Then, for each HI source we compute the minimum, $n_{\rm min}$, and maximum, $n_{\rm max}$ frequency channel it spans in the $\pm2\sigma_i$ limit. Consequently, the normalization is defined as:
\begin{equation}
    N_i=\int_{z(n_{\rm min})}^{z(n_{\rm max}+1)} \exp\left(-\frac{(z'-z_i)^2}{2\sigma_i^2}\right)\, \text{d}z'.
\end{equation}

In Fig.~\ref{fig:line_broadening}, we show the histograms of $W_{50}$ values and of the number of frequency channels that the HI sources span, i.e., $\Delta n=n_{\rm max}-n_{\rm min}+1$. Under the approximation of the HI sources being Dirac delta functions in frequency, each source is confined to a single frequency channel. With the more realistic approach adopted here, the HI sources exhibit a more complex spectral behaviour, with most of them spanning multiple frequency channels. From Fig.~\ref{fig:line_broadening} we see that the mode is nine, and the majority of sources do not span more than thirty channels. Introducing line broadening in the simulations allows us to test how foreground removal performs in a scenario where HI sources exhibit considerable frequency correlation,  which could degrade its performance.

\subsection{Extragalactic radio sources} \label{sec: extragalactic point sources}
The most prominent sources of foreground radio emission on small scales are extragalactic point sources, also referred to as radio-continuum sources, which in the radio wavelength range emit smoothly as a power law in frequency. We use \texttt{T-RECS}~\citep{TRECS_2019MNRAS, TRECS_2023MNRAS} to model the continuum sources as a combination of radio active galactic nuclei (RL AGN) and star-forming galaxies (SFG) using empirical data-derived methods for realistic cosmological evolution of the luminosity functions, the total intensity number counts, polarised intensity, and clustering properties. For this work, we generate continuum sources with flux densities between $10^{-7}$ to $1$ Jy. $L$-band observations of the DEEP2 field show that the field does not contain extremely bright extragalactic sources. However,  we choose not to impose any flux cut on the radio-continuum source population to allow retention of the typical level of radio-continuum source contamination expected in a representative patch of the sky in the simulation. As in the case of the HI sources, the continuum sources are approximated as point sources. 

Simulation-based studies on low-redshift 21-cm cosmology so far have generated the HI and continuum sources independently~\citep{Paul_2021MNRAS,Zhaoting_2023MNRAS, Chen_SKA_2023MNRAS}. However, since the HI-emitting galaxies also produce continuum emission, the emissions are correlated. This work explores, for the first time, whether this spatial correlation significantly affects foreground removal. In the \texttt{T-RECS} code, the continuum and HI sources are initially simulated independently. The continuum sources are simulated between redshifts $z=0\text{--}8$, while the HI sources are only simulated for the observed frequencies corresponding to redshifts $z=0.444\text{--}0.476$. Each continuum source is assigned an HI mass based on the relation between star formation rate (SFR) and HI mass from equation 11 of \citet{TRECS_2023MNRAS}. To introduce the spatial cross-correlation, \texttt{T-RECS} runs a nearest-neighbour algorithm over the HI and continuum sources to assign continuum counterparts to the HI sources, modifying the positions and redshifts of the continuum sources in the catalogue. Although this cross correlation occurs in the full continuum redshift range, we are only affected in the HI sources redshift range, as it is where the signal lies.

\begin{figure}
    \centering
    \includegraphics[width=\columnwidth]{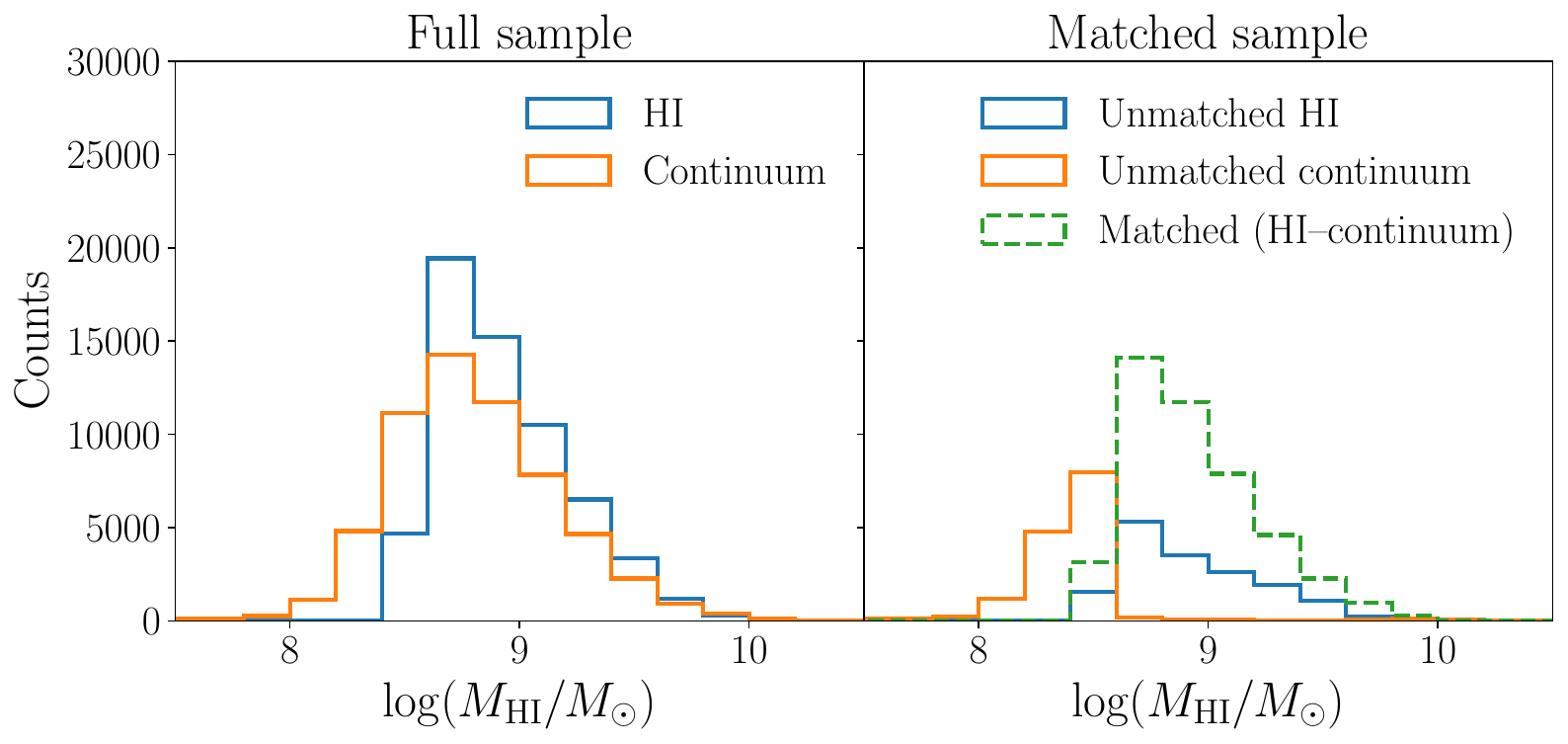}
    \caption{
    Left panel: histograms of the distribution of HI mass values for the HI (blue) and the radio-continuum (orange) source catalogues before introducing any spatial correlation. Right panel: histograms of the unmatched HI (blue), unmatched radio-continuum (orange), and matched HI and radio-continuum sources (dashed green) after running the cross-matching algorithm. Both panels correspond to the redshift range of $z=0.444$--$0.476$.}
    \label{fig:cross-match}
\end{figure}

In Fig.~\ref{fig:cross-match}, we show the histograms of the HI mass values for the HI and continuum source catalogues before and after cross-matching for the redshift range of $z=0.444\text{--}0.476$. As explained in \citet{TRECS_2023MNRAS}, the HI mass assigned to each continuum source does not come from sampling an HIMF but from a SFR-HI mass relation. Consequently, we have to check that the distribution of HI mass for the continuum and HI sources is similar before cross-matching. From Fig.~\ref{fig:cross-match} we can see that the HI and continuum source distributions are similar. This corresponds to the minimum fluxes of $10^{-7}$ Jy for the continuum sources and $1$ Jy\,Hz for the HI sources. Using the aforementioned method and these flux limits, we find that the percentage of HI sources with a continuum source counterpart reaches 73.5\%.

Finally, radio-continuum sources correlate with large-scale structure, producing clustering. This clustering is simulated by associating each continuum source with a dark matter sub-halo of the P-millennium N-body cosmological simulation \citep{Baugh_2019MNRAS}, following the same procedure as for the HI sources. Within the redshift of interest, $z=0.444\text{--}0.476$, the continuum sources are associated with dark matter sub-halos after the HI and continuum source catalogues have been cross-correlated. Outside this redshift range, only the continuum sources are associated with dark matter sub-halos.

\subsection{Diffuse Galactic foregrounds: synchrotron and free-free} \label{sec: diffuse}
Low-frequency observations targeting the HI signal are affected by the diffuse emission from our Galaxy. At frequencies around 1 GHz, the total intensity (Stokes I) diffuse galactic emission is dominated by synchrotron and free-free emissions. High-resolution interferometric observations reveal the small-scale structure of the diffuse Galactic emission. Investigating these small-scale structures with mock interferometric data requires simulating the diffuse emission at arcsecond resolution. Currently available full-sky maps of Galactic synchrotron and free-free are at the degree scale~\citep{Planck2018_CS_2020, pysm3_2025}, hence finer structures have to be added artificially to these maps. In this work, we generate small-scale Gaussian structures modulated by non-Gaussian large scales, leading to non-Gaussian diffuse foregrounds at small scales  \citep{pysm3_2025}. We simulate the diffuse foreground components (synchrotron and free-free) with \texttt{pysm3}\footnote{\url{https://github.com/galsci/pysm}}~\citep{pysm2_2017, Zonca_2021, pysm3_2025}, a Python-based software package that provides models and templates for simulating full-sky astrophysical foreground emission.

For synchrotron emission, we have chosen the \texttt{s5} model, which models the spectral energy distribution (SED) based on a power law scaling with a spatially varying spectral index derived from the analysis of Haslam 408~MHz \citep{Haslam_Remazeilles_2015}, \textit{WMAP} 23~GHz 7-year data \citep{WMAP_7years} and S-PASS \citep{SPASS}. The non-Gaussian small scales are generated using the Logarithm of the Polarization Fraction Tensor formalism \citep{pysm3_2025}. The map, generated in \texttt{HEALPix}\footnote{\url{https://healpix.sourceforge.io/}}~\citep{gorski} format, has a resolution of $N_{\rm side}=8192$, corresponding to a pixel area of (25.77 arcsec)$^2$. 

\texttt{Pysm3} can only model the Galactic free-free emission using the \texttt{f1} model. It describes the SED of the unpolarized free-free emission with a constant power-law index of $-2.14$ using the amplitudes from the 30 GHz \texttt{Commander} fit to \textit{Planck}-2015 data \citep{Commander_Planck_2015,Draine_free_free}. Unfortunately, the native resolution of the \texttt{f1} model is $N_{\rm side}=512$, which corresponds to a pixel area of (6.87 arcmin)$^2$, much coarser than MeerKAT's angular resolution. Therefore, we upgrade the resolution of the map to $N_{\rm side}=8192$, including non-Gaussian small scales modulated by the large scale template using the procedure detailed in Appendix~\ref{App: small-scale free-free}. 

\begin{figure*}
    \centering
    \includegraphics[width=\textwidth]{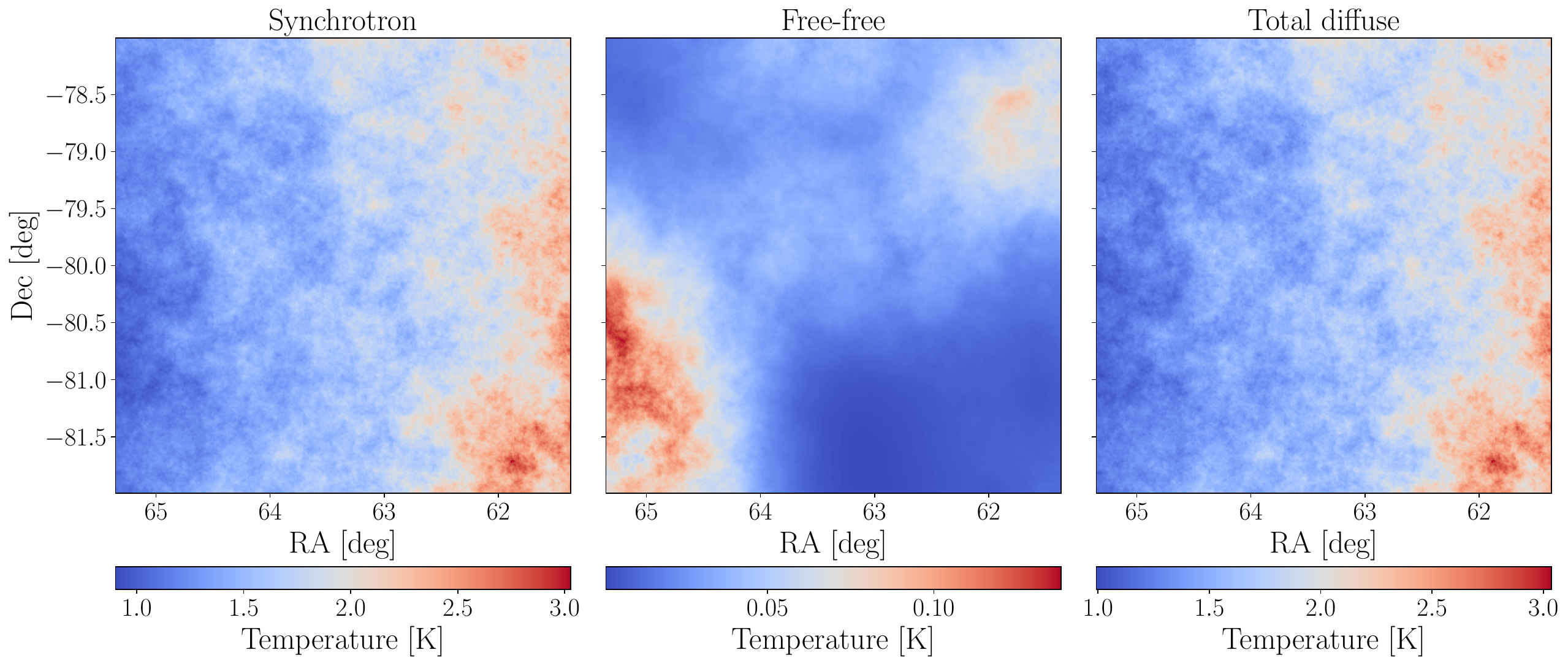}
    \caption{
    Simulated synchrotron (left panel), free-free (central panel) and total diffuse (right panel) emission at the central frequency of 972.85 MHz. The centre of the field corresponds to the DEEP2 pointing, with celestial coordinates $\text{RA}=63.36$ deg and $\text{Dec}=-80.00$ deg, with an extension of $4\times 4$ deg$^2$ and resolution of ($30$ arcsec)$^2$.
    }
    \label{fig:diffuse}
\end{figure*}

For computational efficiency, the synchrotron and free-free frequency maps are projected into a 2D Cartesian grid with a resolution of ($30$ arcsec)$^2$ and $4\times 4$ deg$^2$ extension. Fig.~\ref{fig:diffuse}, shows the synchrotron, free-free and total diffuse emission at the central frequency channel at 972.85 MHz. The diffuse emission is dominated by synchrotron, with an almost negligible contribution coming from free-free emission in the DEEP2 field.

\subsection{Instrument} \label{sec:instrument}

In this section, we describe the simulation of the visibilities of the different components, namely HI, and foregrounds, which comprise diffuse emission and extragalactic point sources. We use \texttt{OSKAR}\footnote{\url{https://github.com/OxfordSKA/OSKAR}} \citep{OSKAR}, which simulates the digital beamforming for the aperture array of an SKA-sized antenna. However, it can also be used to simulate dish arrays by approximating the primary beam as Gaussian. \texttt{OSKAR} uses the Radio Interferometer Measurement equation from \citet{Hamaker_1996} to generate the visibilities. It takes a sky model, observation strategy, and telescope configuration as inputs and generates the visibilities, storing them in a measurement set file. Since our analysis focuses solely on intensity--the component containing the HI signal--we have simulated only Stokes I.
 
The visibilities, $V(u,v,w,f)$, are the complex Fourier modes of the sky brightness distribution, $I(l,m,f)$. They depend on the baseline coordinates $(u,v,w)$, while the sky brightness distribution depends on the sky coordinates in the celestial sphere $l$ and $m$. Both depends on the observing frequency $f$. The $w$-term describes the deviation of the array from a perfect plane, and it becomes more important for off-zenith observations, such as DEEP2. Under the flat-sky approximation, the $w$ dependency can be neglected. This assumes small field of view and sufficiently small $w$ values~\citep{WSCLEAN_Offringa_2014MNRAS}. In this work, we neglect the $w$-term when generating the \texttt{OSKAR} simulations, as the main goal of the paper is to evaluate the performance of foreground removal methods. Although this assumption is not fully realistic, it ensures that the simulated visibilities exactly satisfy the flat-sky visibility equation. A proper treatment of the $w$-term is left for future work. Then, the flat-sky visibilities can be expressed as \citep{Thomson_synthesis_2017}:
\begin{equation} \label{vis-flat-sky}
\begin{aligned}
    V(u,v,f) = \int& \frac{\text{d}l\, \text{d}m}{n}\ I(l,m,f)A(l,m,f)e^{-2\pi i(lu+mv)},
\end{aligned}
\end{equation}
where $A(l,m,f)$ is the primary beam response.

In this work, we use the layout of the MeerKAT telescope to simulate the observed visibilities for both MeerKAT and SKA-Mid. We simulate the observations of 64 antennas assuming a circular Gaussian primary beam with a FWHM of $88.8$ arcmin at 996.65 MHz. Although the main lobe of the primary beam fits well to a Gaussian profile, MeerKAT's primary beam has secondary peaks surrounding the main lobe, named sidelobes \citep{Mauch_PB}. We defer a detailed analysis of the impact of sidelobes on such observations and their effect on foreground cleaning to future work.

\begin{figure}
    \centering
    \includegraphics[width=\columnwidth]{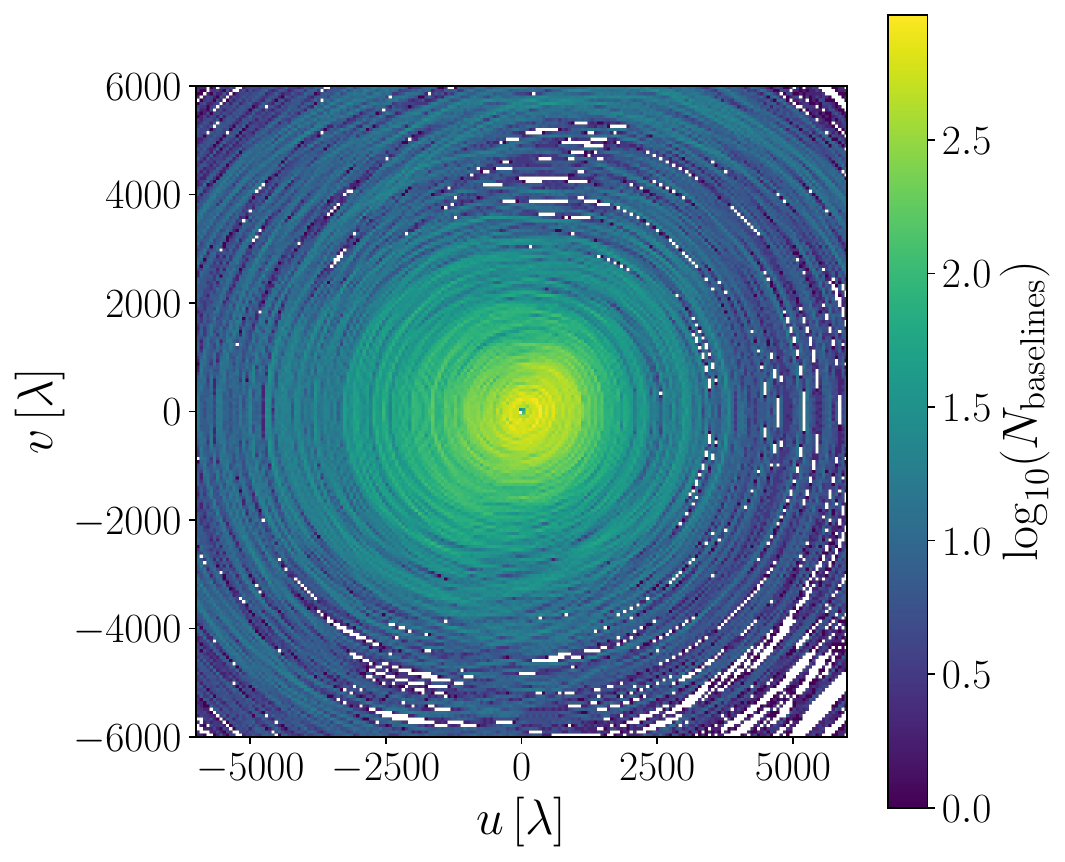}
    \caption{
    The $uv$ coverage for the simulated visibilities. The baseline coordinates $u$ and $v$ are expressed in units of the wavelength of the central channel at 972.85 MHz. Each cell has a size of $\Delta u=\Delta v=60\lambda$, and its value represent the number of baselines, $N_{\rm baselines}$, falling within that cell.}
    \label{fig:uv_tracks}
\end{figure}

We simulate the DEEP2 field by setting the pointing centre to $\text{RA}=63.36$ deg and $\text{Dec}=-80.00$ deg with a 12 h tracking. The starting time in UTC is 07-07-2018 21:40:20.7, allowing the field to be observed continuously over the available observing period. MeerKAT has a time resolution of 8 s, so we set the correlator time-average duration to that value. For computational efficiency, we set the total number of time-steps to 720, which means that we simulate a baseline per minute. In Fig.~\ref{fig:uv_tracks} we show the $uv$ coverage of the 12 h simulation. Owing to the declination of the selected field, the full accessible $uv$-plane has dense coverage, which is ideal for high sensitive observations like HI intensity mapping.

\begin{figure*}
    \centering
    \includegraphics[width=\textwidth]{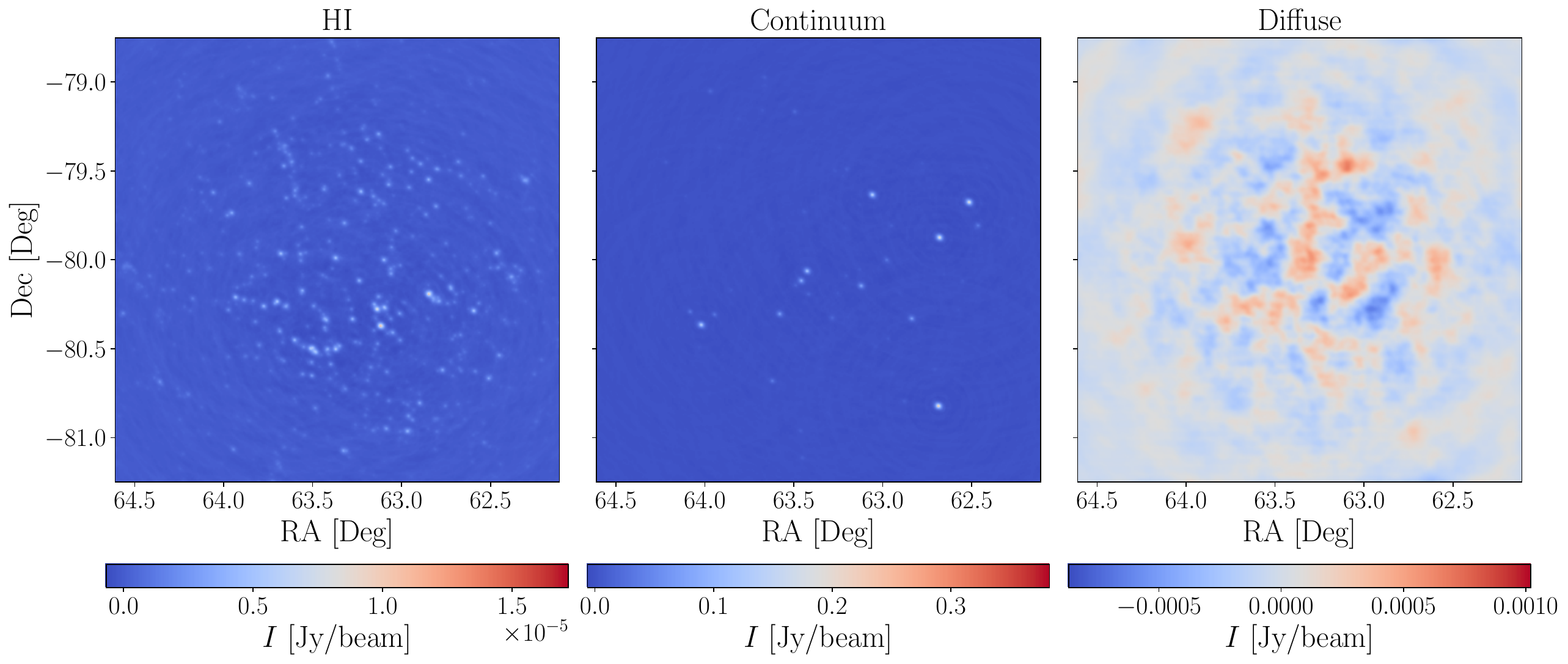}
    \caption{
    Stokes $I$ images at the central channel at 972.85 MHz of the HI sources (left panel), radio-continuum sources (middle panel), and diffuse Galactic emission (right panel) from the \texttt{OSKAR} simulations. The images were generated from the simulated visibilities using a Fast Fourier Transform (FFT) algorithm, and correspond to a field of view of $2.5\times 2.5$ deg$^2$.}
    \label{fig:imaging}
\end{figure*}

We simulate each astrophysical component contributing to interferometric visibility independently with \texttt{OSKAR}, allowing us to study the impact of foreground cleaning in each component and to evaluate the performance. We can get the combined sky visibility, $V_{\rm sky}$:
\begin{equation} \label{eq:Vsky}
    V_{\rm sky} = V_{\rm HI} + V_{\rm diff} + V_{\rm cont},
\end{equation}
where $V_{\rm HI}$ are the visibilities for the HI signal, $V_{\rm diff}$ are the visibilities for diffuse galactic emission, and $V_{\rm cont}$ are the visibilities for extragalactic point source emission. In Fig.~\ref{fig:imaging} we present the resulting images for each of the sky components. We can observe that both radio-continuum sources and the diffuse Galactic emission dominate the HI signal by several orders of magnitude. A closer inspection of the radio-continuum source image shows that the brightest sources, around 1 Jy, dominate the field, rendering most of the fainter sources effectively invisible.

Finally, we get the observed visibility, $V_{\rm obs}$, as the sum of the combined sky visibilities, and thermal noise per visibility, $V_{\rm noise}$:
\begin{equation} \label{eq:Vobs}
    V_{\rm data} = V_{\rm sky} + V_{\rm noise},
\end{equation}
where $V_{\rm noise}$ is a complex number whose real and imaginary parts are drawn from a Gaussian distribution $\mathcal{N}(0,\sigma_{\rm N}^2/2)$. The thermal noise per visibility standard deviation, $\sigma_{\rm N}$, is given by \cite{Thomson_synthesis_2017}:
\begin{equation} \label{std noise}
    \sigma_{\rm N} = \frac{2k_{\rm B}T_{\rm sys}}{A_{\rm e}\sqrt{\delta t \, \delta f}},
\end{equation}
where $k_{\rm B}$ is the Boltzmann constant, $T_{\rm sys}$ is the receiver system temperature, $A_{\rm e}$ is the effective aperture of the dish, $\delta f=104.5$ kHz is the channel bandwidth, and $\delta t=60$ s is the time per step. We use $A_{\rm e}/T_{\rm sys}=6.22$ m$^2\,$K$^{-1}$ \citep{Specs, Paul_a_first_detection}.

We analysed two different noise scenarios: 100 h MeerKAT noise, and 5000 h SKA-Mid-like noise. Instead of simulating the full observation time, we rescale the noise by the appropriate factors. For the $\rm 100~h$ MeerKAT noise, we divide equation~\eqref{std noise} by an effective noise-reduction factor of $\sqrt{100/12}=2.887$ like in~\citet{Zhaoting_2023MNRAS}. For the $\rm 5000~h$ SKA-Mid-like observation, we derive an effective noise-reduction factor of 70. To estimate this, we first simulate the $uv$ coverage corresponding to a $\rm 12~h$ SKA-Mid observation. We then generate Gaussian random noise with a standard deviation given by equation~\eqref{std noise}, scaled by a factor of $\sqrt{5000/12} = 20.412$ to account for the longer total integration time. From these simulations, we compute the ratio of the noise power spectra for the SKA-Mid and MeerKAT configurations, which defines the effective noise reduction. This estimate is approximate, as we are neglecting the $k$-scale dependency. Moreover, we do not account for differences in the primary beam of the $\rm 15~m$ SKA-Mid antennas or for variations in $A_{\rm e}/T_{\rm sys}$ between the two instruments, modelling as they were MeerKAT antennas. However, the impact of these differences is expected to be small.
    
\section{Power spectrum estimation} \label{sec:power spectrum estimation}
In this section, we describe the statistical estimator used to extract the HI signal power spectrum from the simulated visibilities. In this paper, we work in Fourier space, which is the natural domain for visibilities. Previous works related to 21 cm interferometric IM with MeerKAT also used a visibility approach \citep{Paul_MIGHTEE_2021MNRAS, Mazumder_2025MNRAS, Paul_a_first_detection}. In fact, \citet{Zhaoting_2023MNRAS} showed that incomplete $uv$ coverage introduces additional biases at the power spectrum level when working with images from simulated MeerKAT interferometric data. These biases also depend on the weighting scheme used in the imaging. As we will explain in detail later, working in Fourier space has several advantages over image space, including the separation of spectrally smooth foregrounds from the HI signal and increasing sensitivity by coherently combining measurements of the same Fourier mode.

In interferometric 21 cm IM, two conceptual frameworks must be reconciled: cosmology and radio interferometry. These domains differ not only in the variables they employ, but also in the Fourier conventions they adopt \citep{Morales_2004}. The HI brightness temperature field, $T(\boldsymbol{r})$, is expressed in terms of the comoving distance vector $\boldsymbol{r}=(r_x,r_y,r_z)$. The Fourier transform of the HI brightness, $\widetilde{T}(\boldsymbol{k})$, is given by
\begin{equation}
    \widetilde{T}(\boldsymbol{k}) = \int\frac{\text{d}r^3}{\mathbb{V}}T(\boldsymbol{r})e^{-i\boldsymbol{k}\cdot \boldsymbol{r}},
\end{equation}
where $\boldsymbol{k}$ is the Fourier conjugate of $\boldsymbol{r}$ and $\mathbb{V}$ is the cosmological volume. The HI 3D power spectrum, $P(\boldsymbol{k})$, is given by:
\begin{equation} \label{power spectrum}
    \mathbb{V}\langle \widetilde{T}(\boldsymbol{k})\widetilde{T}^\ast(\boldsymbol{k}')\rangle = P(\boldsymbol{k})\delta_\text{D}^3(\boldsymbol{k}-\boldsymbol{k}'),
\end{equation}
where $\delta_\text{D}^3$ is the 3D Dirac delta. The field remains statistically homogeneous, although line-broadening effects break isotropy.

A radio interferometer observes the visibilities, which are the complex Fourier modes of the sky brightness distribution. Starting from equation~\eqref{vis-flat-sky}, we convert the sky brightness distribution $I$ to temperature $T$ and perform a Fourier transform along the frequency axis. This yields to the delay-transformed visibilities $\bar{V}(u,v,\eta)$:
\begin{equation} \label{vis-small-field-weq0}
\begin{aligned}
    \bar{V}(u,v,\eta) = \left(\frac{2k_{\rm B}}{\lambda^2}\right)\int& \text{d}l\, \text{d}m\,\text{d}f\  T(l,m,f)A(l,m,f)\\
    &\times e^{-2\pi i(lu+mv+f\eta)},
\end{aligned}
\end{equation}
where $\eta$ is the Fourier conjugate of frequency and $\lambda$ is the observed HI wavelength at the mean redshift $z_0$.

The objective is to connect the power spectrum measured by a radio interferometer to the HI 3D power spectrum. The power spectrum estimated from the visibilities is called the delay power spectrum, $P_\text{d}=\langle |\bar{V}(u,v,\eta)|^2 \rangle$, defined as the autocorrelation of the delay-transformed visibility for a given baseline $(u,v)$ and $\eta$, and it is connected to the HI power spectrum by the following equation \citep{Parsons_2014ApJ, Zhaoting_2023MNRAS}:
\begin{equation} \label{eq:delay ps}
    P_\text{d}\equiv P_{\rm d}(\boldsymbol{k})=\left(\frac{2k_{\rm B}}{\lambda^2} \right)^2\frac{\mathbb{V}^2}{X^4Y^2}\int \frac{\text{d}^3\boldsymbol{k}'}{(2\pi)^3}\
    |A(\boldsymbol{k}-\boldsymbol{k}')|^2
    P(\boldsymbol{k}').
\end{equation}
where $X=D_c(z_0)$ is the comoving distance at redshift $z_0$, $Y=\lambda_{21}(1+z_0)^2/H(z_0)$ is the line-of-sight comoving distance per frequency, and $\mathbb{V}$ is the integrated volume. The parameter $Y$ depends on $\lambda_{21}$, the rest wavelength of the 21-cm emission, and the Hubble parameter at redshift $z_0$, i.e., $H(z_0)$. Finally, we can decompose $\boldsymbol{k}=(\boldsymbol{k}_\perp, k_\parallel)$ in two components: one perpendicular to the LOS, $\boldsymbol{k}_\perp$, and other parallel to the LOS, $k_\parallel$. Both can be related to $u$, $v$ and $\eta$ by a change of variables:
\begin{equation} \label{eq: kpar_kperp}
    \boldsymbol{k}_\perp=\left(\frac{2\pi u}{X}, \frac{2\pi v}{X}\right),\  k_\parallel = \frac{2\pi\eta}{Y}.
\end{equation}

The roles of the primary beam from equation~\eqref{eq:delay ps} are changing the normalization of the power spectrum, and introducing mode mixing in $\boldsymbol{k}$ space. We approximate MeerKAT's primary beam as a Gaussian and assume no frequency dependence for our analysis. Under these assumptions, mode mixing is only produced in $\boldsymbol{k}_\perp$ and not along $k_\parallel$ \citep{Zhaoting_2023MNRAS}.

\subsection{Gridded power spectrum estimator} \label{subsec: gridded PS}
The volume of data produced by a radio interferometer scales approximately as the square of the number of antennas, making the direct processing of visibility data from modern facilities such as MeerKAT and SKA-Mid computationally demanding. For HI IM analyses, this challenge can be mitigated by averaging visibilities onto a regular grid in the $uv$ plane, yielding gridded visibility cubes:
\begin{equation} \label{eq: gridding}
    V_{\rm grid}^j=\frac{1}{N_j}\sum_{\boldsymbol{u}\in \mathcal{D}_j}V(\boldsymbol{u}),
\end{equation}
where $V_{\rm grid}^j=V_{\rm grid}(u_j,v_j,f_j)$ are the gridded visibilities, centred at $(u_j,v_j,f_j)$, $\mathcal{D}_j$ is defined as the set of baselines $(u,v,f)$ belonging to the grid $j$, and $N_j$ is the number of visibilities that fall into the $j$-th bin. From the gridded visibilities, we can obtain the delay transformed visibilities for a given $uv$ grid by doing a Fast Fourier Transform (FFT):
\begin{equation}\label{eq: delay vis}
    \widetilde{V}_{\rm grid}^{j}=\delta f\sum_{i=1}^{N_{\rm ch}} e^{-2\pi i\eta_i f_i}V_{\rm grid}(u_j,v_j,f_i)\Phi(f),
\end{equation}
where $\widetilde{V}_{\rm grid}^{j}=\widetilde{V}_{\rm grid}(u_j,v_j,\eta_j)$, $N_{\rm ch}$ are the number of frequency channels, and $\Phi(f)$ is the normalised Blackman-Harris tapering function introduced to avoid edge effects in Fourier transforms \citep{Blackman_Harris_1958, Harris_1978}.

From the delay-transformed gridded visibilities, we compute the gridded power spectrum and connect it to the cosmological power spectrum, which is our final goal. From the gridded visibilities that are defined in equations \eqref{eq: gridding} and \eqref{eq: delay vis}, the delay power spectrum is computed as the square of the gridded delay visibilities\footnote{For the derivation of equation~\eqref{eq: gridded PS vis}, we first apply a Fourier transforming along frequency, and $uv$ gridding afterwards. These two operations commute, so equation~\eqref{eq: gridded PS vis} is mathematically correct, although in practice we are doing the operation in reverse order.}: 
\begin{equation} \label{eq: gridded PS vis}
    \left\langle\left|\widetilde{V}_{\rm grid}^{j}\right|^2\right\rangle=\frac{1}{N_j^2}\sum_{\boldsymbol{u}_1, \boldsymbol{u}_2\in\mathcal{D}_j}
    \left\langle \widetilde{V}(\boldsymbol{u}_1)\widetilde{V}^\ast(\boldsymbol{u}_2)\right\rangle,
\end{equation}
where $\left\langle \widetilde{V}(\boldsymbol{u}_1)\widetilde{V}^\ast(\boldsymbol{u}_2)\right\rangle$ corresponds to the visibility covariance. The visibility covariance can be expressed as a function of the cosmological power spectrum we want to infer. As derived in Appendix~\ref{ap: norm}, at zeroth order of the Taylor expansion of the cosmological power spectrum,
\begin{equation} \label{eq: vis cov}
\left\langle \widetilde{V}(\boldsymbol{u}_1)
\widetilde{V}^\ast(\boldsymbol{u}_2) \right\rangle=
\left(\frac{2k_{\rm B}}{\lambda^2}\right)^2
\frac{\delta f N_{\rm ch}\pi\sigma^2}{X^2Y}e^{-\pi^2\sigma^2|\boldsymbol{u}_1-\boldsymbol{u}_2|^2}P(\boldsymbol{k}_{\perp,1}, k_\parallel),
\end{equation}
where $\boldsymbol{u}_1=(u_1,v_1)$, $\sigma$ is the standard deviation of the Gaussian primary beam, and $\boldsymbol{k}_{\perp,1}=2\pi\boldsymbol{u}_1/X$. The exponential term captures the visibility decorrelation produced by the Gaussian primary beam, and can be accounted in the normalisation correction. Combining equations~\eqref{eq: gridded PS vis} and \eqref{eq: vis cov} we can compute the 3D delay power spectrum, $P_{\rm d}^j$, in terms of $(\boldsymbol{k}_\perp, k_\parallel)$ following the result derived in Appendix~\ref{ap: norm}:
\begin{equation} \label{eq: gridded PS cosmological}
    P_{\rm d}^j \equiv P_{\rm d}(\boldsymbol{k}_\perp^j, k_\parallel^j) = \left(\frac{\lambda^2}{2k_{\rm B}}\right)^2\frac{X^2Y}{\delta f N_{\rm ch}\pi\sigma^2}\frac{N_j^2}{S_j(\mathcal{D}_j)}\left|\widetilde{V}_{\rm grid}^{j}\right|^2,
\end{equation}
where 
\begin{equation} \label{eq: S_j red}
S_j(\mathcal{D}_j)=\sum_{\boldsymbol{u}_1,\boldsymbol{u}_2\in\mathcal{D}_j}
e^{-\pi^2\sigma^2\left|\boldsymbol{u}_1-\boldsymbol{u}_2\right|^2}.
\end{equation}

From equation \eqref{eq: S_j red}, we see that in the limit of small grid sizes ($\Delta u \ll 50\lambda$), the exponential terms corresponding to $\boldsymbol{u}_1\neq\boldsymbol{u}_2$ are approximately one. Thus, the sum is equal to the number of terms of the sum, $S_j(\mathcal{D}_j)\approx N_j^2$, making the normalisation independent of the positions of the baselines. Under this limit, equation~\eqref{eq: gridded PS cosmological} reduces to the usual normalisation correction from \citet{Parsons_2012ApJ,Parsons_2014ApJ} corresponding to the Gaussian primary beam case. For medium to large grid sizes, $S_j(\mathcal{D}_j)< N_j^2$. If this extra term is not taken into account, it leads to a biased estimate. In the derivation of equation~\eqref{eq: gridded PS cosmological}, we make several approximations and assumptions, including a first-order Taylor expansion of the power spectrum, treating the power spectrum and its gradient as constant within each grid cell, and assuming Gaussian primary beams. As a consequence, it is expected to work better for small grid sizes. For a detailed mathematical derivation and validation against simulations see Appendix~\ref{ap: norm}.

Thus, the grid size plays an important role in the computed power spectrum amplitude. In addition to decorrelation, there are other effects which are influenced by the grid size. One effect is mode mixing, which smooths features in the power spectrum. As explained in equation~\eqref{eq:delay ps}, the primary beam is convolved with the cosmological power spectrum, producing mode mixing. This effect can be largely mitigated by taking grid sizes much larger than the beam size in Fourier space. In that situation, different cells can be considered as independent, and mode mixing will be contained inside the grid. Furthermore, it will allow us to compute errors at the power spectrum level using the sampling variance under the independence assumption of the different $k$-bins. As derived in Appendix~\ref{ap: norm}, for a grid size larger than $\Delta u=47\lambda$ the grids are independent. Therefore, by choosing a grid size of $\Delta u=60\lambda$ we address mode mixing and error computation. The visibility decorrelation effect is approximately corrected by the modified normalisation correction, as the grid sizes considered are not sufficiently small for the correction to be exact. In conclusion, the estimation of the gridded power spectrum is a complex problem that requires a more thorough investigation, which we defer to future work. Since the primary goal of this paper is foreground removal--and we are neither working with real data nor aiming for parameter estimation--we compare the cleaned HI power spectrum directly to the input HI power spectrum from \texttt{OSKAR}. Therefore, any small bias in the normalisation can be safely neglected.

For analysis purposes, it is more useful to compute the 2D and 1D temperature power spectrum, which are just weighted averages of the 3D power spectrum. It is computed using the following equation,
\begin{equation} \label{eq: gridded power spectrum}
    \hat{P}_{\rm d}^\alpha = \frac{\sum_j w_j^\alpha P_{\rm d}^j}{\sum_j w_j^\alpha},
\end{equation}
where $\hat{P}_{\rm d}^\alpha$ is the delay power spectrum in the band power $\alpha$, $w_j^\alpha$ is a selection function which is one if the 3D delay power spectrum $j$ falls within the grid $\alpha$ and zero otherwise. For the 2D or cylindrical averaged power spectrum, the $\alpha$ grids are annulus in $\boldsymbol{k}_\perp$ space and $\hat{P}_{\rm d}^\alpha=\hat{P}_{\rm d}(|\boldsymbol{k}_\perp^\alpha|, k_\parallel^\alpha)$. For the 1D or spherically averaged power spectrum, the $\alpha$ grids are spherical shells in the $\boldsymbol{k}$ space, and $\hat{P}_{\rm d}^\alpha=\hat{P}_{\rm d}(|\boldsymbol{k}^\alpha|)$.

The error on the power spectrum, $\Delta\hat{P}_{\rm d}^\alpha$, can be calculated from the sampling variance \citep{Mazumder_2025MNRAS,Paul_a_first_detection}. This assumes that different bins are uncorrelated, i.e., that the covariance matrix is diagonal. As seen in Appendix~\ref{ap: norm}, visibilities can be considered independent when they are separated by more than $50\lambda$. Thus, as mentioned earlier, choosing a grid size of $\Delta u=\Delta v=60\lambda$ is the appropriate choice. The error on the power spectrum is computed with the following equation:
\begin{equation}
    \Delta\hat{P}_{\rm d}^\alpha = \sqrt{\frac{\sum_j (w_j^\alpha)^2 (P_{\rm d}^j-\hat{P}_{\rm d}^\alpha)^2}{\left(\sum_j w_j^\alpha\right)^2}}.
\end{equation}

\subsection{Validation of the power spectrum estimator}
\begin{figure}
    \centering
    \includegraphics[width=\columnwidth]{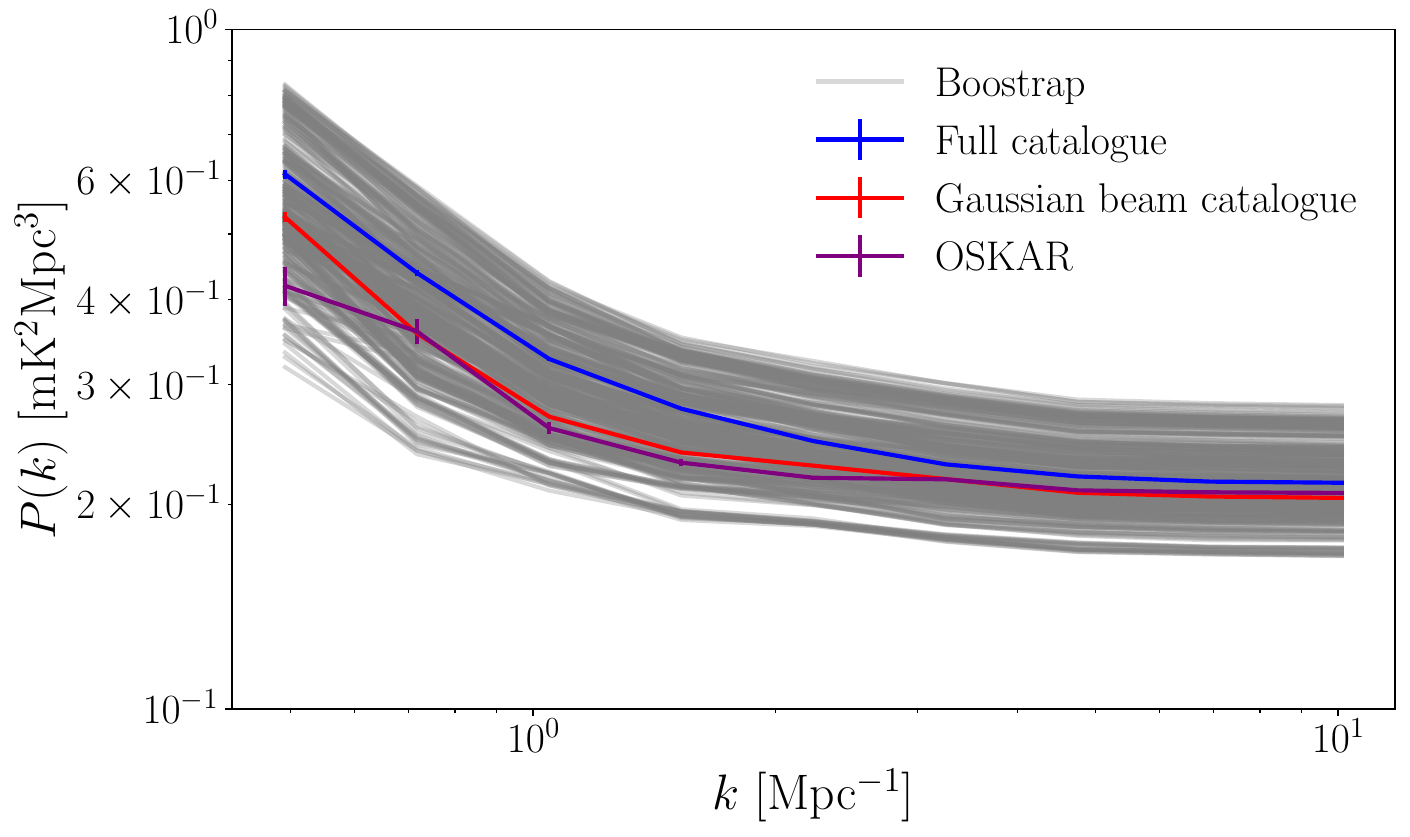}
    \caption{The HI power spectra for the case without line-broadening effects for the full catalogue (blue), the catalogue windowed with the Gaussian primary beam (red), and for the \texttt{OSKAR} simulation. In grey, the results of the HI power spectrum from the catalogue using a bootstrap over 400 different centres with the same field-of-view as MeerKAT.
    }
    \label{fig:HI_bootstrap}
\end{figure}

An important validation is to verify that the power spectrum computed directly from the catalogue matches the power spectrum from the interferometric simulations. In Fig.~\ref{fig:HI_bootstrap}, we compare the HI power spectrum computed from the \texttt{OSKAR} simulation with that obtained from the catalogue using the Gaussian primary beam, finding good agreement between them. The power spectrum from the full catalogue exhibits more power than that from the \texttt{OSKAR} simulation. The reason is that we are working with tiny cosmological volumes of $2.8\times 10^5$ Mpc$^3$, such that modes larger than the volume lead to systematic shifts in the measured power spectrum. To show this, in Fig.~\ref{fig:HI_bootstrap} we perform a bootstrap by computing the power spectrum in 400 squares with a field of view identical to MeerKAT, but at different locations in the catalogue.

These systematic biases show deviations with respect to the cosmological mean, which are totally unaccounted for by the sampling variance errors. One solution is to model the shifts as an additional contribution to the covariance called the Super Sample Covariance~\citep[SSC;][]{Pamuk_2026}, which would increase the error bars to be compatible with the cosmological mean. The other solution is to increase the observing area, for instance, by increasing the number of pointings as explained by \citet{Zhaoting_2023MNRAS}. As we do not carry out parameter estimation in this paper, we neglect the impact of this effect on the error bars.

\subsection{Power spectra of the sky model}
\begin{figure*}
    \centering
    \includegraphics[width=\textwidth]{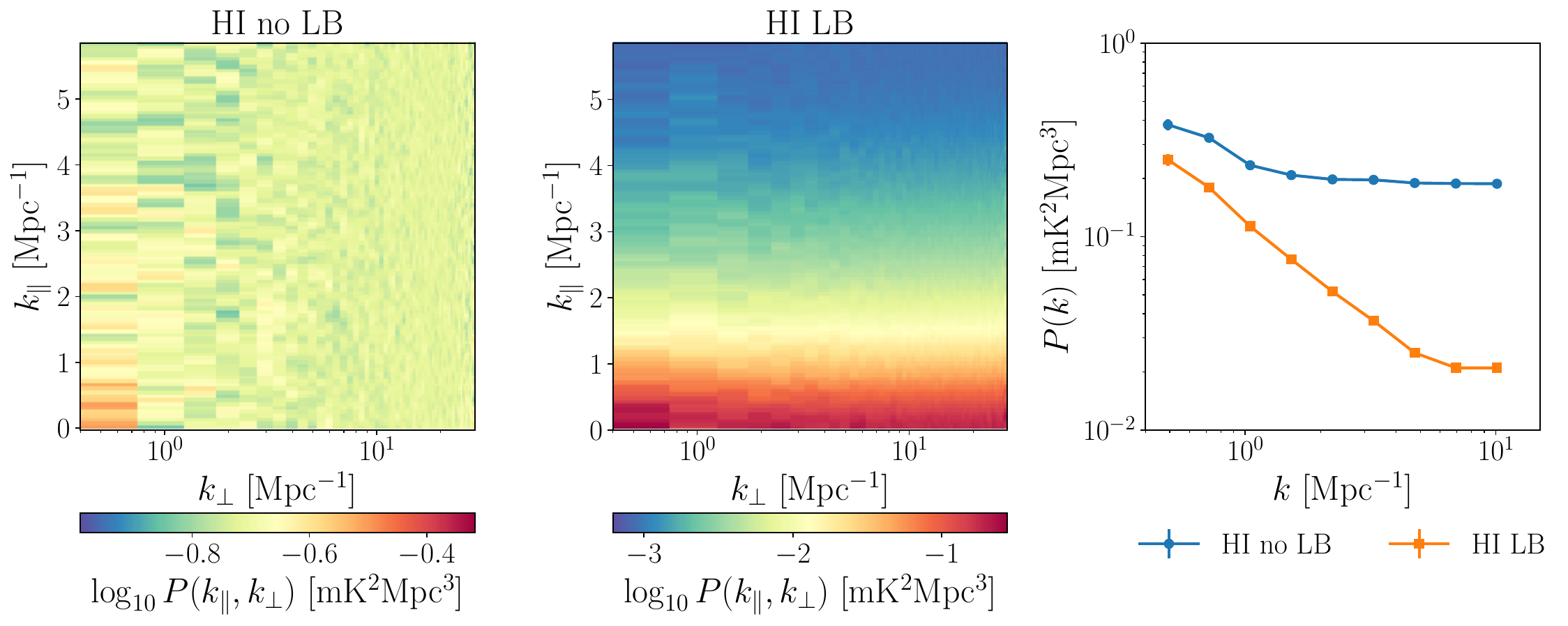}
    \caption{Left panel: 2D power spectrum of the HI signal without line-broadening effects (no LB). Middle panel: 2D power spectrum of the HI signal with line-broadening effects (LB). Right panel: comparison of the 1D power spectrum of the HI signal without (blue circles) and with line-broadening effects (orange squares).
    }
    \label{fig:HI_PS}
\end{figure*}

Using the methodology presented in Section~\ref{subsec: gridded PS}, we compute the 1D and 2D power spectrum of the different components: HI sources, diffuse Galactic foregrounds, radio-continuum sources, and noise. In Fig.~\ref{fig:HI_PS}, we have plotted the 1D and 2D HI power spectrum with and without line-broadening effects. Comparing the left and middle panels, we see that line broadening breaks spherical symmetry in the power spectrum. Line broadening due to the rotation and orientation of the HI galaxies is an example of the Fingers of God effect, which produces an elongation of the structures in the line-of-sight direction. At the 2D power spectrum level, it produces suppression along the $k_\parallel$ direction. The right panel of Fig.~\ref{fig:HI_PS} clearly illustrates that the impact of line-broadening effects is important at small scales ($k>0.5$ Mpc$^{-1}$). We observe a distinct departure in the shapes of the 1D power spectrum for the cases with and without line broadening.  The turning point observed around $k=6~\text{Mpc}^{-1}$ in the 1D power spectrum of HI in the presence of line-broadening effects is due to the frequency resolution of the experiment, which imposes a cut at a maximum $k_\parallel$ of $5.82$ Mpc$^{-1}$, as explained in \cite{Zhixing_2024MNRAS}.

\begin{figure*}
    \centering
    \includegraphics[width=\textwidth]{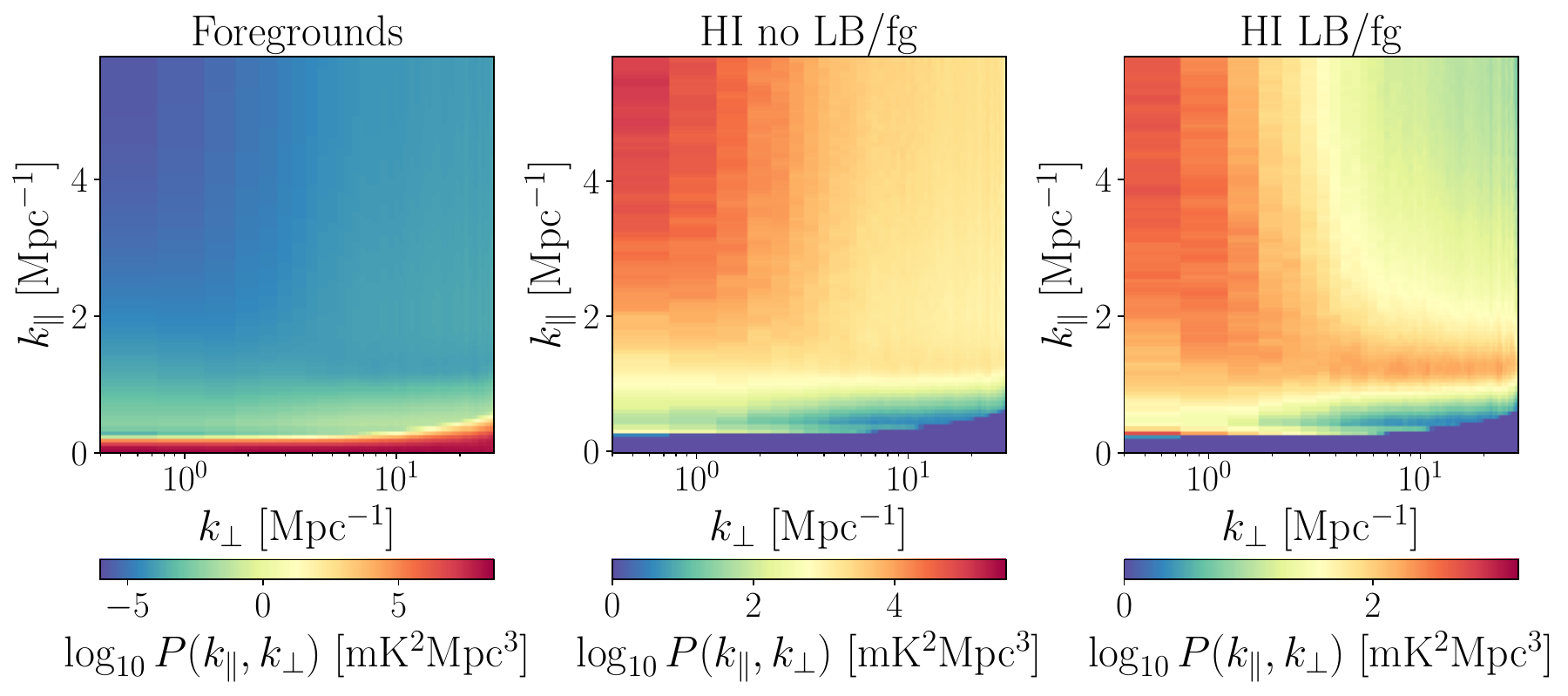}
    \caption{Left panel: 2D power spectrum of the total foregrounds including diffuse Galactic foregrounds (synchrotron and free-free) and radio-continuum sources. Middle and right panels: 2D power spectrum of the ratio between the HI signal and the foregrounds. The middle panel shows the HI case without line-broadening effects (no LB), and the right panel shows the HI case with line-broadening effects (LB).
    }
    \label{fig:foreground_wedge}
\end{figure*}

In the left panel of Fig.~\ref{fig:foreground_wedge}, we show the total foreground power including point sourced and diffuse emission. The foregrounds being spectrally smooth, the majority of the power is located at low $k_\parallel$. However, interferometers are inherently chromatic experiments, producing mode mixing along $k_\parallel$. This effect is more prominent at high $k_\perp$, where there is a larger dispersion along $k_\parallel$ as seen in the plot. This foreground-dominated region is the foreground wedge \citep{Datta2010, EoR_window_2014PhRvD}. In the middle and right panels of Fig.~\ref{fig:foreground_wedge}, we show the ratio between the HI signal (without and with line-broadening effects) and the foregrounds. This clearly points to the existence of a window free from foreground contamination, which can be used to reconstruct the HI signal. However, extensive studies show that several factors like instrumental systematics \citep[see e.g.][]{Joseph2018, 2022ApJ...941..207K, 2024MNRAS.534.3349C}, telescope environment \citep[see e.g.][]{offringa2015, Trott2018, 2025JCAP...02..058P} and data processing limitations \citep[see e.g.][]{2022ApJ...929..104C, 2022MNRAS.515.4020M, 2025ApJ...979..191C} can reduce the number of usable modes for HI signal recovery.

\begin{figure}
    \centering
    \includegraphics[width=\columnwidth]{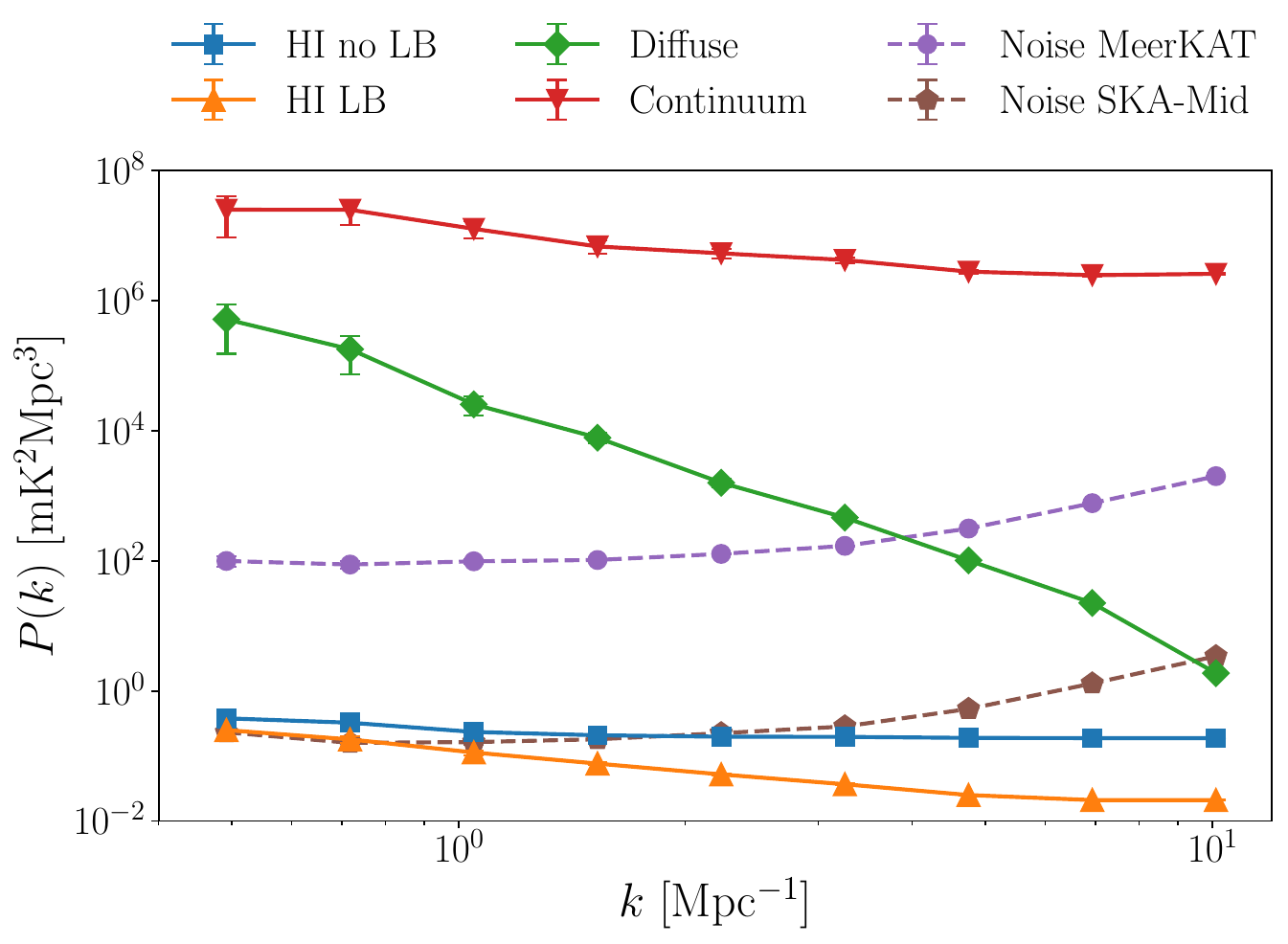}
    \caption{1D power spectrum of the different simulated components: HI without line-broadening effects (no LB; blue squares), HI with line-broadening effects (LB; orange upward triangles), diffuse Galactic emission (green diamonds), radio-continuum sources (red downward triangles), MeerKAT noise (purple circles), and SKA-Mid-like noise (brown pentagons). The error bars are multiplied by a factor $3$ for visualization purposes.
    }
    \label{fig:1d_PS}
\end{figure}

Finally, in Fig.~\ref{fig:1d_PS} we compare the amplitude of the different components simulated. We can see that the diffuse Galactic foregrounds and radio-continuum sources are several orders of magnitude brighter than the HI signal. The noise for the MeerKAT case, described in Section~\ref{sec:instrument}, is also two orders of magnitude higher than the signal. Recent analysis on MeerKAT data required generating data splits and cross-correlating them to suppress the noise contribution \citep{Paul_a_first_detection}. For the SKA-Mid-like case, the noise is reduced by more than two orders of magnitude compared to the MeerKAT case. The amplitude of the HI signal is still uncertain, showing differences with respect to simulations. For instance, recent measurements by the CHIME collaboration at redshift $z\sim 1$, showed that the amplitude of the HI signal is underestimated by the IlustrisTNG simulations with a significance of $3.1\sigma$ for TNG100 and $4.0\sigma$ for TNG300 \citep{CHIME_interp_2026arXiv}. The results presented in this work are shown with respect to the HI model described in \ref{sec:HI model}, but the broad implications are independent of the model selected.

\section{Foreground mitigation strategies in visibility space} 
\label{sec:foreground_cleaning_methods}
In this section, we present a novel foreground mitigation technique for reconstructing the HI signal: the Generalised Visibility Internal Linear Combination (GVILC). This method combines Principal Component Analysis (PCA) for foreground subspace identification, with a ILC to suppress foreground residuals in the data. In Section~\ref{sec: GVILC}, we introduce GVILC for application to the ungridded visibilities. In Section~\ref{sec: GVILC_grid_vis}, we extend this framework to operate on gridded visibilities. Finally, in Section~\ref{sec: foreground avoidance}, we present a hybrid approach that combines foreground avoidance with GVILC.

The simulated visibilities, $\boldsymbol{V}_{\rm data}$, form a 2D matrix with dimensions $N_{\rm ch}\times N_{\rm u}$, where $N_{\rm ch}$ is the number of frequency channels, and $N_{\rm u}=N_{\rm bl} N_{\rm steps}$ is the product of the number of baselines, $N_{\rm bl}$, and the number of time-steps, $N_{\rm steps}$. It is a combination of HI signal ($\boldsymbol{V}_{\rm HI}$), diffuse Galactic emission ($\boldsymbol{V}_{\rm diff}$), radio-continuum sources ($\boldsymbol{V}_{\rm cont}$), and noise ($\boldsymbol{V}_{\rm noise}$) as shown in equation~\eqref{eq:Vobs}.

Before applying any foreground removal method, we separate the angular scales by dividing the visibilities into different annular regions in Fourier space. This enables us to perform scale-dependent cleaning, accounting for the difference in the foreground wedge effect on angular modes. We define $\boldsymbol{V}_{\rm data}^{(s)}$ as a 2D matrix containing all the visibilities falling into annulus $s$. It has dimensions $N_{\rm ch}\times N_{\rm u}^{(s)}$, where $N_{\rm u}^{(s)}$ are the number of visibilities inside an annulus $s$ for a single frequency channel. The foreground cleaning process can be described in terms of a mixing matrix, $\boldsymbol{B}^{(s)}\in\mathbb{C}^{N_{\rm ch}\times N_{\rm ch}}$, which projects the highest variance modes associated with the diffuse Galactic foregrounds and radio-continuum sources. Then, the cleaned visibilities for annulus $s$, $\boldsymbol{\hat{V}}_{\rm cleaned}^{(s)}$, can be written as,
\begin{equation}
\begin{aligned}
    \boldsymbol{\hat{V}}_{\rm cleaned}^{(s)} &=
    \boldsymbol{V}_{\rm data}^{(s)}-\boldsymbol{B}^{(s)} \boldsymbol{V}_{\rm data}^{(s)}=\boldsymbol{A}^{(s)} \boldsymbol{V}_{\rm data}^{(s)} =\\
    &=\boldsymbol{\hat{V}}_{\rm HI}^{(s)} + \boldsymbol{\hat{V}}_{\rm diff}^{(s)} + \boldsymbol{\hat{V}}_{\rm cont}^{(s)} + \boldsymbol{\hat{V}}_{\rm noise}^{(s)},
\end{aligned}
\end{equation}
where $\boldsymbol{A}^{(s)}=\boldsymbol{I}-\boldsymbol{B}^{(s)}\in \mathbb{C}^{N_{\rm ch}\times N_{\rm ch}}$ is the foreground cleaning matrix for annulus $s$, and $\hat{V}^{(s)}$ are the projected visibilities for each component for annulus $s$. In the foreground cleaning step, we expect to minimise the diffuse Galactic foregrounds and the radio-continuum source residuals, $\boldsymbol{\hat{V}}_{\rm diff}^{(s)} + \boldsymbol{\hat{V}}_{\rm cont}^{(s)}$, while preserving the majority of HI signal.

\begin{figure}
    \centering
    \includegraphics[width=\columnwidth]{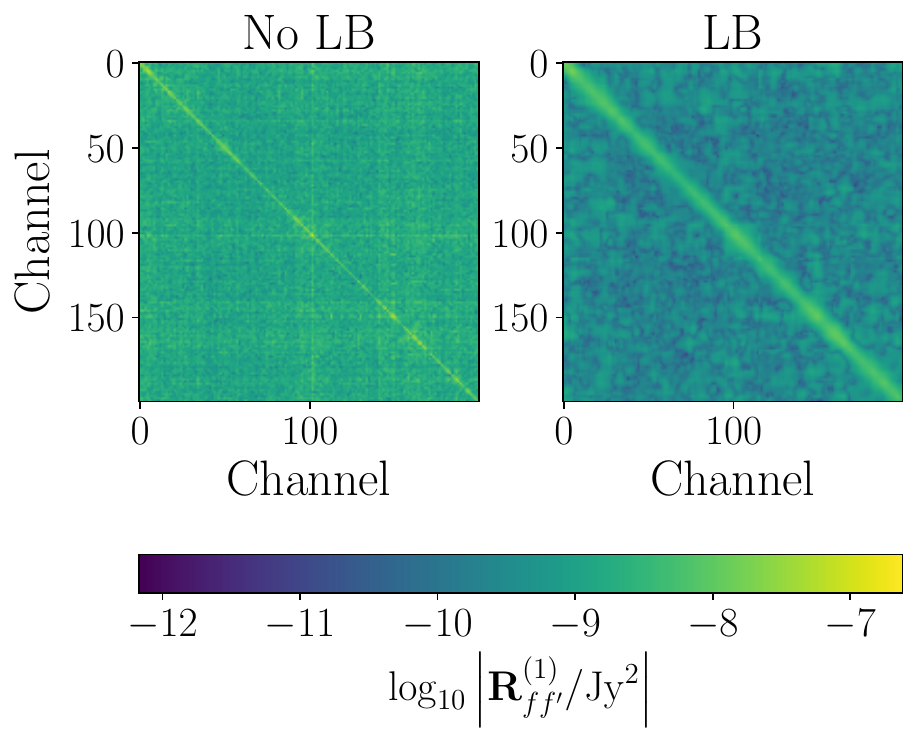}
    \caption{HI frequency-frequency covariances computed from the ungridded visibilities in the first annulus centred at $k_\perp=0.66$ Mpc$^{-1}$. Left panel: HI signal without line broadening (No LB). Right panel: HI signal with line broadening (LB).
    }
    \label{fig:HI_cov}
\end{figure}

To obtain the matrix $\boldsymbol{A}^{(s)}$, we need to compute the frequency-frequency covariance matrix for each annulus $s$. The annuli have variable widths to keep the number of visibilities per annulus constant and to ensure sufficient data for covariance-matrix convergence. In the foreground removal over the ungridded visibilities, we have 50000 visibilities per annulus (29 bins), and over the gridded visibilities we have 5000 visibilities per annulus (8 bins)\footnote{This is true except for the last bin, which has fewer visibilities.}. We find these values to be appropriate to have a converged covariance matrix. We can estimate the frequency-frequency covariance matrix for a pair of frequencies $(f,f')$, $\boldsymbol{R}_{ff'}^{(s)}$, from the data \citep{Dillon_2015PhRvD},
\begin{equation} \label{eq:freq-freq cov}
    \boldsymbol{R}_{ff'}^{(s)}=\sum_{(u,v)\in \mathcal{D}^{(s)}} 
    \frac{\big (V(u,v,f)-\langle V(f)\rangle\big )
    \big (V(u,v,f')-\langle V(f')\rangle\big )^\ast}{N^{(s)}-1},
\end{equation}
where $N^{(s)}$ are the number of points inside the annulus $s$, $\mathcal{D}^{(s)}$ is the set of points inside annulus $s$, and $\langle V(f)\rangle$ is the average of all $(u,v)$ points at frequency $f$ inside annulus $s$. In Fig.~\ref{fig:HI_cov}, we show the HI signal frequency-frequency covariance. As evident from the left panel, the covariance matrix is diagonal without line-broadening effects, while there are significant correlations among fifteen nearby frequency channels in the case with line-broadening effects. In this paper, we explore whether these off-diagonal correlations degrade the performance of the foreground removal methods.

\subsection{Generalized Visibility Internal Linear Combination} \label{sec: GVILC}
In this section, we present the Generalized Visibility Internal Linear Combination (GVILC) method, a new PCA-based method that uses data-driven statistical criteria to determine the number of modes to remove, followed by an ILC step to minimise the variance of the cleaned visibilities. It is based on the Generalised Needlet Internal Linear Combination (GNILC) method, which uses spatial and frequency information to separate components of the observed data \citep{Remazeilles_GNILC_2011MNRAS}. It was originally devised for extracting the total emission of the Galactic foregrounds from CMB data and used to deliver several data products for the \textit{Planck} mission \citep{Planck2015_CS_XLVIII_2016, Planck2018_CS_2020}. GNILC has also been adapted to single-dish HI intensity mapping at low redshift \citep{Olivari_GNILC_2016MNRAS} and applied to simulations of the BINGO experiment \citep{Bingo_GNILC_2022} and SKA-mid \citep{SKA_MID_GNILC2025}. In this paper, we extend GNILC for interferometric HI intensity mapping.

In principle, GNILC can be directly applied to the imaging products from the interferometric observations. However, as explained previously, intensity mapping using interferometric images for low redshifts introduces additional complications. Moreover, since most data observations we work with have small areas and are generally located away from the galactic plane, the foreground variability is not strong enough to be as important here as it is for large observing volumes like full-sky CMB or single-dish HI maps.

The GVILC method presented here is designed to operate on the ungridded visibilities, where the noise variance is constant over all the frequencies and visibilities. This method can be easily extended to the case of frequency-dependent noise. Our GVILC method follows a prescription similar to that described in \cite{Olivari_GNILC_2016MNRAS, Bingo_GNILC_2022}. GVILC is applied independently in each annulus, but the annulus dependence is omitted from the derivation for simplicity.

First, we estimate the HI plus noise prior frequency-frequency covariance matrix, $\boldsymbol{\hat{R}}_{\rm S}=\boldsymbol{\hat{R}}_{\rm HI}+\boldsymbol{\hat{R}}_{\rm noise}$, to run a constrained PCA driven by the local signal-plus-noise-to-foreground ratio. The signal covariance, $\boldsymbol{\hat{R}}_{\rm HI}$, is computed from a different realisation of the HI field, while the noise covariance, $\boldsymbol{\hat{R}}_{\rm noise}$, is calculated analytically as $\boldsymbol{\hat{R}}_{\rm noise}=\sigma_{\rm N}^2\boldsymbol{I}$. HI signal and noise have a very similar behaviour, and they are difficult to disentangle; for that reason, we decided to jointly reconstruct them. Today, SKA precursors like MeerKAT are noise dominated, so the prior is dominated by the noise covariance $\boldsymbol{\hat{R}}_{\rm S}\approx\boldsymbol{\hat{R}}_{\rm noise}$.

Using the prior covariance, we transform the visibilities:
\begin{equation}
    \boldsymbol{V}_{\rm data}^{\rm W}=\boldsymbol{\hat{R}}_{\rm S}^{-1/2}\boldsymbol{V}_{\rm data},
\end{equation}
where $\boldsymbol{V}_{\rm data}^{\rm W}$ are the transformed (whitened) visibilities. In that way, the covariance matrix of the transformed visibilities takes the following form,
\begin{equation}\label{eq: cov GVILC}
    \boldsymbol{\hat{R}}_{\rm S}^{-1/2}\boldsymbol{R}\boldsymbol{\hat{R}}_{\rm S}^{-1/2}=
    \boldsymbol{\hat{R}}_{\rm S}^{-1/2}\boldsymbol{R}_{\rm F}\boldsymbol{\hat{R}}_{\rm S}^{-1/2}+
    \boldsymbol{\hat{R}}_{\rm S}^{-1/2}\boldsymbol{R}_{\rm S}\boldsymbol{\hat{R}}_{\rm S}^{-1/2},
\end{equation}
where $\boldsymbol{R}=\boldsymbol{R}_{\rm F}+\boldsymbol{R}_{\rm S}$ is the data covariance matrix, implicitly receiving contributions from the foreground (continuum sources and diffuse Galactic emission), and signal-plus-noise, $\boldsymbol{R}_{\rm S}$, covariance matrices. If the prior is close to the real signal-plus-noise covariance matrix, then equation~\eqref{eq: cov GVILC} is reduced to,
\begin{equation}\label{eq: cov GVILC red}
    \boldsymbol{\hat{R}}_{\rm S}^{-1/2}\boldsymbol{R}\boldsymbol{\hat{R}}_{\rm S}^{-1/2}=
    \boldsymbol{\hat{R}}_{\rm S}^{-1/2}\boldsymbol{R}_{\rm F}\boldsymbol{\hat{R}}_{\rm S}^{-1/2}+
    \boldsymbol{\tilde{I}},
\end{equation}
where $\tilde{\boldsymbol{I}}$ is close to the identity matrix. Now, performing a spectral decomposition of equation~\eqref{eq: cov GVILC red},
\begin{equation} \label{eq: GVILC decomposition}
    \boldsymbol{\hat{R}}_{\rm S}^{-1/2}\boldsymbol{R}\boldsymbol{\hat{R}}_{\rm S}^{-1/2}=\boldsymbol{U}_{\rm F}\boldsymbol{D}_{\rm F}\boldsymbol{U}_{\rm F}^\dagger+
    \boldsymbol{U}_{\rm S}\boldsymbol{U}_{\rm S}^\dagger,
\end{equation}
where $\boldsymbol{D}_{\rm F}=\text{diag}(\lambda_1+1,\ldots, \lambda_m+1)$, and $\lambda_1>\lambda_2>\dots >\lambda_{m}>0$ are the eigenvalues associated with the foreground subspace ordered decreasingly. The decomposition in equation~\eqref{eq: GVILC decomposition} allows us to differentiate the foreground and signal subspace. The first $m$ eigenvalues larger than one correspond to the foreground subspace, and the last $N_{\rm ch}-m$ eigenvalues close to one correspond to the signal-plus-noise subspace. Therefore, the $N_{\rm ch}\times m$ matrix $\boldsymbol{U}_{\rm F}$ contains the eigenvectors associated to the foregrounds subspace, and the $N_{\rm ch}\times (N_{\rm ch}-m)$ matrix $\boldsymbol{U}_{\rm S}$ contains the eigenvectors associated to the signal-plus-noise subspace.

In the next step, we establish a method to determine the dimension $m$ of the foreground subspace. In PCA, the dimension of the foreground subspace is determined arbitrarily, with the expectation that the number of components chosen to be subtracted is close to the optimal. In GVILC, this number can be estimated from the data using equation~\eqref{eq: GVILC decomposition} and statistical selection methods. We propose two methods:
\begin{enumerate}
    \item \textbf{Akaike Information Criterion (AIC)}. Assuming that the data in the annulus follows a Gaussian distribution within each frequency channel (with channel-dependent variance), determining the number of components of the foreground subspace, $m_{\rm AIC}$, consists of solving,
    \begin{equation}
    m_{\rm AIC}=\arg\min_{m}\left\{
    2m + \sum_{i=m+1}^{N_{\rm ch}} \left( \mu_i - \log \mu_i - 1 \right)
    \right\},
    \end{equation}
    where $m\in\{1,\dots,N_{\rm ch}\}$ and $\mu_m$ are the eigenvalues of $\boldsymbol{\hat{R}}_{\rm S}^{-1/2}\boldsymbol{R}\boldsymbol{\hat{R}}_{\rm S}^{-1/2}$. The penalty term $2m$ tries to prevent excessive signal loss from happening. However, as we show in Section~\ref{sec:results}, in a noise-dominated scenario the penalty term leads to an under-cleaning situation where the foreground residuals are larger than the HI signal we try to reconstruct. It is expected to work much better on signal-dominated cases as shown by \cite{Olivari_GNILC_2016MNRAS}, where a full derivation of AIC in this context can be found.

    \item \textbf{Marchenko-Pastur Criterion (MPC)}. The signal-plus-noise covariance matrix after whitening is constructed from data that follow a zero-mean and unit-variance distribution, such that its eigenvalue distribution converges asymptotically to a Marchenko-Pastur (MP) distribution~\citep{Marchenko_1967, Lopez_de_Prado_2020}. The MP distribution has compact support on $[\lambda_-, \lambda_+]$ where
    \begin{equation}
    \lambda_{\pm} = \left(1\, \pm \, \sqrt{\frac{N_{\rm ch}}{N_{\rm vis}-1}}\right)^2,
    \end{equation}
    where $N_{\rm vis}$ is the number of visibilities inside the annulus, and we have used that $\sigma^2=1$. The $N_{\rm vis}-1$ factor arises from the loss of one degree of freedom when estimating and subtracting the sample mean before computing the covariance matrix. The eigenvalues smaller than $\lambda_+$ are considered part of the signal-plus-noise subspace, allowing to do a clean distinction between the foreground and the signal-plus-noise subspaces. The MPC determines the dimension of the foreground subspace, $m_{\rm MPC}$, as
    \begin{equation}
        m_{\rm MPC}=\max\left\{m:\mu_m > \lambda_+\right\}.
    \end{equation}
    
    This criterion is also particularly useful for detecting mismatches between the data and the prior. If the distribution of eigenvalues with values around one of the frequency-frequency covariance does not follow an MP distribution, it is indicative of issues regarding systematics or the prior. A more detailed explanation of the MP distribution is provided in the Appendix~\ref{ap:MP_dist}.
\end{enumerate}

The last step is to perform an $(N_{\rm ch}-m)$-dimensional ILC to the observed data projected onto the signal-plus-noise subspace identified by the PCA + AIC/MPC step. The objective is to minimise the total variance of the visibilities while preserving the HI signal and noise, i.e., minimising the foreground component leaking into the signal-plus-noise subspace. The foreground cleaning matrix for GVILC is defined as~\citep{Olivari_GNILC_2016MNRAS},
\begin{equation} \label{eq:ILC_filter}
    \boldsymbol{A} = \boldsymbol{\hat{S}}\left(\boldsymbol{\hat{S}}^{\dagger}\boldsymbol{R}^{-1}\boldsymbol{\hat{S}}\right)^{-1}\boldsymbol{\hat{S}}^{\dagger}\boldsymbol{R}^{-1},
\end{equation}
where the matrix $\boldsymbol{\hat{S}}$ is connected to the projected signal-plus-noise prior covariance by $\boldsymbol{\hat{R}}_{\rm S}^{\rm p}=\boldsymbol{\hat{S}}\boldsymbol{\hat{S}}^{\dagger}$. To compute that, first we need to use the projector into the signal-plus-noise subspace defined as $\boldsymbol{\Pi}_{\rm S}=\boldsymbol{U}_{\rm S}\boldsymbol{U}_{\rm S}^\dagger$, and to apply this to the whitened data to get the signal-plus-noise visibilities in the whitened space,
\begin{equation}
    \boldsymbol{V}_{\rm S}^{\rm W} = \boldsymbol{\Pi}_{\rm S}\boldsymbol{V}_{\rm HI+noise}^{\rm W},
\end{equation}
which after going back to the original space we obtain:
\begin{equation}
    \boldsymbol{V}_{\rm S} = \boldsymbol{\hat{R}}_{\rm S}^{1/2}\boldsymbol{\Pi}_{\rm S}\boldsymbol{\hat{R}}_{\rm S}^{-1/2}\boldsymbol{V}_{\rm HI+noise},
\end{equation}
where this operation is the traditional projection of the data into the signal-plus-noise subspace, as we were doing in PCA but with the prior covariance matrix to go back to the original space. We then compute the covariance of the projected signal-plus-noise covariance, $\boldsymbol{\hat{R}}_{\rm S}^{\rm p}$, using the prior covariance matrix, $\text{Cov}(\boldsymbol{V}_{\rm HI+noise})=\boldsymbol{\hat{R}}_{\rm S}$:
\begin{equation}
\begin{aligned}
    \boldsymbol{\hat{R}}_{\rm S}^{\rm p} = \text{Cov}(\boldsymbol{V}_{\rm S}) &= \boldsymbol{\hat{R}}_{\rm S}^{1/2}\boldsymbol{\Pi}_{\rm S}\boldsymbol{\hat{R}}_{\rm S}^{-1/2}\text{Cov}(\boldsymbol{V}_{\rm HI+noise})\boldsymbol{\hat{R}}_{\rm S}^{-1/2}\boldsymbol{\Pi}_{\rm S}\boldsymbol{\hat{R}}_{\rm S}^{1/2}=\\
    &=\boldsymbol{\hat{R}}_{\rm S}^{1/2}\boldsymbol{\Pi}_{\rm S}^2\boldsymbol{\hat{R}}_{\rm S}^{1/2}=\boldsymbol{\hat{R}}_{\rm S}^{1/2}\boldsymbol{U}_{\rm S}\boldsymbol{U}_{\rm S}^\dagger\boldsymbol{\hat{R}}_{\rm S}^{1/2},
\end{aligned}
\end{equation}
where we have used the projector property $\boldsymbol{\Pi}_{\rm S}^2=\boldsymbol{\Pi}_{\rm S}$, and we conclude that $\boldsymbol{\hat{S}}=\boldsymbol{\hat{R}}_{\rm S}^{1/2}\boldsymbol{U}_{\rm S}$. Finally, if we substitute $\boldsymbol{\hat{S}}$ in equation~\eqref{eq:ILC_filter} and use the orthonormality condition,
\begin{equation}
    \boldsymbol{\hat{R}}_{\rm S}^{1/2}(\boldsymbol{U}_{\rm S}\boldsymbol{U}_{\rm S}^\dagger+\boldsymbol{U}_{\rm F}\boldsymbol{U}_{\rm F}^\dagger)\boldsymbol{\hat{R}}_{\rm S}^{-1/2}=\boldsymbol{I},
\end{equation}
we get,
\begin{equation}
\begin{aligned}
    \boldsymbol{A} &= \boldsymbol{\hat{R}}_{\rm S}^{1/2}\boldsymbol{U}_{\rm S}\left(\boldsymbol{U}_{\rm S}^\dagger\boldsymbol{\hat{R}}_{\rm S}^{1/2}\boldsymbol{R}^{-1}\boldsymbol{\hat{R}}_{\rm S}^{1/2}\boldsymbol{U}_{\rm S}\right)^{-1}\boldsymbol{U}_{\rm S}^\dagger\boldsymbol{\hat{R}}_{\rm S}^{1/2}\boldsymbol{R}^{-1}\\
    &= \boldsymbol{\hat{R}}_{\rm S}^{1/2}\boldsymbol{\Pi}_{\rm S}\boldsymbol{\hat{R}}_{\rm S}^{-1/2}\\
    & + \boldsymbol{\hat{R}}_{\rm S}^{1/2}\boldsymbol{U}_{\rm S}\left(\boldsymbol{U}_{\rm S}^\dagger\boldsymbol{\hat{R}}_{\rm S}^{1/2}\boldsymbol{R}^{-1}\boldsymbol{\hat{R}}_{\rm S}^{1/2}\boldsymbol{U}_{\rm S}\right)^{-1}\boldsymbol{U}_{\rm S}^\dagger\boldsymbol{\hat{R}}_{\rm S}^{1/2}\boldsymbol{R}^{-1}\boldsymbol{\hat{R}}_{\rm S}^{1/2}\boldsymbol{U}_{\rm F}\boldsymbol{U}_{\rm F}^\dagger\boldsymbol{\hat{R}}_{\rm S}^{-1/2},
\end{aligned}
\end{equation}
where the $\boldsymbol{A}$ matrix can be decomposed in two terms. The first term is the conventional projection onto the signal-plus-noise subspace determined by the PCA, whereas the second term, arising from the ILC, suppresses the leakage of foreground components into the signal-plus-noise subspace. The key is the factor $\boldsymbol{U}_{\rm S}^\dagger\boldsymbol{\hat{R}}_{\rm S}^{1/2}\boldsymbol{R}^{-1}\boldsymbol{\hat{R}}_{\rm S}^{1/2}\boldsymbol{U}_{\rm F}$ which is equal to zero in the case of perfect separation between the signal and noise subspaces, and different from zero otherwise. In the case of perfect separation, the signal-plus-noise and foreground subspaces are orthogonal, $\boldsymbol{U}_{\rm S}^\dagger \boldsymbol{U}_{\rm F} = \boldsymbol{0}$, and the covariance matrices are block-diagonal in this basis, such that there is no coupling between the two subspaces.

\subsection{GVILC on the gridded visibilities} \label{sec: GVILC_grid_vis}
Motivated by the decreased noise levels and the reduced computational cost of the gridded visibilities, we decided to adapt GVILC to work on the gridded visibilities. The main difference with the raw visibilities is that the noise variance is no longer constant and changes from one gridded visibility to another. To account for this heterogeneity of variance, we apply the whitening with a visibility-visibility prior covariance matrix instead of a frequency-frequency prior covariance matrix. This method assumes that we are in a noise-dominated regime, such as MeerKAT DEEP2 observation, and the prior visibility-visibility covariance can be reduced to the noise covariance, $\boldsymbol{\hat{Q}}_{\rm noise}$. We note that off-diagonal correlations are suppressed by the primary beam and the large grid size used of $\Delta u = 60\lambda$, so the diagonal approximation is justified. Using this prior, we transform the visibilities:
\begin{equation}
    \boldsymbol{V}_{\rm data}^{\rm W}=\boldsymbol{V}_{\rm data}\boldsymbol{\hat{Q}}_{\rm noise}^{-1/2},
\end{equation}
where $\boldsymbol{V}_{\rm data}^{\rm W}$ are the transformed visibilities (whitened) and
\begin{equation}
    \boldsymbol{\hat{Q}}_{\rm noise} = \text{diag}(\sigma_{\rm N}^2/N_1,\sigma_{\rm N}^2/N_2, \ldots, \sigma_{\rm N}^2/N_{\rm N_{\rm vis}}),
\end{equation}
where $(N_1, N_2,\ldots,N_{\rm N_{\rm vis}})$ are the number of visibilities per grid, and $N_{\rm vis}$ is the number of visibilities inside the annulus. In that way, the frequency-frequency covariance matrix of the transformed visibilities takes the following form,
\begin{equation}\label{eq: cov GVILC gridded}
    \text{Cov}(\boldsymbol{V}_{\rm data}^{\rm W})=\frac{1}{N_{\rm vis}-1} \boldsymbol{X}_{\rm data}\boldsymbol{\hat{Q}}_{\rm noise}^{-1}\boldsymbol{X}_{\rm data}^\dagger,
\end{equation}
where $\boldsymbol{X}_{\rm data}=\boldsymbol{V}_{\rm data} - \langle\boldsymbol{V}_{\rm data}\rangle$. The frequency-frequency covariance can be expanded in the different components of the visibilities in the following way,
\begin{equation}\label{eq: cov GVILC gridded expanded}
\begin{aligned}
    \text{Cov}(\boldsymbol{V}_{\rm data}^{\rm W})&=\frac{1}{N_{\rm vis}-1}\Big[ (\boldsymbol{X}_{\rm HI}+\boldsymbol{X}_{\rm noise})\boldsymbol{\hat{Q}}_{\rm noise}^{-1}(\boldsymbol{X}_{\rm HI}+\boldsymbol{X}_{\rm noise})^\dagger\\
    &\quad + \boldsymbol{X}_{\rm FG}\boldsymbol{\hat{Q}}_{\rm noise}^{-1}\boldsymbol{X}_{\rm FG}^\dagger +(\boldsymbol{X}_{\rm HI}+\boldsymbol{X}_{\rm noise})\boldsymbol{\hat{Q}}_{\rm noise}^{-1}\boldsymbol{X}_{\rm FG}^\dagger\\
    &\qquad +\boldsymbol{X}_{\rm FG}\boldsymbol{\hat{Q}}_{\rm noise}^{-1}(\boldsymbol{X}_{\rm HI}+\boldsymbol{X}_{\rm noise})^\dagger \Big],
\end{aligned}
\end{equation}
where in the first term, if the prior is close to the true signal-plus-noise covariance matrix, the entries of 
\begin{equation} \label{eq: HI_plus_noise cov term}
    (\boldsymbol{X}_{\rm HI}+\boldsymbol{X}_{\rm noise})\boldsymbol{\hat{Q}}_{\rm noise}^{-1/2},
\end{equation}
are approximately distributed as independent and identically distributed standard normal variables. This happens because noise dominates and is not correlated across frequencies. In a signal dominated scenario, off-diagonal correlations introduced by line broadening would break this approximation. In that situation, we have that the expectation value of the covariance matrix is,
\begin{equation}
    \left\langle\frac{1}{N_{\rm vis}-1}(\boldsymbol{X}_{\rm HI}+\boldsymbol{X}_{\rm noise})\boldsymbol{\hat{Q}}_{\rm noise}^{-1}(\boldsymbol{X}_{\rm HI}+\boldsymbol{X}_{\rm noise})^\dagger\right\rangle = \boldsymbol{\tilde{I}},
\end{equation}
where $\tilde{\boldsymbol{I}}$ is close to the identity matrix. The last two terms of equation~\eqref{eq: cov GVILC gridded expanded} are approximately zero. This is because the foregrounds and the dominant noise component are uncorrelated, and the correlation between the foregrounds and the subdominant HI signal, although present in the case with matched HI and radio-continuum sources, is compatible with zero. 

Simplifying equation~\eqref{eq: cov GVILC gridded expanded}, we get the following expression:
\begin{equation}\label{eq: cov GVILC grid red}
    \text{Cov}(\boldsymbol{V}_{\rm data}^{\rm W})=
    \frac{1}{N_{\rm vis}-1}\boldsymbol{X}_{\rm FG}\boldsymbol{\hat{Q}}_{\rm noise}^{-1}\boldsymbol{X}_{\rm FG}^\dagger+\boldsymbol{\tilde{I}}.
\end{equation}

Now, performing a spectral decomposition of equation~\eqref{eq: cov GVILC grid red},
\begin{equation} \label{eq: GVILC grid decomposition}
    \text{Cov}(\boldsymbol{V}_{\rm data}^{\rm W})=\boldsymbol{U}_{\rm F}\boldsymbol{D}_{\rm F}\boldsymbol{U}_{\rm F}^\dagger+
    \boldsymbol{U}_{\rm S}\boldsymbol{U}_{\rm S}^\dagger,
\end{equation}
where $\boldsymbol{D}_{\rm F}=\text{diag}(\lambda_1+1,\ldots, \lambda_m+1)$, and $\lambda_1>\lambda_2>\dots >\lambda_{m}>0$ are the eigenvalues associated with the foreground subspace ordered decreasingly. Both the AIC and MPC methods described in Section~\ref{sec: GVILC} can be used in the gridded case to determine the dimension of the foreground subspace. Although the noise is heterogeneous, the whitening step makes the visibilities identically distributed, allowing both methods to be applied.

Finally, we compute the foreground cleaning matrix $\boldsymbol{A}$ using the ILC in equation~\eqref{eq:ILC_filter}. To do so, we compute the matrix $\boldsymbol{\hat{S}}$ which is connected to the propagated signal-plus-noise covariance by $\boldsymbol{\hat{R}}_{\rm S}^{\rm p}=\boldsymbol{\hat{S}}\boldsymbol{\hat{S}}^{\dagger}$. We thus use the projector into the signal-plus-noise subspace defined as $\boldsymbol{\Pi}_{\rm S}=\boldsymbol{U}_{\rm S}\boldsymbol{U}_{\rm S}^\dagger$, and apply this to the whitened data to get the signal-plus-noise visibilities in the whitened space,
\begin{equation}
    \boldsymbol{V}_{\rm S}^{\rm W} = \boldsymbol{\Pi}_{\rm S}\boldsymbol{V}_{\rm HI+noise}^{\rm W},
\end{equation}
which, after going back to the original space, we get the following:
\begin{equation}
    \boldsymbol{V}_{\rm S} = \boldsymbol{\Pi}_{\rm S}\boldsymbol{V}_{\rm HI+noise}\boldsymbol{\hat{Q}}_{\rm noise}^{-1/2}\boldsymbol{\hat{Q}}_{\rm noise}^{1/2}=\boldsymbol{\Pi}_{\rm S}\boldsymbol{V}_{\rm HI+noise}.
\end{equation}

Next, we compute the projected signal-plus-noise prior covariance, $\boldsymbol{\hat{R}}_{\rm S}^{\rm p}$, using the prior frequency-frequency covariance matrix, $\text{Cov}(\boldsymbol{V}_{\rm HI+noise})\approx\boldsymbol{\hat{R}}_{\rm noise}=\sigma_{\rm eff}^2\boldsymbol{I}$:
\begin{equation}
    \boldsymbol{\hat{R}}_{\rm S}^{\rm p} = \text{Cov}(\boldsymbol{V}_{\rm S}) = \boldsymbol{\Pi}_{\rm S}\text{Cov}(\boldsymbol{V}_{\rm HI+noise})\boldsymbol{\Pi}_{\rm S}=\boldsymbol{\Pi}_{\rm S}\boldsymbol{\hat{R}}_{\rm noise}\boldsymbol{\Pi}_{\rm S}=\sigma_{\rm eff}^2\boldsymbol{\Pi}_{\rm S},
\end{equation}
where
\begin{equation}
    \sigma_{\rm eff}^2 = \frac{1}{N_{\rm vis}}\sum_{i=1}^{N_{\rm vis}} \frac{\sigma_{\rm N}^2}{N_i},
\end{equation}
and from which we conclude that $\boldsymbol{\hat{S}}=\sigma_{\rm eff}\boldsymbol{U}_{\rm S}$. The foreground cleaning matrix, given in equation~\eqref{eq:ILC_filter}, is invariant under a multiplication of the matrix $\boldsymbol{\hat{S}}$ by a constant. We therefore use $\boldsymbol{\hat{S}}=\boldsymbol{U}_{\rm S}$ instead.

\subsection{Foreground avoidance} \label{sec: foreground avoidance}
Diffuse foregrounds and radio-continuum sources are smooth in frequency, which means that in the delay power spectrum, they are confined in the low $k_\parallel$ region (foreground wedge). By filtering out these contaminated modes, the HI signal can be recovered. Although this method does not suffer from signal loss like GVILC, foreground avoidance discards modes containing the HI signal, leading to increased error in the reconstructed power spectrum. The foreground avoidance region can be defined with the following inequality,
\begin{equation} \label{eq: fa}
    k_\parallel < Ak_\perp + B,
\end{equation}
where $A$ and $B$ are the foreground avoidance parameters. The $A$ term is connected to the chromatic response of a radio interferometer. For a radio interferometer with a primary beam FWHM $\theta_{\rm B}$, the foregrounds should ideally be confined inside the foreground wedge~\citep{EoR_window_2014PhRvD, Kothari_2024JCAP},
\begin{equation} \label{eq: foreground_wedge}
    k_\parallel < \frac{XH(z_0)\theta_{\rm B}}{c(1+z_0)}k_\perp,
\end{equation}
where for our MeerKAT observation we get the value $k_\parallel < 0.01k_\perp$. We adopt a higher value of $k_\parallel < 0.02k_\perp$ to make sure that any foreground residual is inside the HI window before foreground cleaning. The $B$ term is related to the frequency smoothness of the foregrounds, which made them reside in the low-$k_\parallel$ region. We find that a value of $B=0.25$ is needed in order to avoid the foregrounds in the 2D power spectrum before foreground cleaning. In the context of interferometric HI intensity mapping, the term foreground wedge is commonly used to refer to the foreground-dominated region of Fourier space. We therefore adopt this convention. Strictly speaking, however, the wedge is the characteristic region of power spectrum space where the instrument's chromatic response causes foreground power to leak into higher $k_\parallel$ modes.

Foreground avoidance changes the selection function in equation~\eqref{eq: gridded power spectrum}. For the grids $\alpha$ that verify the relation \eqref{eq: fa}, the selection function $w_j^\alpha$ is zero. A natural extension is to apply foreground avoidance after foreground removal by GVILC, which we name the hybrid approach. As we will show in Section~\ref{sec:results}, the hybrid approach allows for a less aggresive foreground avoidance, granting access to Fourier modes previously contaminated by foregrounds.

\section{Results} \label{sec:results}
In this section, we present the performance of the different foreground removal methods with respect to foreground residuals, signal loss, and modes lost. Additionally, we explore whether physical effects such as line-broadening effects of the HI line and spatial correlation of the HI signal and the radio-continuum sources have any kind of impact on the foreground removal process. Finally, we compare the performance of applying a foreground removal method such as GVILC, with or without foreground avoidance, to a pure foreground avoidance strategy.

\subsection{Cleaned power spectrum} \label{sec:cleaned_power_spectrum}

\begin{figure*}
    \centering
    \includegraphics[width=\textwidth]{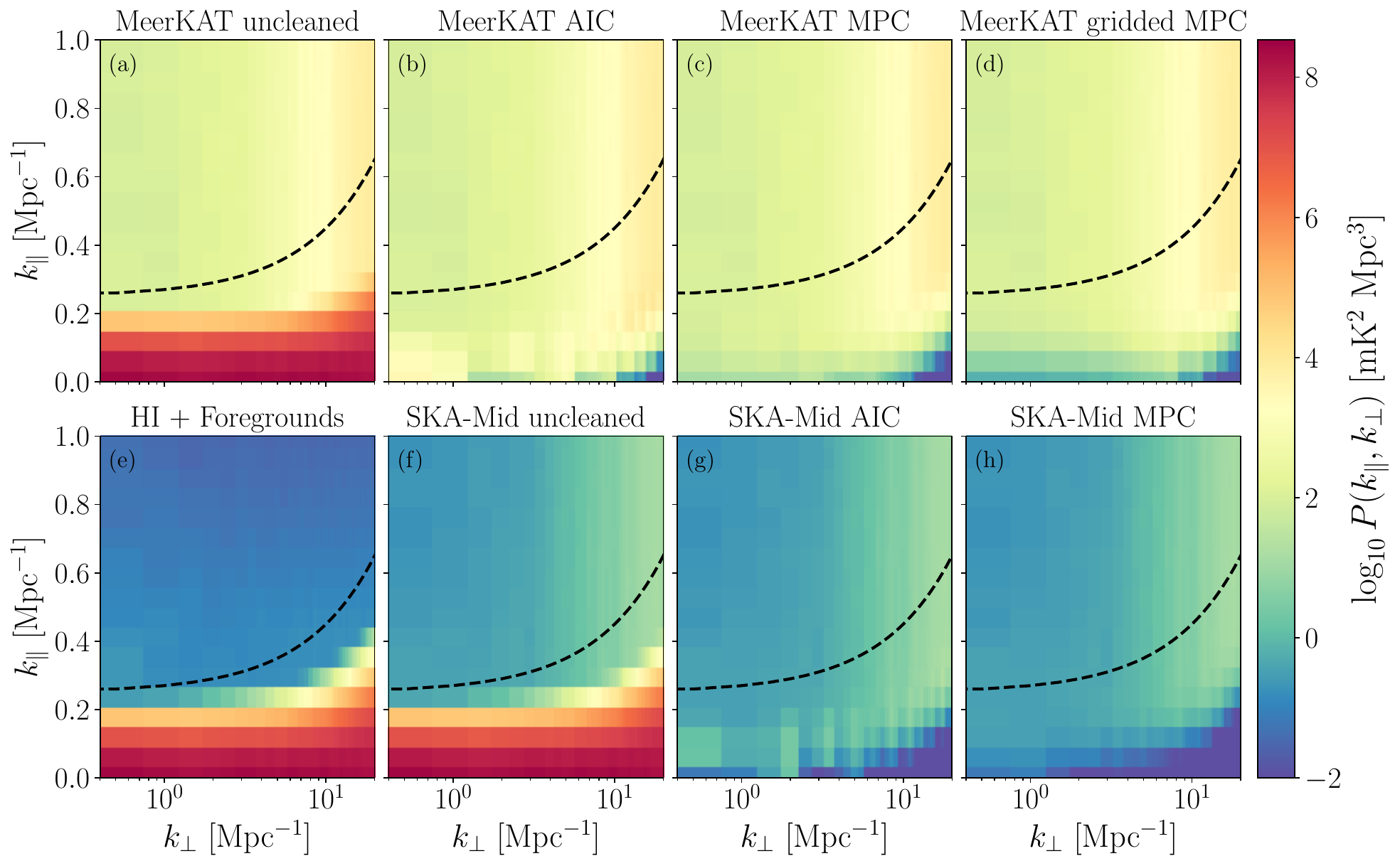}
    \caption{Panels (a)-(d): 2D power spectra for MeerKAT corresponding to the uncleaned and GVILC-cleaned cases. Panel (e): 2D power spectrum of the input HI plus foregrounds. Panels (f)–(h): 2D power spectra for SKA-Mid corresponding to the uncleaned and GVILC-cleaned cases. The dashed black lines correspond approximately to the size of the foreground wedge ($k_\parallel=0.02k_\perp+0.25$). All panels correspond to the case with line-broadening effects and matched HI and radio-continuum sources. The difference in noise levels between MeerKAT and SKA-Mid is apparent outside the foreground wedge.}
    \label{fig:power_spectrum_2D_cleaned}
\end{figure*}

\begin{figure*}
    \centering
    \includegraphics[width=0.8\textwidth]{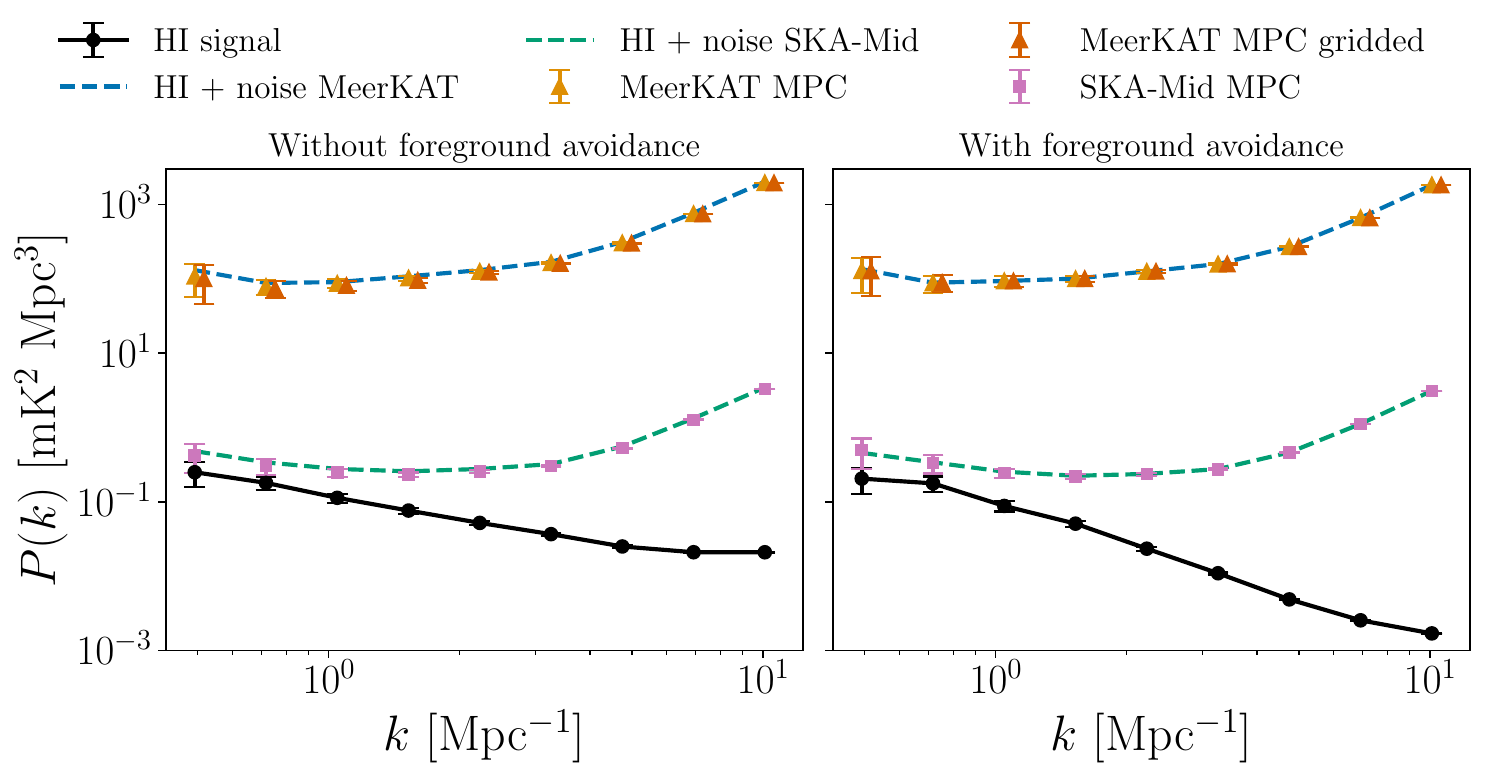}
    \caption{Cleaned 1D power spectra after applying GVILC for the case with line-broadening effects and matched HI and radio-continuum sources. The left panel shows the cleaned power spectra without applying any foreground avoidance strategy. The right panel shows the cleaned power spectra after applying a foreground avoidance strategy of $k_\parallel < 0.25k_\perp$. The black solid line corresponds to the input HI signal, while the dashed lines correspond to the HI signal plus noise both with the same foreground avoidance strategy as the simulations. The error bars are multiplied by a factor $5$ for visualization purposes.
    }
    \label{fig:power_spectrum_1D_cleaned}
\end{figure*} 

In this section, we show the 1D and 2D power spectra of the cleaned visibilities. In Fig.~\ref{fig:power_spectrum_2D_cleaned}, we show the foreground cleaned 2D power spectra for the different experiments and GVILC methods. We do not apply GVILC gridded method to the SKA-Mid case because we are no longer in a noise-dominated regime and its assumptions does not hold. First, looking at GVILC AIC-cleaned power spectra (Fig.~\ref{fig:power_spectrum_2D_cleaned} panels (b) and (g)), we can see that foreground residuals remain in the foreground wedge region which are above the noise level for both MeerKAT and SKA-Mid cases when compared to the power spectrum outside the foreground wedge. In contrast, GVILC MPC-cleaned power spectra (panels (c), (d) and (h)) show no appreciable foreground residuals for both MeerKAT and SKA-Mid, indicating that they lie below the noise level. For the MeerKAT case, we have applied GVILC MPC for both the ungridded and gridded visibilities, finding lower power in the foreground wedge region for the gridded case. We can conclude that GVILC significantly reduces foreground residuals by comparing the GVILC-cleaned power spectra with panels (a) and (f). Another important feature is the bluish regions located inside the foreground wedge. This indicates that we are not only removing the foregrounds, but also the HI signal, leading to an over-cleaning situation. This over-cleaning is not a problem because this part of the $(k_\parallel, k_\perp)$ plane is dominated by foregrounds, so the HI signal is not recoverable and we would not use this part for cosmological/scientific analysis.  

In Fig.~\ref{fig:power_spectrum_1D_cleaned}, the cleaned 1D power spectra are plotted for GVILC MPC applied to the ungridded visibilities from both MeerKAT and SKA-Mid, and to the gridded visibilities for MeerKAT. We decide to only show the GVILC MPC as it is the method showing the best performance in terms of foreground residuals. Looking to the left panel of Fig.~\ref{fig:power_spectrum_1D_cleaned}, we can conclude that GVILC is capable of reconstructing the HI signal plus noise with negligible foreground residuals without relying on foreground avoidance. Signal loss is still present, which can be appreciated in the first two band powers of the left panel when compared with the corresponding dashed lines of the input HI signal plus noise. This can be corrected by applying a foreground avoidance strategy, which enables us to obtain an unbiased estimate of the HI signal-plus-noise power spectrum, as shown in the right panel of Fig.~\ref{fig:power_spectrum_1D_cleaned}. If no foreground avoidance had been applied to the unclean data, important foreground residuals would have been observed. While some modes are completely lost due to foregrounds, GVILC combined with a mild foreground avoidance can still recover previously inaccessible foreground-dominated modes.  However, the goal is to reconstruct the HI signal without noise bias. This is usually done by cross-correlating different data splits to get a noise-free power spectrum estimation \citep{Mazumder_2025MNRAS, Paul_a_first_detection}. For that reason, it is important to analyse the foreground residuals and signal loss, which is done in Sections~\ref{sec:foreground_residual} and \ref{sec:signal_loss}. 

\subsection{Foreground residuals} \label{sec:foreground_residual}

\begin{figure*}
    \centering
    \includegraphics[width=\textwidth]{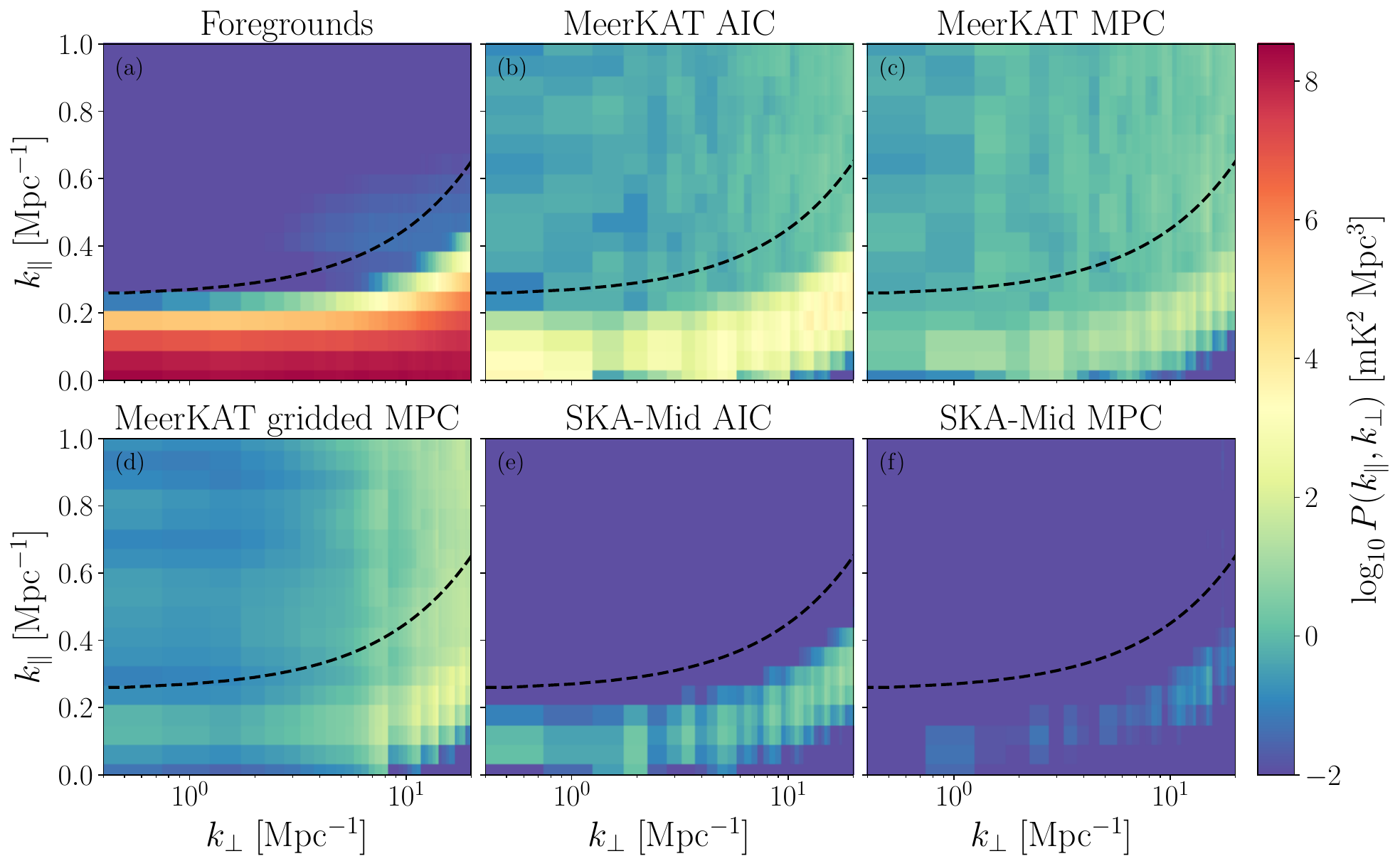}
    \caption{Panel (a): 2D power spectrum of the foregrounds consisting of radio-continuum sources and diffuse Galactic emission. Panels (b)-(f): foreground residuals for the different experiments and GVILC cleaning methods. The dashed black lines correspond approximately to the size of the foreground wedge ($k_\parallel=0.02k_\perp+0.25$). All the panels correspond to the case with line-broadening effects and matched HI and radio-continuum sources.
    }
    \label{fig:foreground_residuals_2D}
\end{figure*}

\begin{figure}
    \centering
    \includegraphics[width=\columnwidth]{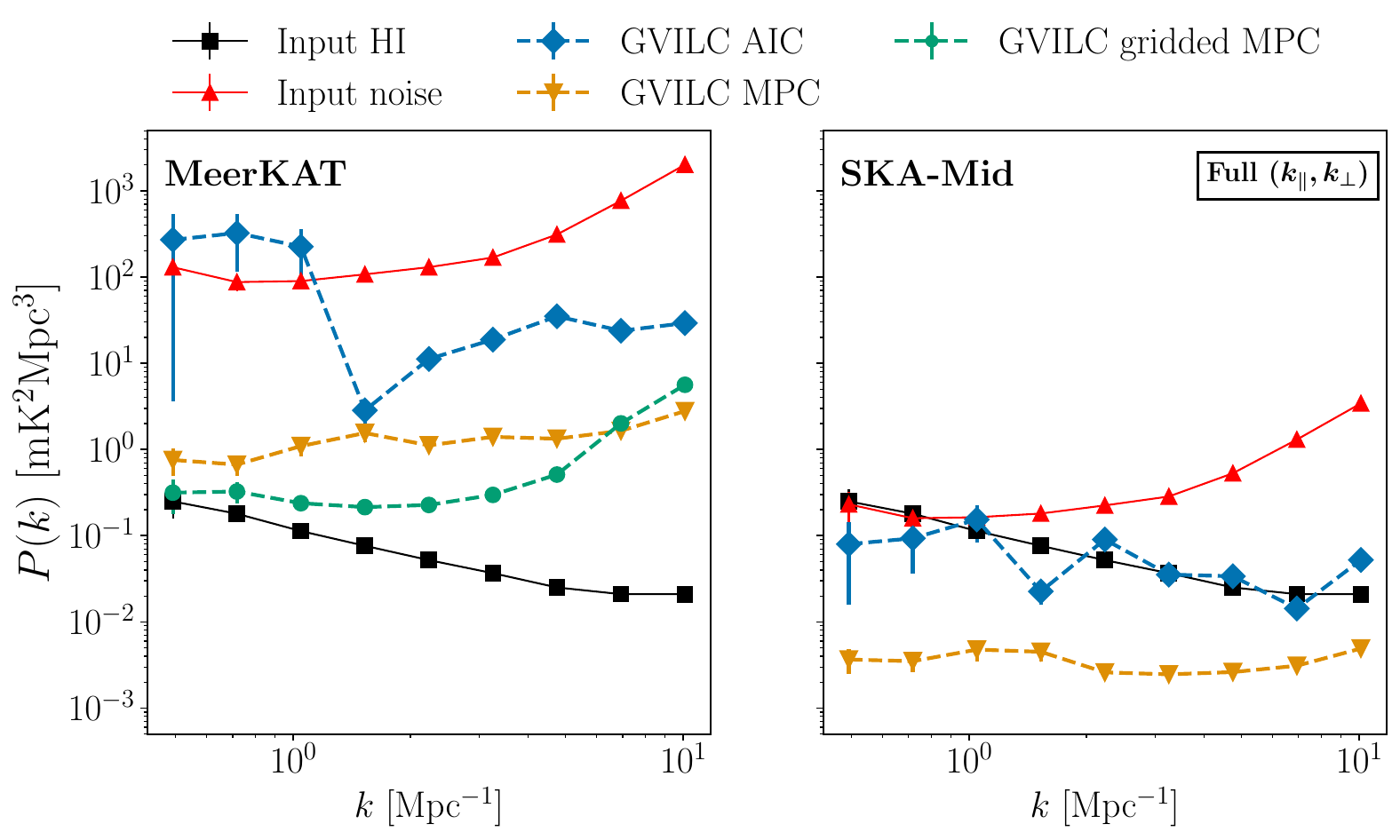}
    \caption{1D power spectrum of the total foreground residuals after applying GVILC for the case with line-broadening effects and matched HI and radio-continuum sources. The left panel corresponds to the MeerKAT simulations and the right panel, to the SKA-Mid simulations. The error bars are multiplied by a factor $5$ for visualization purposes.
    }
    \label{fig:total_foreground_residuals_1D}
\end{figure}

\begin{figure}
    \centering
    \includegraphics[width=\columnwidth]{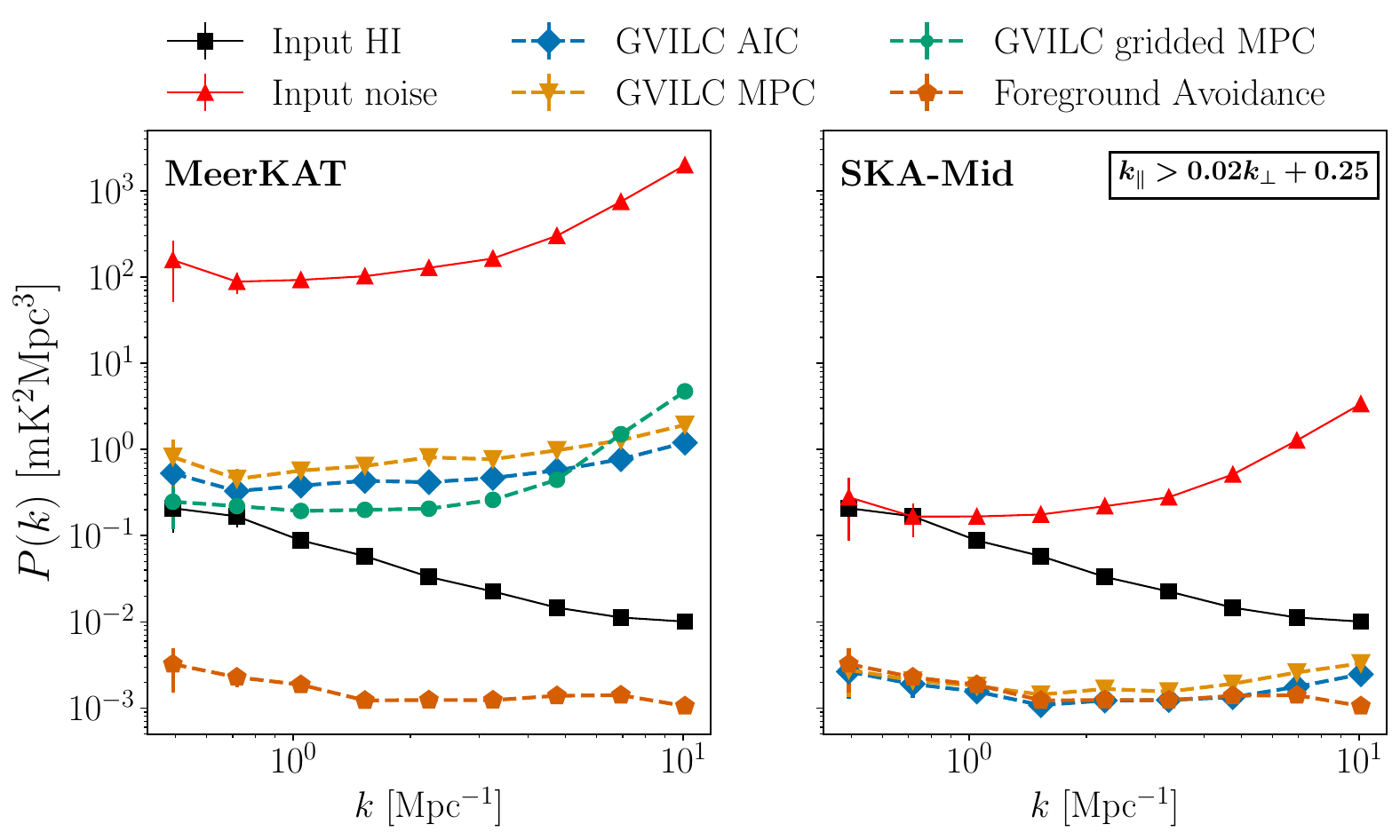}
    \caption{Same as Fig.~\ref{fig:total_foreground_residuals_1D} with additional foreground avoidance ($k_\parallel>0.02k_\perp+0.25$) after cleaning.}
    \label{fig:foreground_residuals_FA_1D}
\end{figure}

In this section, we discuss the residual foregrounds left after running GVILC. We are interested in both the total foreground power and the foreground power inside the HI window, i.e. outside the foreground wedge region. First, in Fig.~\ref{fig:foreground_residuals_2D} we show the 2D power spectra of the foreground residuals after applying GVILC for the different methods and experiments, corresponding to the case with line-broadening effects and matched HI and radio-continuum sources. By comparing panels (a) with panels (b)-(f), we can see that the foreground power inside the foreground wedge has decreased significantly. For MeerKAT, the foreground wedge is suppressed around 4 to 5 orders of magnitude, while for SKA-Mid the foreground wedge has almost disappeared with a reduction of 6 to 10 orders of magnitude. GVILC MPC leads to the smallest foreground residuals for both MeerKAT and SKA-Mid. However, for the MeerKAT experiment cases (panels (b)-(d)), we observe that the HI window has foreground residuals, which are not present in the input foreground power spectrum (panel (a)). The input foregrounds are smooth in frequency, but the foreground residuals are not. This causes a spread over high $k_\parallel$ regions. Nevertheless, this spread is suppressed when running GVILC for the SKA-Mid experiment. The high-noise levels present in the MeerKAT case are perturbing the eigenvectors, breaking the smoothness of the foreground residuals and contaminating the HI window. This effect is driven by the noise levels, and removing less modes does not reduce this contamination.

To determine how impactful the foreground residuals are for HI signal recovery, we plot the 1D power spectra of the total foregrounds and the foregrounds outside the foreground wedge. Figure \ref{fig:total_foreground_residuals_1D} shows the 1D power spectrum of the projected foreground residuals for MeerKAT and SKA-Mid in the case with line-broadening effects and matched HI and radio-continuum sources. In both panels, GVILC AIC yields the largest residuals, exceeding the noise level for MeerKAT and reaching the HI signal level for SKA-Mid. This behaviour arises because GVILC AIC suppresses foregrounds only until their power falls below the noise level in the ungridded visibility domain, while strongly penalising the removal of additional modes. Consequently, this method tends to under-clean the data, leaving significant residual foreground contamination. To mitigate this effect, we introduce the MPC selection criterion, which is more aggressive than AIC. As shown in both panels, GVILC MPC reduces the residual foreground power below the noise level, and in the SKA-Mid case even below the HI signal. Since the performance of GVILC depends on the noise level, applying GVILC MPC to gridded visibilities, where the effective noise is reduced, further improves the cleaning, yielding lower foreground residuals than the standard GVILC MPC implementation on ungridded visibilities.

Figure \ref{fig:foreground_residuals_FA_1D} shows the 1D power spectra of the foreground residuals after applying foreground avoidance $(k_\parallel=0.02k_\perp+0.25)$. In the MeerKAT panel, the foreground residual power is below the noise but still exceeds the amplitude of the HI signal for almost all $k$. In the same panel, we also show the results of foreground avoidance without any prior cleaning, for which the foreground residuals are below the HI signal. On the contrary, for the SKA-mid case the foreground power inside the HI window is below the HI signal, and with similar amplitude for the different methods.

\begin{figure}
    \centering
    \includegraphics[width=\columnwidth]{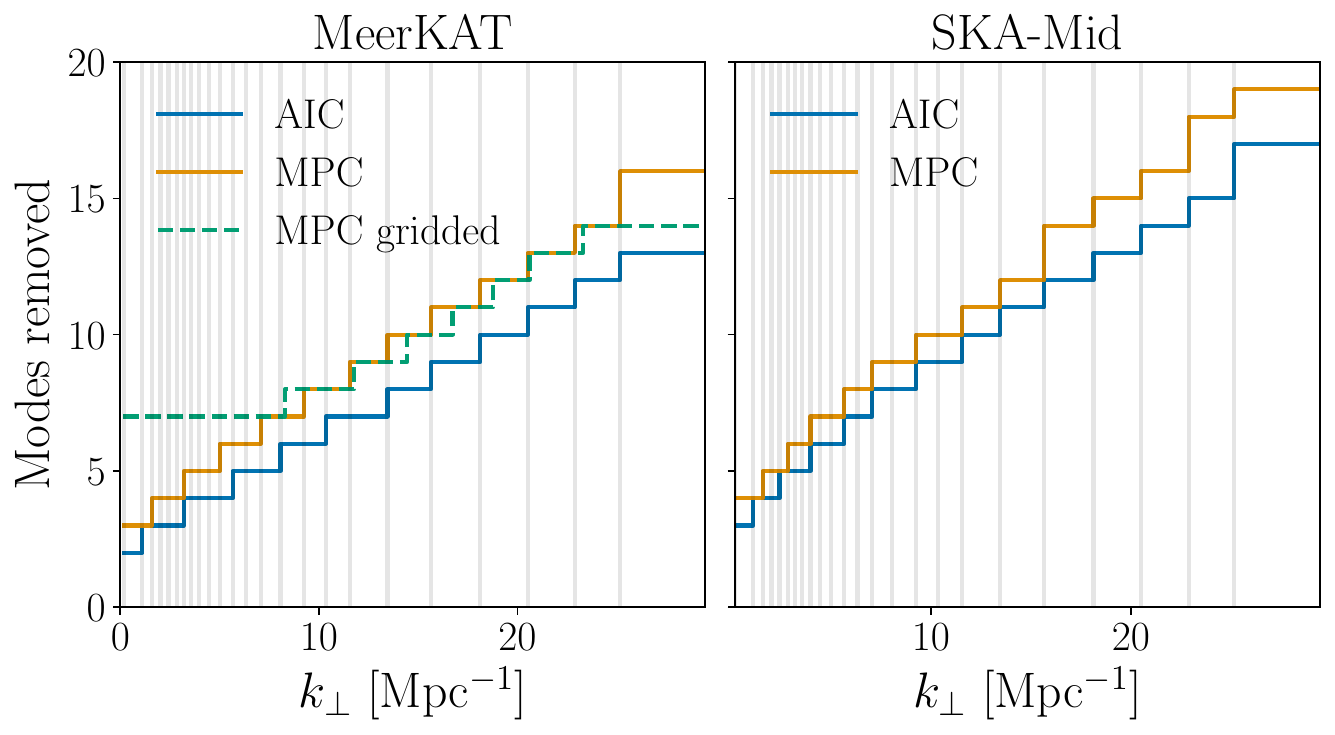}
    \caption{Number of modes removed as a function of $k_\perp$ for the different GVILC methods and experiments. The vertical lines correspond to the edges of the annuli built on the ungridded visibilities. For the gridded case, the step sizes correspond to the annuli sizes of MPC gridded.
    }
    \label{fig:modes_removed}
\end{figure}

The level of foreground residuals is directly related to the number of modes removed. In Fig.~\ref{fig:modes_removed}, we show the number of modes removed as a function of foreground removal method and $k_\perp$ scale. MPC removes more modes than AIC for both MeerKAT and SKA-Mid, as the penalty term in AIC leads to an under-cleaning situation in the noise-dominated regime. MPC is more robust in that regime by using the statistical distribution of the eigenvalues to distinguish between signal-plus-noise and the foregrounds subspaces. Looking at the MeerKAT panel and comparing GVILC MPC applied to gridded and ungridded visibilities, we can see the effect of the different annulus sizes. The large grid size of $\Delta u=60\lambda$ results in much larger annuli than in the ungridded case, reducing the resolution of the foreground removal process. The number of modes removed for SKA-Mid is higher than for MeerKAT, as the lower noise level enables GVILC to better identify the foreground subspace. When comparing GVILC results for the two complexities of the HI model, we observe no difference in the number of modes removed. This indicates that the number of modes removed is primarily driven by the foreground and noise levels.

\subsection{Signal loss} \label{sec:signal_loss}

\begin{figure*}
    \centering
    \includegraphics[width=0.7\textwidth]{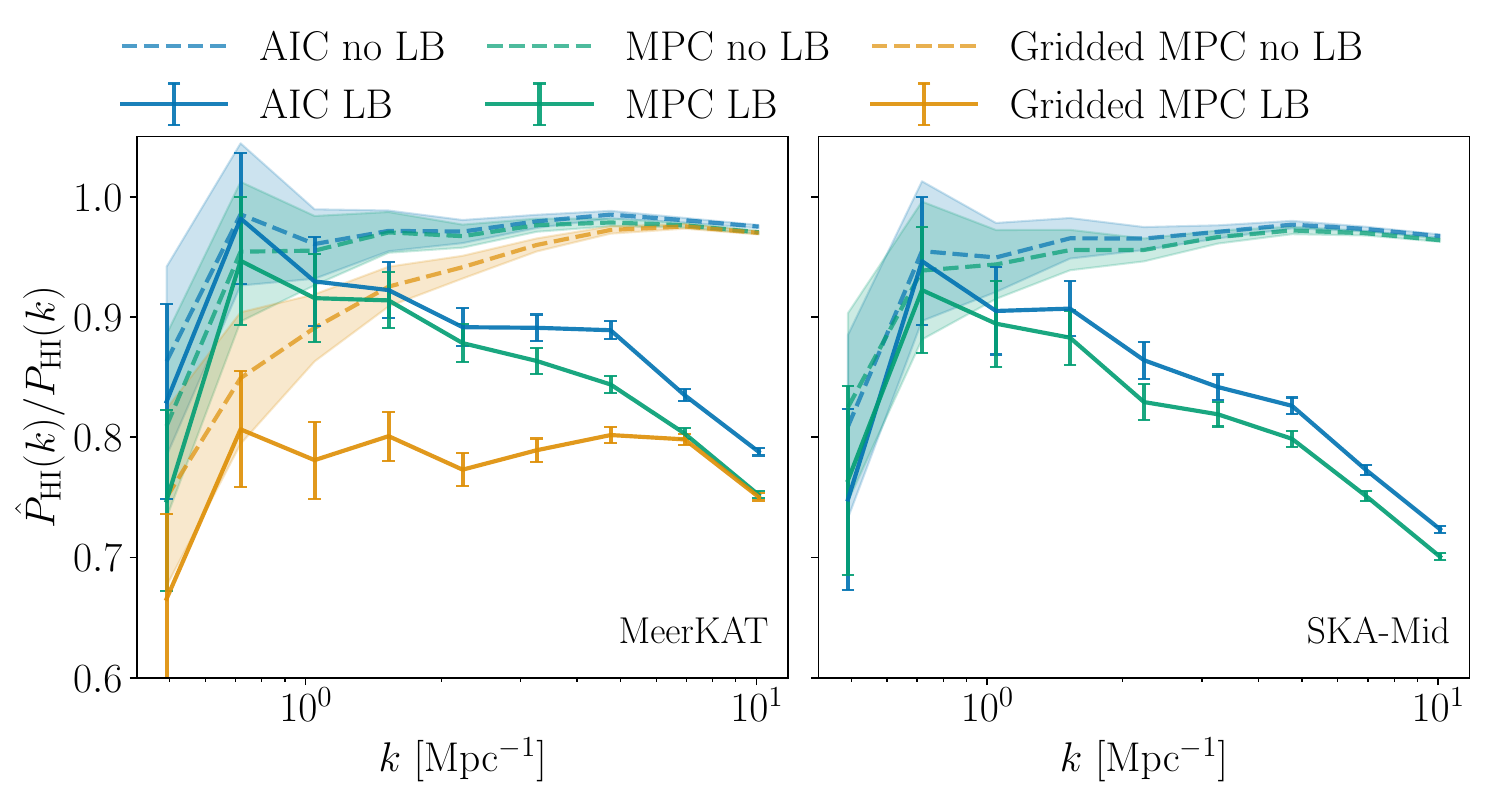}
    \caption{Ratio of the projected HI signal 1D power spectrum after cleaning and before cleaning, i.e., the signal preserved after foreground cleaning. The left and right panel show the results for MeerKAT and SKA-Mid, respectively. The dashed lines indicate the signal preserved for the case without line-broadening effects and unmatched HI and radio-continuum sources, and the solid lines correspond to the case with line-broadening effects and matched HI and radio-continuum sources.
    }
    \label{fig:power_spectrum_1D_signal_preserved}
\end{figure*}

\begin{figure*}
    \centering
    \includegraphics[width=\textwidth]{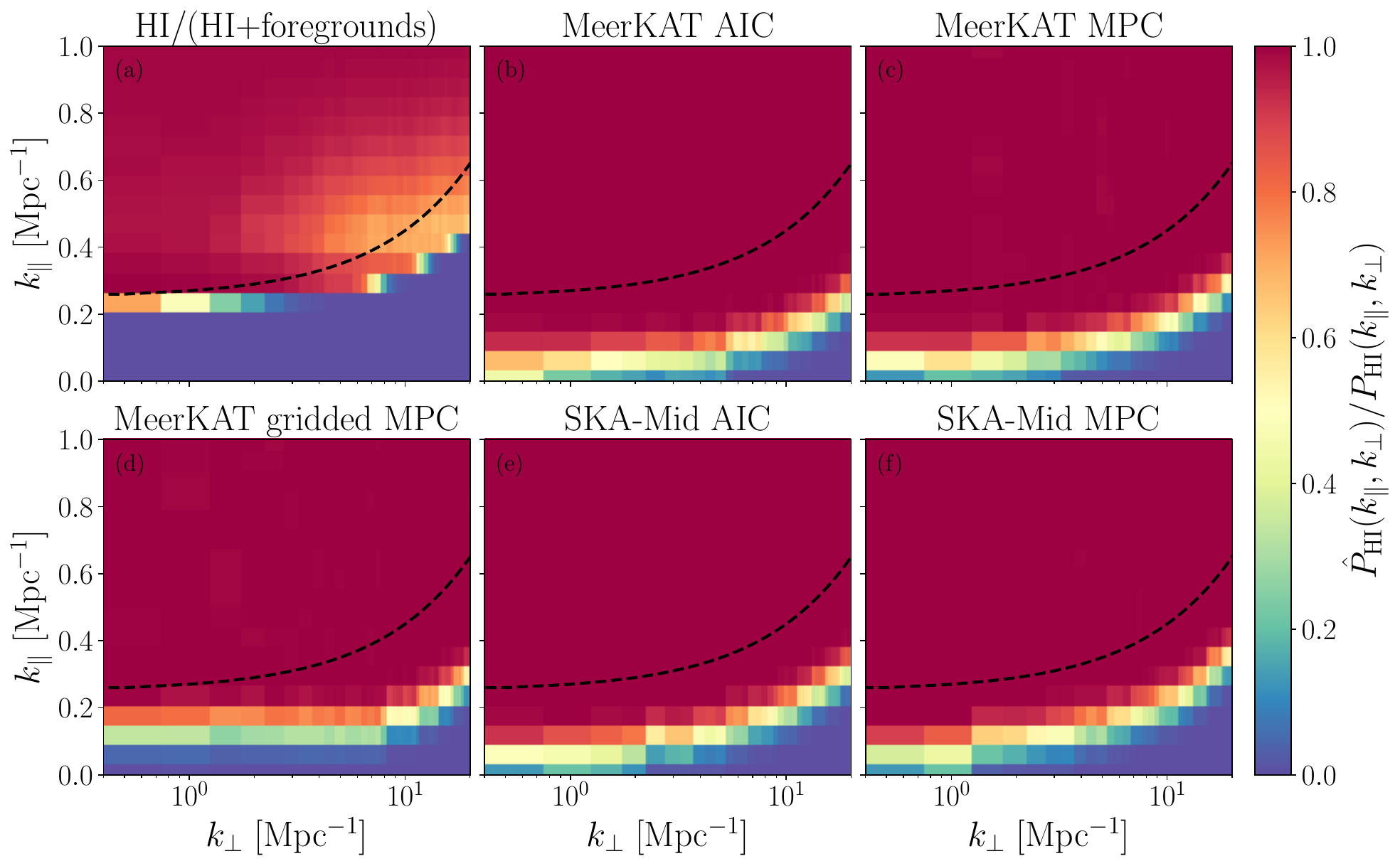}
    \caption{Panel (a): ratio between HI and HI plus total foregrounds. Panels (b)-(f): ratio of the projected HI signal 2D power spectrum after and before foreground cleaning, i.e., the signal preserved. We show the signal preserved for the different experiments and GVILC cleaning methods. The dashed black lines correspond approximately to the size of the foreground wedge ($k_\parallel=0.02k_\perp+0.25$). All the panels correspond to the case with line-broadening effects and matched HI and radio-continuum sources. 
    }
    \label{fig:power_spectrum_2D_signal_preserved}
\end{figure*}

Signal loss is another important aspect to take into account. A more aggressive foreground removal leads to lower foreground residuals as well as to greater signal loss. From Fig.~\ref{fig:power_spectrum_1D_signal_preserved} we notice that the signal preserved varies between $67\%$ and $98\%$. GVILC AIC leads to the smallest signal loss, with a small difference with respect to GVILC MPC. Running GNILC on the gridded visibilities, although benefiting from lower foreground residuals, also has increased signal loss. In order to have the full picture, we have to look into the 2D power spectrum because signal loss is not isotropic. As shown in Fig.~\ref{fig:power_spectrum_2D_signal_preserved}, signal loss dominates inside the foreground wedge being approximately zero outside of it. In fact, the foreground wedge is bigger than the signal loss region. By applying a small foreground avoidance, we can largely mitigate signal loss, while accessing modes which were initially foreground dominated. Another interesting information that can be extracted from Fig. \ref{fig:power_spectrum_1D_signal_preserved} is the signal loss comparison between the two different HI signal complexities. By visual inspection, including line-broadening effects and matched HI and radio-continuum sources results in larger signal loss than not including them, for both experiments and all foreground-removal methods. The main reason is that signal loss is not isotropic, and from Fig.~\ref{fig:HI_PS} we can observe the difference in distribution of the power along the $(k_\parallel, k_\perp)$ plane. In the case with line-broadening effects, the HI signal has a bigger overlap with the foreground wedge compared to the case without line-broadening effects. We also test other causes of the increased signal loss, either due to removing correlations between frequencies due to line-broadening effects, or due to removing the correlated part of the HI with the radio-continuum source emission. However, the tests disfavour this idea. First, by comparing the HI covariance before and after cleaning we do not observe a greater reduction of power on the off-diagonal elements compared to the diagonal ones. Secondly, the measured cross spectrum between radio-continuum and HI sources is compatible with zero, even without instrumental noise.

\subsection{Mean Squared Error} \label{sec:MSE}
\begin{figure}
    \centering
    \includegraphics[width=\columnwidth]{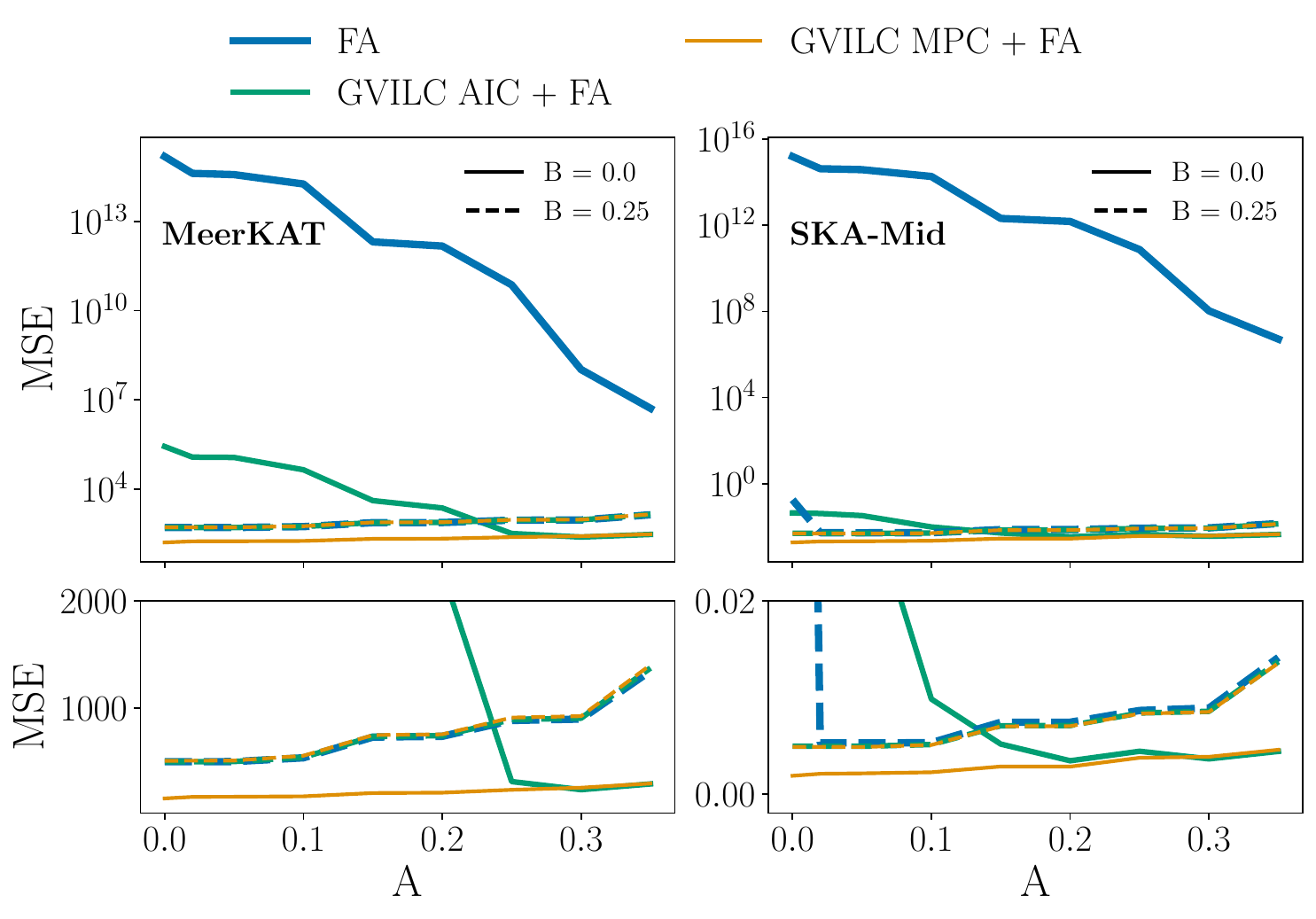}
    \caption{Mean Squared Error (MSE) as a function of foreground cleaning method and foreground avoidance strategy. The results for the MeerKAT experiment are shown in the left panel and the SKA-Mid results, in the right panel. The bottom panels are just zoomed versions of the top panels. The parameters $A$ and $B$ correspond to the foreground avoidance parameters defined in equation~\eqref{eq: fa}. FA corresponds to foreground avoidance applied to the uncleaned simulations, and GVILC AIC + FA and GVILC MPC + FA correspond to the hybrid methods combining foreground cleaning and avoidance. In the MeerKAT panel, GVILC MPC gridded + FA overlaps with GVILC MPC + FA, so it is not shown for visualisation purposes.
    }
    \label{fig:MSE}
\end{figure}

In this section, we compare the performance of the different foreground mitigation methods using the Mean Squared Error (MSE). With this metric, we are sensitive to signal loss, foreground residuals, and mode loss. We define the MSE in terms of the 1D power spectrum and the foreground avoidance strategy as:
\begin{equation} \label{eq:MSE}
\begin{aligned}
\text{MSE}(A,B) = \sum_{i=1}^n \Big[&
\left(\hat{P}_i(A,B)-\hat{P}_i^{\rm HI,input}(A,B)
-\hat{P}_i^{\rm noise}(A,B)\right)^2\\
&+ \Delta\hat{P}_i^2(A,B)\Big].
\end{aligned}
\end{equation}
where $A$ and $B$ are the foreground avoidance parameters defined in equation~\eqref{eq: fa}, $\hat{P}_i(A, B)$ is the reconstructed power spectrum after applying the foreground avoidance, $\hat{P}_i^{\rm HI, input}(A,B)$ is the input HI power spectrum after foreground avoidance, $\hat{P}_i^{\rm noise}(A,B)$ is the projected noise power spectrum after foreground avoidance, and $\Delta\hat{P}_i^2(A, B)$ is the variance of the reconstructed power power spectrum after foreground avoidance. The MSE is sensitive to biases in our reconstructed power spectrum, i.e., signal loss and foreground residuals, and the variance, which is connected to mode loss.

In Fig.~\ref{fig:MSE}, the MSE is presented for different foreground avoidance strategies, experiments and cleaning methods. For MeerKAT, the MSE is dominated by the variance term due to the noise contribution. Having larger foreground residuals in GVILC MPC without applying any foreground avoidance ($A=0.0$, $B=0.0$) is preferred with respect to applying foreground avoidance to the uncleaned data. The reason is that decreasing the variance has a bigger impact than decreasing the bias on the MSE. This is a positive remark about GVILC in this high noise regime, even though there are foreground residuals inside the HI window, it is overall performing better than foreground avoidance. Although not shown in Fig.~\ref{fig:MSE}, we observe similar performance between applying GVILC on the ungridded and gridded visibilities. For SKA-Mid, hybrid GVILC AIC and GVILC MPC outperforms foreground avoidance. More importantly, in both MeerKAT and SKA-mid scenarios, GVILC MPC is very stable for the different foreground avoidance strategies applied. For both cases, if a sufficiently aggressive foreground avoidance strategy is chosen, hybrid GVILC and pure foreground avoidance matches.

\section{Conclusions} \label{sec:conclusions}
In this paper, we introduce the Generalised Visibility Internal Linear Combination (GVILC), a new non-parametric foreground subtraction method for HI intensity mapping directly applicable to both ungridded and gridded interferometric visibilities. GVILC relies on a Principal Component Analysis to remove the highest variance components from the data using a prior of the HI signal-plus-noise covariance matrix. We have used data-driven statistical measures, namely Akaike Information Criterion (AIC) and Marchenko-Pastur Criterion (MPC) to determine the number of modes to be removed. The cleaned solution is obtained as an Internal Linear Combination of the visibilities, minimising the variance of the cleaned visibilities.

We demonstrate the effectiveness of GVILC in removing foregrounds using simulated visibilities from a deep integration for MeerKAT and SKA-Mid-like telescopes, mimicking the observation in the reported detection in \cite{Paul_a_first_detection}. The simulated visibilities include HI signal, thermal noise and foregrounds of both galactic (synchrotron and free-free) and extragalactic (radio-continuum sources) origin. Our approach uses a more realistic HI signal by including line-broadening effects and correlation of foregrounds with cosmological HI signal via the matching of the spatial emission of the HI and radio-continuum sources.

When applying the GVILC method to MeerKAT and SKA-Mid-like simulations, the cleaned data show a significant reduction of foreground residuals in the 1D and 2D power spectrum. At the same time, the preserved signal ranges between 67\% and 98\%, depending on the GVILC method applied, the $k$ scale and, the presence of line-broadening effects and matched HI and radio-continuum sources. This signal loss can be corrected using a minimal foreground avoidance strategy, making an unbiased recovery of the HI signal achievable with GVILC. We find that the inclusion of line-broadening effects and matched HI and radio-continuum sources increases signal loss after foreground removal, without affecting the number of modes removed by GVILC. For both MeerKAT and SKA-Mid, we find that GVILC provides a better solution in terms of mean squared error statistic than traditional foreground avoidance methods. Therefore, GVILC has proved to be robust against different noise levels, and stable in combination with foreground avoidance.

In this paper, we also demonstrate that the primary beam produces a $uv$ grid-size dependent suppression of the reconstructed power spectrum if not taken into account. We show that the primary beam induces a decorrelation at the visibility covariance level. We derive an analytical expression for the grid-size dependent modification of the normalisation correction, which we verify with our simulations. 

This work demonstrates the feasibility of GVILC for foreground removal and HI signal recovery directly from interferometric visibilities. A detailed treatment of instrumental systematics, such as side-lobes and RFI contamination, is beyond the scope of this study. Future work will apply the method to real MeerKAT observations to assess its performance in the presence of instrumental systematics and flagged data.

\section*{Acknowledgements}
The authors thank J. L. Bernal, Z. Chen, S. Pamuk, and S. Paul for useful discussions, P. Diego-Palazuelos for help in generating the free-free template and A. Acebr\'on for comments on the draft. MRG acknowledges financial support from the Formación del Profesorado Universitario program of the Spanish Ministerio de Ciencia, Innovación y Universidades, and thanks the Jodrell Bank Centre for Astrophysics for hospitality during the early stages of this work. MRG, MR, DH and PV have been supported by MICIU/AEI and by FEDER (UE) under the projects with references PID2022-140670NA-I00, PID2022-139223OB-C21, and PID2025-173927NB-I00. AM thanks the UK Research and Innovation Future Leaders Fellowship for supporting a part of this research [grant MR/V026437/1]. LW is a UK Research
and Innovation Future Leaders Fellow [grant MR/V026437/1].  MR also acknowledges support from the RadioForegroundsPlus Project HORIZONCL4-2023-SPACE-01, GA 101135036. We acknowledge Santander Supercomputacion support group at the University of Cantabria who provided access to the supercomputer Altamira Supercomputer at the Institute of Physics of Cantabria (IFCA-CSIC), member of the Spanish Supercomputing Network, for performing simulations/analyses. We acknowledge the use of the \texttt{healpy}~\citep{healpy_2019}, \texttt{astropy}~\citep{astropy:2013, astropy:2018, astropy:2022}, \texttt{pandas}~\citep{mckinney_proc_scipy_2010}, \texttt{CASA}~\citep{CASA_2022PASP}, \texttt{scipy}~\citep{scipy_2020}, \texttt{numpy}~\citep{numpy_harris2020array}, and \texttt{matplotlib}~\citep{Matplotlib_2007} software packages.

\section*{Data Availability}

Data underlying this paper will be shared on reasonable request to the corresponding author.
 



\bibliographystyle{mnras}
\bibliography{biblio} 




\appendix

\appendix
\section{Simulating small-scale free-free emission} \label{App: small-scale free-free}

In this section, we detail how we generate the small scales for the free-free template from \texttt{pysm3}. As explained in Section~\ref{sec: diffuse}, we need to upgrade the free-free template from resolution $N_{\rm side}=512$ to $N_{\rm side}=8192$. The \texttt{f1} model is based on the 30 GHz Commander free-free template from \textit{Planck} 2015 \citep{Commander_Planck_2015}, which has an angular resolution of 1 degree, and has small scales added up to an angular resolution of 6.87 arcminute ($N_{\rm side}=512$). We follow similar procedure as that described by \cite{pysm2_2017} to simulate free-free small scales:

\begin{enumerate}
\item Simulate full-sky free-free template using \texttt{pysm3} at a frequency $\nu_0=30$ GHz at the native resolution of $N_{\rm side}=512$. This map is smoothed with a 1 degree FWHM Gaussian beam and then upgraded to $N_{\rm side}=8192$ using \texttt{HealPIX} functions \texttt{map2alm} and \texttt{alm2map}, leading to a large scale template $M_{\nu_0}^{\rm LS}(\boldsymbol{\hat{n}})$.

\item Generate a Gaussian realisation at $N_{\rm side}=8192$, $M_{\nu_0}^{\rm SS}(\boldsymbol{\hat{n}})$, using the \texttt{HealPIX} function \texttt{anafast} and the small-scales angular power spectrum, $C_{\ell, \nu_0}^{\rm SS}$, up to $\ell=24575$,

\begin{equation}\label{extrapolation}
    C_{\ell, \nu_0}^{\rm SS} = \ell^\gamma \left [1-e^{-\ell(\ell+1)\sigma_{\rm temp}^2}\right ],
\end{equation}
where $\gamma=-3$, and $\sigma_{\rm temp}=25.48$ arcmin. The $\gamma$ value differs from the one appearing in~\cite{pysm2_2017}, but we find good agreement with the former value when comparing the power spectra of the template with ours.

\item Generate the modulation template for the small scales introduced by the large scales, $R_{\nu_0}(\boldsymbol{\hat{n}})$. The modulation is anisotropic, being stronger in the Galactic plane and weaker outside of it:
\begin{equation}
R_{\nu_0}(\boldsymbol{\hat{n}}) = \frac{\langle M_{\nu_0}^{\rm LS}(\boldsymbol{\hat{n}})\rangle}{4\sigma^{\rm SS}_{\nu_0}}\left (\frac{\hat{M}_{\nu_0}^{\rm LS}(\boldsymbol{\hat{n}})}{\langle \hat{M}_{\nu_0}^{\rm LS}(\boldsymbol{\hat{n}})\rangle}\right )^\alpha,
\end{equation}
where $\alpha$ is set to 1.25 in order to recover the correct power law behaviour at low multipoles, and 
\begin{equation}
    \sigma^{\rm SS}_{\nu_0}=\sqrt{\sum^{\ell_1}_{\ell=0}\frac{C_{\ell, \nu_0}^{\rm SS}(2\ell+1)}{4\pi}}.
\end{equation}

\item The new template with the small scales is the sum of the large- and the modulated small-scales contributions,
\begin{equation}
\widetilde{M}_{\nu_0}(\boldsymbol{\hat{n}}) = \hat{M}_{\nu_0}^{\rm LS}(\boldsymbol{\hat{n}}) + R_{\nu_0}(\boldsymbol{\hat{n}})M_{\nu_0}^{\rm SS}(\boldsymbol{\hat{n}}).
\end{equation}

\item The negative pixels in the template ($<0.1\, \%$) are changed to the average with positive neighbours. The remaining negative pixels are replaced with the large-scale template.
\end{enumerate}

\begin{figure}
    \centering
    \includegraphics[width=\columnwidth]{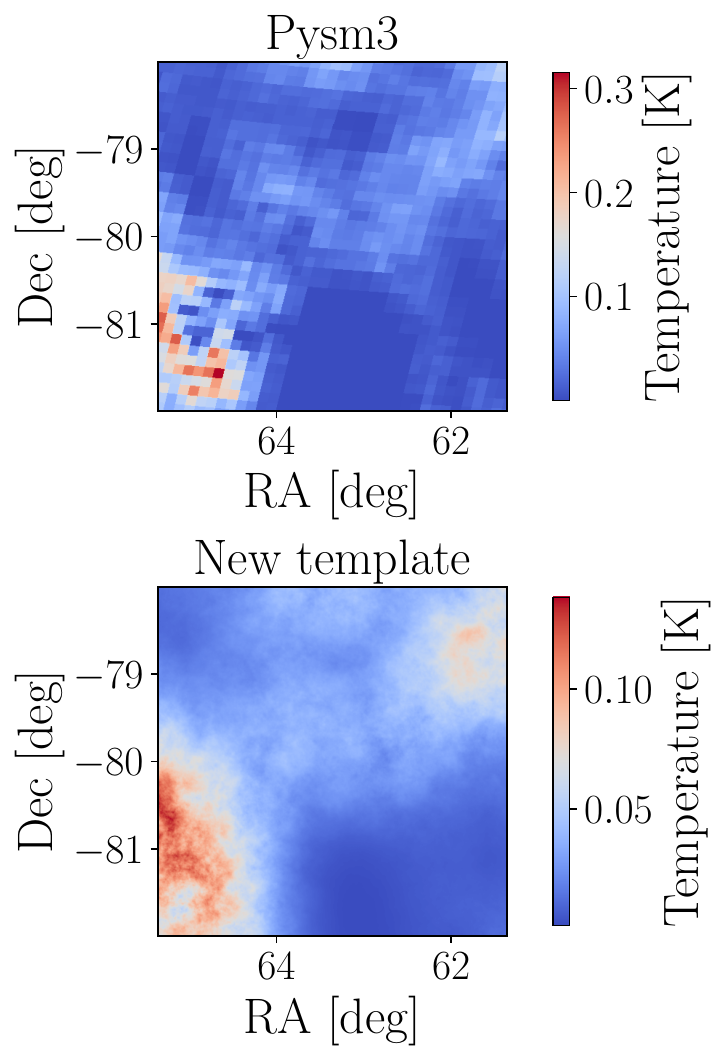}
    \caption{
    Simulated free-free at the central frequency of 972.85 MHz. Top panel: low-resolution \texttt{pysm3} simulation with a pixel size of ($30$ arcsec)$^2$. Bottom panel: new free-free template with pixel size of ($6.87$ arcmin)$^2$. The centre of the field corresponds to the DEEP2 pointing, with celestial coordinates $\text{RA}=63.36$ deg and $\text{Dec}=-80.00$ deg, with an extension of $4\times 4$ deg$^2$.
    }
    \label{fig:free_free}
\end{figure}

\section{Impact of the Gaussian primary beam on the power spectrum} \label{ap: Gaussian PB}

The primary beam induces two effects at the power spectrum level: mode-mixing and decorrelation effects. In the situation of a narrow frequency range of observations, the frequency dependence of the beam can be neglected, $A(l,m,f)\cong A(l,m)$. In that situation,
\begin{equation} \label{eq: PB_no_freq_dep}
    A(\boldsymbol{k})=\frac{X^2\Delta X}{\mathbb{V}}\delta_{\rm D}(k_\parallel)A_{\perp}(\boldsymbol{k}_\perp),
\end{equation}
where $X=D_c(z_0)$ is the comoving distance at redshift $z_0$, $\Delta X=Y\delta fN_{\rm ch}$ is the line-of-sight comoving distance, $\boldsymbol{k}_\perp=(k_x, k_y)$ is the scale in the angular plane (perpendicular to the line-of-sight), and $k_\parallel=k_z$ is the scale along the line-of-sight.

The primary beam of MeerKAT can be modelled as a Gaussian beam \citep{Mauch_PB}, where $A(l,m,f)\cong A(l,m)=\exp(-(l^2+m^2)/2\sigma^2)$. As shown by equation~\eqref{eq:delay ps}, we need the Fourier transformed beam. To get that, first we need to convert the effective primary beam from sky coordinates into cosmological coordinates, 
\begin{equation}
    A_{\perp}(\boldsymbol{r}_\perp) = \exp\left [-\frac{|\boldsymbol{r}_\perp|^2}{2\sigma^2X^2}\right ].
\end{equation}
Then, we can apply Fourier transform the primary beam:
\begin{equation} \label{eq: PB kperp}
\begin{aligned}
    A_{\perp}(\boldsymbol{k}_\perp)&=\frac{1}{X^2}\int \text{d}^2\boldsymbol{r}_\perp \ A_{\perp}(\boldsymbol{r}_\perp)e^{-i\boldsymbol{k_\perp}\cdot\boldsymbol{r_\perp}}\\
    &=2\sigma^2\pi\exp\left[-\frac{\sigma^2X^2|\boldsymbol{k}_\perp|^2}{2}\right].
\end{aligned}
\end{equation}

\subsection{Normalisation of the gridded power spectrum estimator} \label{ap: norm}
In this section, we derive the normalisation correction for the gridded power spectrum estimator. From the gridded visibilities that are defined in equations~\eqref{eq: gridding} and \eqref{eq: delay vis}, the delay power spectrum is computed as the squared of the gridded delay visibilities
\begin{equation} \label{eq: gridded power spectrum vis}
    \left\langle\left|\widetilde{V}_{\rm grid}^{j}\right|^2\right\rangle=
    \frac{1}{N_j^2}\sum_{\boldsymbol{u}_1, \boldsymbol{u}_2\in\mathcal{D}_j}
    \left\langle \widetilde{V}(\boldsymbol{u}_1)\widetilde{V}^\ast(\boldsymbol{u}_2)\right\rangle,
\end{equation}
where $\widetilde{V}_{\rm grid}^{j}$ are the gridded delay transformed visibilities given by equation~\eqref{eq: delay vis}, $N_j$ are the number of visibilities inside grid $j$, and $\widetilde{V}(\boldsymbol{u})$ are the delay transformed visibilities. The next step for deriving the normalisation correction is to compute the expression for the visibility covariance \citep{Zhaoting_2023MNRAS},
\begin{equation} \label{eq: covariance vis.}
\begin{aligned}
    \left\langle \widetilde{V}(\boldsymbol{u}_1)\widetilde{V}^\ast(\boldsymbol{u}_2)\right\rangle &=\left(\frac{2k_{\rm B}}{\lambda^2} \right)^2\frac{\Delta X}{Y^2}\int \frac{\text{d}^2\boldsymbol{k}_\perp'}{(2\pi)^2}\
    A_\perp(\boldsymbol{k}_{\perp, 1}-\boldsymbol{k}_\perp')\\
    &\times A_\perp^\ast(\boldsymbol{k}_{\perp,2}-\boldsymbol{k}_\perp')
    P(\boldsymbol{k}_\perp', k_\parallel)=\left(\frac{2k_{\rm B}}{\lambda^2} \right)^2\frac{\Delta X \sigma^4}{Y^2}\\
    &\times \int \text{d}^2\boldsymbol{k}_\perp'\
    e^{-\beta|\boldsymbol{k}_{\perp,1}-\boldsymbol{k}'_\perp|^2}
    e^{-\beta|\boldsymbol{k}_{\perp,2}-\boldsymbol{k}'_\perp|^2}
    P(\boldsymbol{k}_\perp', k_\parallel),
\end{aligned}
\end{equation}
where $\boldsymbol{u}=(u,v,\eta)$ are the baseline coordinates, $\beta=X^2\sigma^2/2$, $\boldsymbol{k}_{\perp,1}=\frac{2\pi}{X}(u_1, v_1)$, and $k_\parallel=2\pi\eta/Y$. Afterwards, we perform the Taylor expansion the cosmological HI power spectrum up to first order centred at $\boldsymbol{k}_{\perp,1}$:
\begin{equation} \label{eq: Taylor PS}
    P(\boldsymbol{k}_\perp', k_\parallel) = 
    P(\boldsymbol{k}_{\perp,1}, k_\parallel)+\nabla_{\boldsymbol{k}_\perp} P(\boldsymbol{k}_{\perp,1}, k_\parallel)\cdot(\boldsymbol{k}_\perp'-\boldsymbol{k}_{\perp,1})+\cdots,
\end{equation}
where we assume that the HI power spectrum is a smooth function and that the width in Fourier space of primary beams larger than 30 arcmin is much smaller than the scales at which the HI power spectrum varies~\citep{Parsons_2012ApJ}. Substituting equation~\eqref{eq: Taylor PS} into equation~\eqref{eq: covariance vis.}:
\begin{equation} \label{eq: cov + taylor}
\begin{aligned}
&\left\langle \widetilde{V}(\boldsymbol{u}_1) 
\widetilde{V}^\ast(\boldsymbol{u}_2) \right\rangle
=
\left(\frac{2k_{\rm B}}{\lambda^2}\right)^2
\frac{\Delta X\sigma^4}{Y^2} \\
&\quad \times
\Bigg\{
P(\boldsymbol{k}_{\perp,1}, k_\parallel)
\int \text{d}^2\boldsymbol{k}_\perp'\,
e^{-\beta|\boldsymbol{k}_{\perp,1}-\boldsymbol{k}'_\perp|^2}
    e^{-\beta|\boldsymbol{k}_{\perp,2}-\boldsymbol{k}'_\perp|^2}
\\
&\qquad +
\nabla_{\boldsymbol{k}_\perp}
P(\boldsymbol{k}_{\perp,1}, k_\parallel)
\cdot
\int \text{d}^2\boldsymbol{k}_\perp'\,
(\boldsymbol{k}_\perp'-\boldsymbol{k}_{\perp,1})e^{-\beta|\boldsymbol{k}_{\perp,1}-\boldsymbol{k}'_\perp|^2}\\
&\qquad\times e^{-\beta|\boldsymbol{k}_{\perp,2}-\boldsymbol{k}'_\perp|^2}\Bigg\},
\end{aligned}
\end{equation}
where both integrals have an analytic solution. We start by simplifying equation~\eqref{eq: cov + taylor} using the following relationship,
\begin{equation}
\begin{aligned}
&|\boldsymbol{k}_{\perp,1}-\boldsymbol{k}'_\perp|^2+|\boldsymbol{k}_{\perp,2}-\boldsymbol{k}'_\perp|^2\\
&\qquad = 2\left|\boldsymbol{k}'_\perp-\frac{\boldsymbol{k}_{\perp,1}+\boldsymbol{k}_{\perp,2}}{2} \right|^2+\frac{1}{2}\left|\boldsymbol{k}_{\perp,1}-\boldsymbol{k}_{\perp,2} \right|^2,
\end{aligned}
\end{equation}
which transform equation~\eqref{eq: cov + taylor} into:
\begin{equation} \label{eq: cov + taylor 2}
\begin{aligned}
&\left\langle \widetilde{V}(\boldsymbol{u}_1)
\widetilde{V}^\ast(\boldsymbol{u}_2) \right\rangle
=
\left(\frac{2k_{\rm B}}{\lambda^2}\right)^2
\frac{\Delta X\sigma^4}{Y^2}e^{-\frac{\beta}{2}|\boldsymbol{k}_{\perp,1}-\boldsymbol{k}_{\perp,2}|^2} \\
&\quad \times
\Bigg\{
P(\boldsymbol{k}_{\perp,1}, k_\parallel)
\int \text{d}^2\boldsymbol{k}_\perp'\,
    e^{-2\beta\left|\boldsymbol{k}'_\perp-\frac{\boldsymbol{k}_{\perp,1}+\boldsymbol{k}_{\perp,2}}{2} \right|^2}
\\
&\qquad +
\nabla_{\boldsymbol{k}_\perp}
P(\boldsymbol{k}_{\perp,1}, k_\parallel)
\cdot
\int \text{d}^2\boldsymbol{k}_\perp'\,
(\boldsymbol{k}_\perp'-\boldsymbol{k}_{\perp,1}) e^{-2\beta\left|\boldsymbol{k}'_\perp-\frac{\boldsymbol{k}_{\perp,1}+\boldsymbol{k}_{\perp,2}}{2} \right|^2}\Bigg\}.
\end{aligned}
\end{equation}
Thus, the two integrals have the following solutions,
\begin{equation} \label{eq: integrals1}
\begin{aligned}
    &\int \text{d}^2\boldsymbol{k}_\perp'\,
    e^{-2\beta\left|\boldsymbol{k}'_\perp-\frac{\boldsymbol{k}_{\perp,1}+\boldsymbol{k}_{\perp,2}}{2} \right|^2}=
    \int \text{d}^2\boldsymbol{q}\,
    e^{-2\beta\left|\boldsymbol{q} \right|^2}\\
    &\qquad =\left(\int \text{d}q_x\,
    e^{-2\beta q_x^2}\right)^2=\frac{\pi}{2\beta},
\end{aligned}
\end{equation}
\begin{equation} \label{eq: integrals2}
\begin{aligned}
    &\int \text{d}^2\boldsymbol{k}_\perp'\,
    (\boldsymbol{k}_\perp'-\boldsymbol{k}_{\perp,1}) e^{-2\beta\left|\boldsymbol{k}'_\perp-\frac{\boldsymbol{k}_{\perp,1}+\boldsymbol{k}_{\perp,2}}{2} \right|^2}\\
    &\qquad =\int \text{d}^2\boldsymbol{q}\,
    \left(\boldsymbol{q}+\frac{\Delta \boldsymbol{k}_\perp}{2}\right)e^{-2\beta\left|\boldsymbol{q} \right|^2}\\
    & \qquad =\int \text{d}^2\boldsymbol{q}\,\,
    \boldsymbol{q}\,e^{-2\beta\left|\boldsymbol{q} \right|^2}+
    \frac{\Delta \boldsymbol{k}_\perp}{2}\int \text{d}^2\boldsymbol{q}\,
    e^{-2\beta\left|\boldsymbol{q} \right|^2}\\ 
    &\qquad =\boldsymbol{0}+\frac{\Delta \boldsymbol{k}_\perp}{2}\frac{\pi}{2\beta}=\frac{\Delta \boldsymbol{k}_\perp}{2}\frac{\pi}{2\beta},
\end{aligned}
\end{equation}
where, for each integral, we perform the same change of variables $\boldsymbol{q}=\boldsymbol{k}'_\perp-\frac{\boldsymbol{k}_{\perp,1}+\boldsymbol{k}_{\perp,2}}{2}$, and $\Delta \boldsymbol{k}_\perp=\boldsymbol{k}_{\perp,2}-\boldsymbol{k}_{\perp,1}$. Finally, substituting the equations~\eqref{eq: integrals1} and \eqref{eq: integrals2} in equation~\eqref{eq: cov + taylor 2}, we obtain the final result for the visibility covariance:
\begin{equation} \label{eq: final cov}
\begin{aligned}
\left\langle \widetilde{V}(\boldsymbol{u}_1)
\widetilde{V}^\ast(\boldsymbol{u}_2) \right\rangle
&=
\left(\frac{2k_{\rm B}}{\lambda^2}\right)^2
\frac{\delta f N_{\rm ch}\pi\sigma^2}{X^2Y}e^{-\frac{\beta}{2}|\boldsymbol{k}_{\perp,1}-\boldsymbol{k}_{\perp,2}|^2} \\
&\quad \times
\Bigg\{
P(\boldsymbol{k}_{\perp,1}, k_\parallel) + \frac{1}{2}
\nabla_{\boldsymbol{k}_\perp}
P(\boldsymbol{k}_{\perp,1}, k_\parallel)
\cdot \Delta\boldsymbol{k}_\perp\Bigg\}.
\end{aligned}
\end{equation}

This result demonstrates that a Gaussian primary beam induces decorrelation between visibilities. This decorrelation is captured by the exponential term, which suppresses the covariance as the separation between the two baseline positions increases. For the MeerKAT case, we can compute the correlation length, $\mathrm{FWHM}_{\rm k}$, to determine when these suppression effects become important. We use the FWHM of the primary beam at the mean redshift $z_0$:
\begin{equation}
    \frac{1}{2\sigma_{\rm k}^2}=\frac{X^2\sigma^2}{4}\Rightarrow \mathrm{FWHM}_{\rm k} = \frac{8\sqrt{2}\ln 2}{X\cdot\mathrm{FWHM}}=0.16\ \text{Mpc}^{-1},
\end{equation}
which implies that visibilities separated more than $\Delta u=47.17\lambda$, can be considered independent \citep{Bull_2015ApJ}. Using a coarser gridding than this value is very useful for the analysis part, because the power spectrum errors can be computed using the sampling variance of the gridded visibilities, i.e. the covariance matrix is approximately diagonal. In addition, coarse gridding reduces the coupling between neighbouring $uv$-modes introduced by the beam convolution.

As a consistency test, we can compute the diagonal terms of the visibility covariance matrix using equation~\eqref{eq: final cov}, 
\begin{equation} \label{eq: diag cov}
\left\langle\left| \widetilde{V}(\boldsymbol{u})
\right|^2 \right\rangle=
\left(\frac{2k_{\rm B}}{\lambda^2}\right)^2
\frac{\delta f N_{\rm ch}\pi\sigma^2}{X^2Y}
P(\boldsymbol{k}_{\perp}, k_\parallel),
\end{equation}
which matches the normalisation correction from \cite{Parsons_2012ApJ,Parsons_2014ApJ} corresponding to the Gaussian primary beam case. Substituting equations~\eqref{eq: diag cov} and \eqref{eq: final cov} into equation~\eqref{eq: gridded power spectrum vis} we get the following result,
\begin{equation}
\label{eq: gridded power spectrum vis 2}
\begin{aligned}
\left\langle \left| \widetilde{V}_{\rm grid}^{j} \right|^2 \right\rangle
&=
\left(\frac{2k_{\rm B}}{\lambda^2}\right)^2
\frac{\delta f\, N_{\rm ch}\, \pi \sigma^2}{X^2 Y N_j^2}
\\[6pt]
&\times
\Bigg\{
\sum_{\boldsymbol{u}_1,\boldsymbol{u}_2\in\mathcal{D}_j}
e^{-\frac{\beta}{2}\left|\boldsymbol{k}_{\perp,1}-\boldsymbol{k}_{\perp,2}\right|^2}
P(\boldsymbol{k}_{\perp,1}, k_\parallel)
\\[4pt]
&\quad +
\frac{1}{2}
\sum_{\boldsymbol{u}_1,\boldsymbol{u}_2\in\mathcal{D}_j}
e^{-\frac{\beta}{2}\left|\boldsymbol{k}_{\perp,1}-\boldsymbol{k}_{\perp,2}\right|^2}
\nabla_{\boldsymbol{k}_\perp}
P(\boldsymbol{k}_{\perp,1}, k_\parallel)
\cdot
\Delta\boldsymbol{k}_\perp
\Bigg\},
\end{aligned}
\end{equation}
where $\boldsymbol{k}_{\perp,1}=\boldsymbol{k}_{\perp,1}(\boldsymbol{u}_1)$, $\boldsymbol{k}_{\perp,2}=\boldsymbol{k}_{\perp,2}(\boldsymbol{u}_2)$, and $\Delta\boldsymbol{k}_\perp=\boldsymbol{k}_{\perp,2}(\boldsymbol{u}_2)-\boldsymbol{k}_{\perp,1}(\boldsymbol{u}_1)$. Assuming that the power spectrum and the gradient of the power spectrum vary negligibly within the grid $j$, we can apply the approximations $P(\boldsymbol{k}_{\perp,1}, k_\parallel)\approx P(\boldsymbol{k}_{\perp}, k_\parallel)\approx P(\boldsymbol{k}_{\perp}^j, k_\parallel)$, and $\nabla_{\boldsymbol{k}_\perp}
P(\boldsymbol{k}_{\perp,1}, k_\parallel)\approx\nabla_{\boldsymbol{k}_\perp}
P(\boldsymbol{k}_{\perp}^j, k_\parallel)$. This allow us to further simplify equation~\eqref{eq: gridded power spectrum vis 2}:
\begin{equation}
\label{eq: gridded power spectrum vis 3}
\begin{aligned}
\left\langle \left| \widetilde{V}_{\rm grid}^{j} \right|^2 \right\rangle
&=
\left(\frac{2k_{\rm B}}{\lambda^2}\right)^2
\frac{\delta f\, N_{\rm ch}\, \pi \sigma^2}{X^2 Y N_j^2}
\\[6pt]
&\times
\Bigg\{
P(\boldsymbol{k}_{\perp}^j, k_\parallel)
\sum_{\boldsymbol{u}_1,\boldsymbol{u}_2\in\mathcal{D}_j}
e^{-\frac{\beta}{2}\left|\boldsymbol{k}_{\perp,1}-\boldsymbol{k}_{\perp,2}\right|^2}
\\[4pt]
&\quad +
\frac{1}{2}\nabla_{\boldsymbol{k}_\perp}
P(\boldsymbol{k}_{\perp}^j, k_\parallel)
\cdot
\sum_{\boldsymbol{u}_1,\boldsymbol{u}_2\in\mathcal{D}_j}
e^{-\frac{\beta}{2}\left|\boldsymbol{k}_{\perp,1}-\boldsymbol{k}_{\perp,2}\right|^2}
\Delta\boldsymbol{k}_\perp
\Bigg\}\\
&=\left(\frac{2k_{\rm B}}{\lambda^2}\right)^2
\frac{\delta f\, N_{\rm ch}\, \pi \sigma^2}{X^2 Y}P(\boldsymbol{k}_{\perp}^j, k_\parallel) \frac{S_j(\mathcal{D}_j)}{N_j^2},
\end{aligned}
\end{equation}
where the gradient term vanishes because, for every pair $(\boldsymbol{u}_1, \boldsymbol{u}_2)$, there is a corresponding pair $(\boldsymbol{u}_2, \boldsymbol{u}_1)$ with the same exponential factor, but opposite $\Delta\boldsymbol{k}_\perp$. These contributions cancel exactly in the sum. Additionally, 
\begin{equation} \label{eq: S_j}
    S_j(\mathcal{D}_j)=\sum_{\boldsymbol{u}_1,\boldsymbol{u}_2\in\mathcal{D}_j}
e^{-\frac{\beta}{2}\left|\boldsymbol{k}_{\perp,1}-\boldsymbol{k}_{\perp,2}\right|^2}=\sum_{\boldsymbol{u}_1,\boldsymbol{u}_2\in\mathcal{D}_j}
e^{-\pi^2\sigma^2\left|\boldsymbol{u}_1-\boldsymbol{u}_2\right|^2}.
\end{equation}

In the limit of small grid sizes ($\Delta u \ll 50\lambda$), the exponential terms corresponding to $\boldsymbol{u}_1\neq\boldsymbol{u}_2$ inside equation~\eqref{eq: S_j} are approximately one. In that situation, the sum is equal to the number of terms of the sum, $S_j(\mathcal{D}_j)\approx N_j^2$. This implies that the normalisation is independent of the positions of the baselines, and equation~\eqref{eq: gridded power spectrum vis 3} reduces to the usual normalisation correction.

\subsection{Testing the normalisation correction against simulations}
\begin{figure*}
    \centering
    \includegraphics[width=\textwidth]{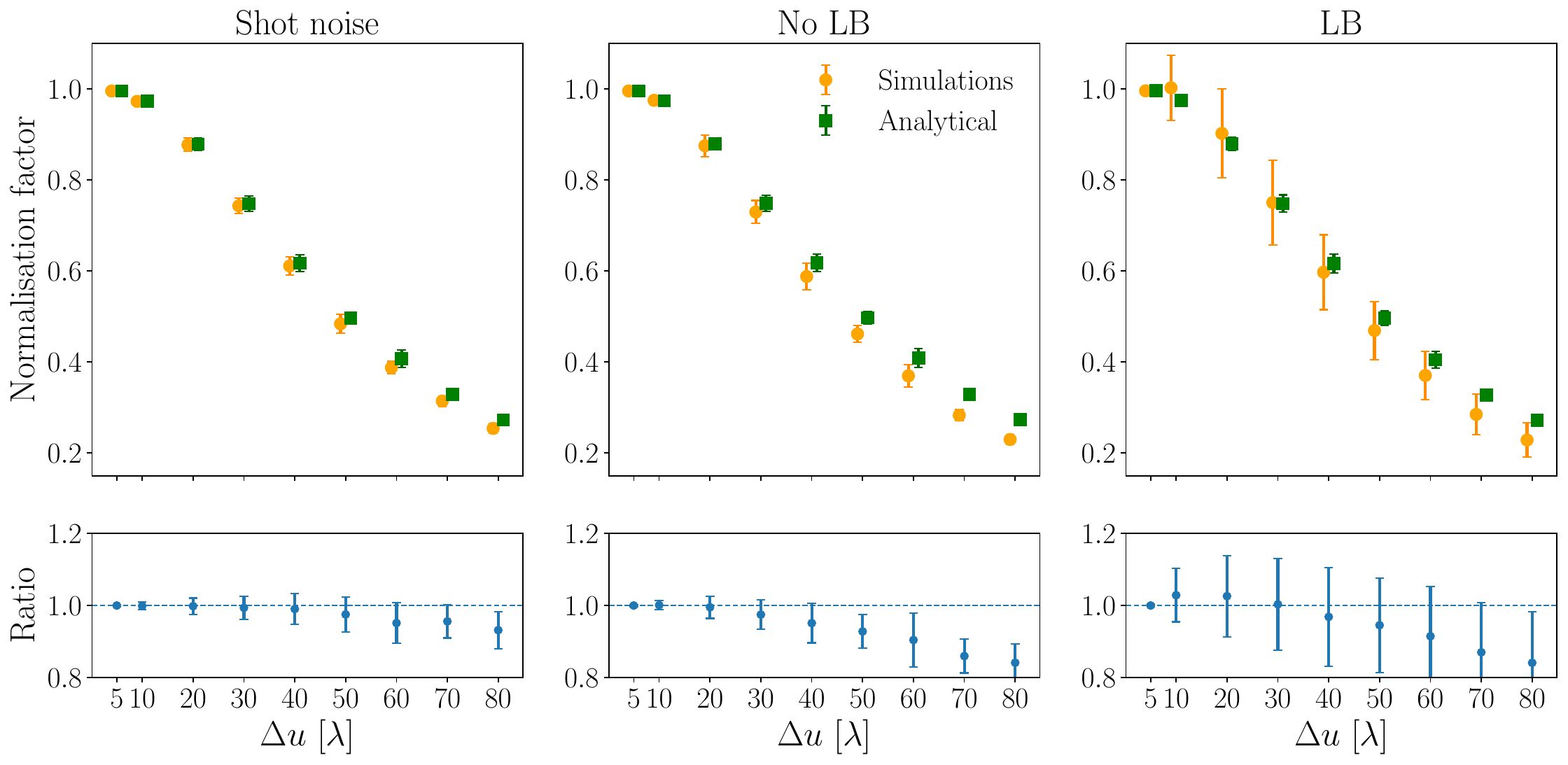}
    \caption{(Top panels) Grid size dependent normalisation computed with the simulation and analytical estimators, for the different HI scenarios as a function of $\Delta u$. (Bottom panels) Ratio between the analytical and simulations based estimators. The different HI scenarios are shot noise (left panel), HI clustering without line broadening (middle panel), and HI clustering with line broadening (right panel). We have excluded the first two bins for both estimators, as they are cosmic variance dominated. The errors for the normalisation factors are computed from the sampling variance of the different bins.}
    \label{fig:norm_factor}
\end{figure*}

\begin{figure}
    \centering
    \includegraphics[width=\columnwidth]{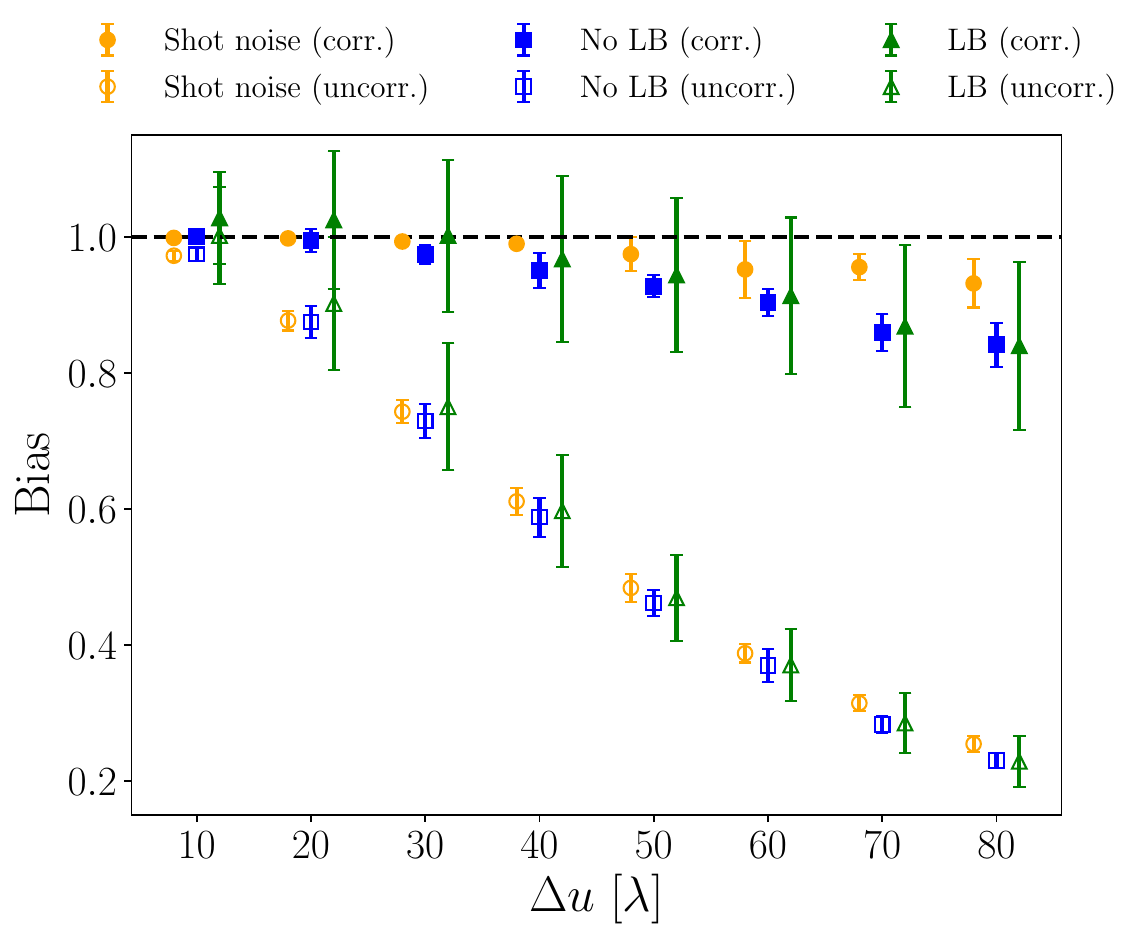}
    \caption{Grid-dependent bias as a function of the grid-size $\Delta u$ for the three scenarios: HI shot noise (orange), HI clustering without line broadening (blue), and  HI clustering with line broadening (green). The filled points correspond to the bias after correction, whereas the empty points correspond to the bias before applying any correction. The error bars are computed from the sampling variance of the different bins.
    }
    \label{fig:bias_norm}
\end{figure}

In this section, we test the performance of the new normalisation correction against simulations in three scenarios: HI shot noise, HI clustering + shot noise and HI clustering + shot noise + line broadening. The objective is to test whether the grid-dependent correction $S_j(\mathcal{D}_j)/N_j^2$ is capturing properly the dependence with the grid size. In that way, we are targeting $\Delta u$-dependent effects and not overall normalisation mismatch. We compute the grid-dependent correction and test for biases after applying the grid-dependent correction, all as a function of the grid size $\Delta u$.

The grid-dependent correction is estimated from simulations, $\text{Norm-simulations}(\Delta u)$, and from our analytical estimator, $\text{Norm-analytical}(\Delta u)$:
\begin{equation} \label{eq: norm-sims}
    \text{Norm-simulations}(\Delta u) = \frac{1}{|\mathcal{D}_\alpha|}\sum_{\alpha\in\mathcal{D}_\alpha}\frac{\hat{P}_{\Delta u,\text{d}}^\alpha}{\hat{P}_{5\lambda,\text{d}}^{\alpha, \text{corr}}},
\end{equation}

\begin{equation} \label{eq: norm-analytical}
    \text{Norm-analytical}(\Delta u) = \frac{1}{|\mathcal{D}_\alpha|}\sum_{\alpha\in\mathcal{D}_\alpha}\frac{\hat{P}_{\Delta u,\text{d}}^\alpha}{\hat{P}_{\Delta u,\text{d}}^{\alpha, \text{corr}}},
\end{equation}
where $\mathcal{D}_\alpha$ are the domain of band powers, $\hat{P}_{\Delta u,\text{d}}^\alpha$ is the 1D power spectrum for the band power $\alpha$ and for grid size $\Delta u$, and $\hat{P}_{\Delta u,\text{d}}^{\alpha, \text{corr}}$ is the 1D power spectrum for which we have applied the grid-dependent correction $S_j(\mathcal{D}_j)/N_j^2$. For the simulations based estimator, we use the $5\lambda$ gridded power spectrum because the grid dependent correction is close to 1 (0.995), meaning that visibility decorrelation has a negligible effect on it (although we still apply a correction for that).

In Fig.~\ref{fig:norm_factor}, we show the grid-dependent correction from simulations using equation~\eqref{eq: norm-sims}, and from the analytical correction using \eqref{eq: norm-analytical}. We find a good agreement between simulations and analytical correction, showing that grid-dependent correction becomes important for $\Delta u>10\lambda$. There is a small mismatch for large grid sizes, which we attribute to the assumptions made in the derivation of the analytical correction, which are the first order Taylor expansion of the cosmological power spectrum, constant power spectrum and gradient of the power spectrum within each cell, and the Gaussian primary beam assumption which is broken due to the limited field of view of $4\times 4$ deg$^2$.

The grid-dependent bias is estimated before and after applying the analytical correction. The former is calculated using equation~\eqref{eq: norm-sims} and the latter, $\text{Bias-corrected}(\Delta u)$, is given by,
\begin{equation}
    \text{Bias-corrected}(\Delta u) = \frac{1}{|\mathcal{D}_\alpha|}\sum_{\alpha\in\mathcal{D}_\alpha}\frac{\hat{P}_{\Delta u,\text{d}}^{\alpha, \text{corr}}}{\hat{P}_{5\lambda,\text{d}}^{\alpha, \text{corr}}},
\end{equation}
where $\mathcal{D}_\alpha$ are the domain of band powers, $\hat{P}_{\Delta u,\text{d}}^\alpha$ is the 1D power spectrum for the band power $\alpha$ and for grid size $\Delta u$, and $\hat{P}_{\Delta u,\text{d}}^{\alpha, \text{corr}}$ is the 1D power spectrum for which we have applied the grid-dependent correction $S_j(\mathcal{D}_j)/N_j^2$.

In Fig.~\ref{fig:bias_norm}, we present the bias as a function of grid size for the three scenarios under study. First, we observe that not correcting for the grid-dependent normalisation can produce large biases, for instance, more than 60 \% biases for $\Delta u=60\lambda$. If not taken properly into account it can bias cosmological parameter estimation significantly. Additionally, our analytical normalisation correction is able to correct for most of the primary beam decorrelation, leading to unbiased results for $\Delta u\leq 30$. Small biases appear at larger grid sizes, where the assumptions underlying the analytical correction are better satisfied for smaller grids. Nevertheless, for $\Delta u=60\lambda$ the bias after correction is less than $10\%$, while before correction it is greater than $60\%$. As the focus of this analysis is foreground removal, small biases are not a primary concern, and we grid our simulations with $\Delta u=60\lambda$. Although beyond of the scope of this work, this study suggests that the best way to correct for decorrelation is to use small grid sizes, and use the analytical normalisation derived in this paper to correct for it. This would imply the necessity to correct for mode mixing, and to estimate the power spectrum errors taking into account correlations between grids, which we defer to future work.

\section{Marchenko-Pastur distribution} \label{ap:MP_dist}

\begin{figure}
    \centering
    \includegraphics[width=\columnwidth]{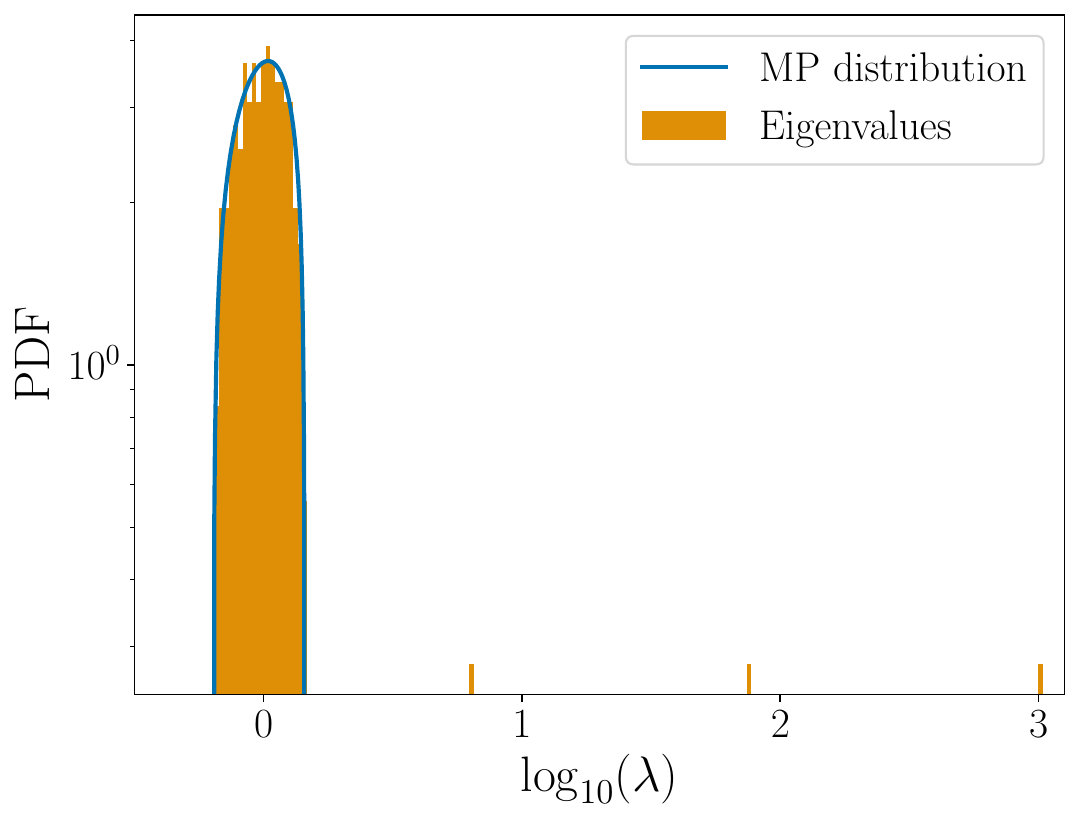}
    \caption{Eigenvalue decomposition of the data covariance matrix (orange) compared to the theoretical Marchenko-Pastur (MP) distribution (blue) for the MeerKAT gridded simulation. Since the histogram is plotted in terms of $\log_{10}\lambda$, the corresponding probability density is given by $f_{\log_{10}\lambda}(\log_{10}\lambda)=\ln (10)\lambda f_\lambda(\lambda)$. Note that 3 eigenvalues larger than $10^3$ are not shown for better visualisation.
    }
    \label{fig:MP_demo}
\end{figure}

In this appendix, we explain in more detail how the Marchenko-Pastur Criterion works when applied to visibility data. Consider a matrix of independent identically distributed random observations, $\boldsymbol{X}=\boldsymbol{V}-\langle\boldsymbol{V}\rangle\in \mathbb{C}^{N_{\rm ch}\times N_{\rm vis}}$, with mean zero and variance $\sigma^2<\infty$. The covariance matrix 
\begin{equation}
    \boldsymbol{C}=\frac{1}{N_{\rm vis}-1}\boldsymbol{X}\boldsymbol{X}^\dagger,
\end{equation}
is complex with dimensions $N_{\rm ch}\times N_{\rm ch}$. The distribution of eigenvalues $\lambda$  of the covariance matrix $\boldsymbol{C}$ asymptotically converge (as \mbox{$N_{\rm vis}-1,N_{\rm ch}\rightarrow\infty$}, with $1<\frac{N_{\rm vis}-1}{N_{\rm ch}}<\infty$) to the Marchenko-Pastur (MP) probability density function,
\begin{equation} \label{eq:MP}
f_\lambda(\lambda)= \left\{
\begin{aligned}
&\dfrac{N_{\rm vis}-1}{N_{\rm ch}}\dfrac{\sqrt{(\lambda_{+}-\lambda)(\lambda-\lambda_{-})}}
      {2\pi \lambda\sigma^2}, &\lambda_{-} \le \lambda \le \lambda_{+},\\
&0, &\text{otherwise},
\end{aligned}
\right.
\end{equation}
where the maximum and minimum expected eigenvalues are 
\begin{equation}
    \lambda_\pm = \sigma^2\left(1\pm\sqrt{\frac{N_{\rm ch}}{N_{\rm vis}-1}}\right)^2,
\end{equation}
and because of whitening, for the GVILC method the variance $\sigma^2=1$. Eigenvalues $\lambda\in [\lambda_-, \lambda_+]$ are consistent with the signal-plus-noise subspace that we used as a covariance matrix prior, and eigenvalues $\lambda\notin [\lambda_-, \lambda_+]$ belong to the foreground subspace. In Fig.~\ref{fig:MP_demo} we show the eigenvalue distribution for the MeerKAT gridded data covariance. We can see from this plot that the eigenvalues around $\lambda=1$ follow a MP distribution bounded by $[\lambda_-, \lambda_+]$. Eigenvalues greater than $\lambda_+=1.44$ are attributed to the foreground subspace, while the rest belong to the signal-plus-noise subspace. The value of $\lambda_+$ depends exclusively on the number of channels and the number of visibilities per annulus, and consequently on the size of the annulus. This method allows us not only to determine the number of components to remove, but also serves as a consistency test for the prior as well as possible systematics present in the data.

\bsp	
\label{lastpage}
\end{document}